\documentclass[11pt,a4paper]{article}

\usepackage{styling}
\usepackage[nomessages]{fp}
\usetikzlibrary{arrows.meta}

\hypersetup{
  pdftitle={Balanced Allocations with Incomplete Information: The Power of Two Queries},
  pdfauthor={Dimitrios Los, Thomas Sauerwald}
  }

\title{Balanced Allocations with Incomplete Information: \\ The Power of Two Queries}

\author[]{Dimitrios Los\thanks{\texttt{dimitrios.los@cl.cam.ac.uk}} }
\author[]{Thomas Sauerwald\thanks{\texttt{thomas.sauerwald@cl.cam.ac.uk}. The author was supported by the ERC grant ``Dynamic March''. Part of this work was done while visiting Hasso-Plattner Institute, Potsdam, Germany.}}
\affil[]{Department of Computer Science \& Technology, University of Cambridge}

\begin{document}

\maketitle

\thispagestyle{empty}

\begin{abstract}
We consider the allocation of $m$ balls into $n$ bins with incomplete information. In the classical \TwoChoice process 
a ball first queries the load of {\em two} randomly chosen bins and is then placed in the least loaded bin. In our setting, each ball also samples two random bins but can only estimate a bin's load by sending  {\em binary queries} of the form ``Is the load at least the median?'' or ``Is the load at least $100$?''. 

For the lightly loaded case $m=\Oh(n)$, Feldheim and Gurel-Gurevich~(2021) showed that with one query it is possible to achieve a maximum load of $\Oh(\sqrt{\log n/\log \log n})$, and they also pose the question whether a maximum load of $m/n+\Oh(\sqrt{\log n/\log \log n})$ is possible for any $m = \Omega(n)$. In this work, we resolve this open problem by proving a lower bound of $m/n+\Omega( \sqrt{\log n})$ for a fixed $m=\Theta(n \sqrt{\log n})$, and a lower bound of $m/n+\Omega(\log n/\log \log n)$ for some $m$ depending on the used strategy. %

We complement this negative result by proving a positive result for multiple queries. In particular, we show that with only {\em two} binary queries per chosen bin, there is an oblivious strategy which ensures a maximum load of $m/n+\Oh(\sqrt{\log n})$ for any $m \geq 1$. 
Further, for any number of $k = \Oh(\log \log n)$ binary queries, the upper bound on the maximum load improves to $m/n + \Oh(k(\log n)^{1/k})$ for any $m \geq 1$. 

This result for $k$ queries has several interesting consequences: 
$(i)$ it implies new bounds for the $(1+\beta)$-process  introduced by Peres, Talwar and Wieder~(2015), $(ii)$ it leads to new bounds for the graphical balanced allocation process on dense expander graphs,
and $(iii)$ it recovers 
and generalizes the bound of $m/n+\Oh(\log \log n)$ on the maximum load achieved by the \TwoChoice process, including the heavily loaded case $m=\Omega(n)$ which was derived in previous works by Berenbrink, Czumaj, Steger and V\"{o}cking (2006) as well as Talwar and Wieder~(2014). %

One novel aspect of our proofs 
is the use of multiple super-exponential potential functions, which might be of use in future work.
\end{abstract}

\thispagestyle{empty}

\newpage

\tableofcontents

\newpage

\setcounter{page}{1}
\newpage

\section{Introduction}

We study balls-and-bins processes where the goal is to allocate $m$ balls (jobs) sequentially into $n$ bins (servers). The balls-and-bins framework a.k.a.~balanced allocations~\cite{ABK99}~is a very popular and simple framework for various resource allocation and storage problems such as load balancing, scheduling or hashing (see surveys~\cite{MRS01,W17} for more details). In most of these settings, the goal is to find a simple allocation strategy that results in an as balanced allocation as possible.

It is a classical result that if each ball is placed in a random bin chosen independently and uniformly (called \OneChoice), then the maximum load is $\Theta( \log n / \log \log n)$ \Whp\footnote{In general, with high probability refers to probability of at least $1 - n^{-c}$ for some constant $c > 0$.} for $m=n$, and $m/n + \Theta( \sqrt{ (m/n) \log n})$ \Whp~for $m \gg n$. 
Azar et al.~\cite{ABK99} (and implicitly Karp et al.~\cite{KLM96}) proved that if each ball is placed in the lesser loaded of {\em two} randomly chosen bins, then the maximum load drops to $\log_2 \log n + \Oh(1)$ \Whp, if $m=n$. This dramatic improvement of \TwoChoice is widely known as ``power of two choices'', and similar ideas have been applied to other problems including routing, hashing and randomized rounding~\cite{MRS01}. 

While for $m=n$ a wide range of different proof techniques 
have been employed, the heavily loaded case $m \gg n$ turns out to be much more challenging. In a seminal paper~\cite{BCSV06}, Berenbrink et al.~proved a maximum load of $m/n+\log_2 \log n +\Oh(1)$ \Whp~using a sophisticated Markov chain analysis. A simpler and more self-contained proof was recently found by Talwar and Wieder~\cite{TW14}, giving a slightly weaker upper bound of $m/n+\log_2 \log n+\Oh(\log\log \log n)$ for the maximum load and at the cost of a larger error probability. 

In light of the dramatic improvement of \TwoChoice (or \DChoice) over \OneChoice, it is important to understand the robustness of these processes. For example, in a concurrent environment, information about the load of a bin might quickly become outdated or communication with bins might be restricted. Also, acquiring always {\em $d \geq 2$} uncorrelated choices might be costly in practice. Motivated by this, Peres et al.~\cite{PTW15} introduced the $(1+\beta)$-process, in which two choices are available with probability $\beta$, and otherwise only one. Thus, the $(1+\beta)$-process interpolates nicely between \TwoChoice and \OneChoice, and surprisingly, a bound on the gap between maximum and average load
of $\Oh(\log n/\beta)$ \whp was shown, which also holds in the heavily loaded case where $m=\Omega(n)$. The $(1+\beta)$-process has been also connected to other processes, including population protocols~\cite{AGR21}, balls-and-bins with weights~\cite{TW07,TW14} and, most notably, graphical balanced allocation~\cite{KP06,PTW15,ANS20,BF21}. In this graphical model, bins correspond to vertices of a graph, and for each ball we sample an edge uniformly at random and place the ball in the lesser loaded bin of the two endpoints. 
Also the results and analysis techniques of the $(1+\beta)$-process have found important connections to population protocols~\cite{AGR21} and balls-and-bins with weights~\cite{TW07,TW14}.

\textbf{Our Model.} In this work, we will investigate the following model. At each step, a ball is allowed to sample two random bins chosen independently and uniformly, however, the load comparison between the two bins will be performed under incomplete information. This may capture scenarios in which it is costly to communicate or maintain the exact load of a bin.

Specifically, we assume that each ball is allowed to send up to $k$ binary queries to each of the two bins, inquiring about their current load. These queries can either be about the absolute load (i.e., is the load at least $100$?), which we call {\em threshold processes}, or about the relative load (i.e., is the load at least the median?), which we call {\em quantile processes}.

We will distinguish between {\em oblivious} and {\em adaptive} allocation strategies. For an {\em adaptive} strategy, the queries may depend on the current load configuration (i.e., the full history of the process), whereas in the {\em oblivious} setting, queries may depend only on the current time-step. %

\textbf{Our Results.} 
For the case of $k=1$ query,  Feldheim and Gurel-Gurevich~\cite{FG18} proved a bound of $\Oh(\sqrt{\log n/\log \log n})$ on the gap (between the maximum and average load) in the lightly loaded case $m=\Oh(n)$. In the same work, the authors suggest that the same bound might be also true in the heavily loaded case~\cite[Problem 1.3]{FG18}. In this work, we disprove this by showing a lower bound of $\Omega(\sqrt{\log n})$ on the gap for $m=\Theta(n \sqrt{\log  n})$ (\cref{thm:new_adaptive_quantile_lower_bound}). We also prove a lower bound of $\Omega( \log n/ \log \log n)$ on the gap, which holds for at least $\Omega(n \log n/ \log \log n)$ of the time-steps in $[1,n \log^2 n]$ (\cref{cor:large_gap_in_nlog2n_interval}). These two lower bounds hold even for the more general class of adaptive strategies. The basic idea behind all these lower bounds is that, as $m \gg n$, one query is not enough to prevent the process from emulating the \OneChoice process on a small scale.

It is natural to ask whether we can get an improved performance by allowing more, say {\em two} queries per bin. We prove that this is indeed the case, establishing a ``power of two-queries'' result. Specifically, we show in \cref{thm:new_multiple_quantiles} that for any $k=\Oh(\log \log n)$, there is an allocation process with $k$ uniform quantiles (i.e., queries only depend on $n$, but not on the time $t$) that achieves for any $m \geq 1$:
\[
 \Pro{ \Gap(m) = \Oh\Big( k \cdot ( \log n )^{1/k} \Big) } \geq 1-n^{-3}.
\]
Comparing this for $k=2$ to the lower bounds for $k=1$, we indeed observe a ``power of two-queries'' effect. For $k = \Theta(\log \log n)$, the gap even becomes $\Oh(\log \log n)$, which matches the \TwoChoice result up to a multiplicative constant~\cite{BCSV06,TW14}. Hence, for large values of $k$, the process approximates \TwoChoice, whereas for $k=1$ it resembles the $(1+\beta)$-process. Indeed, the same upper bound of $\Oh(\log n)$ follows from the analysis of the $(1+\beta)$-process~(\cref{thm:median_quantile_logn_gap_whp}).

We also prove new upper bounds on the gap of the $(1+\beta)$ process with $\beta$ close to $1$ by relating it to a relaxed quantile process (\cref{thm:oneplusbeta_application}). We show that these in turn imply new upper bounds on the graphical balanced allocation on dense expander graphs, making progress towards Open Question 2 in~\cite{PTW15} (\cref{cor:thomas}).

\textbf{Applications and Implications on other Models.}
A direct implementation of the $k$-quantile protocol in practice requires to maintain some \emph{global} information about the load configuration (that is, the exact, or at least, the approximate, values of the quantiles). If this can be achieved, then the results of $k$-quantile for $k \geq 2$ demonstrate that a sub-logarithmic gap is possible -- even with very limited \emph{local} information about the individual bin loads.

In addition, our study of the $k$-quantile process also leads to new results for some previously studied allocation processes. We demonstrate that a $(1+\beta)$-process for $\beta$ close to $1$ is majorized by a (relaxed version of the) $k$-quantile process. For any $\beta=1-o(1)$, this leads to a sub-logarithmic bound on the gap, and if $\beta=1-1/\poly(n)$, we recover the $\Oh(\log \log n)$ gap from the \TwoChoice process. 
Secondly, we use a similar majorization argument to analyze graphical balanced allocation, which has been studied in several works on different graphs \cite{PTW15,KP06,ANS20,BF21}.
Specifically, we prove for dense and strong expander graphs (including random $d$-regular graphs for $d=\poly(n)$) a gap of $\Oh(\log \log n)$.
To the best of our knowledge, these are the first sub-logarithmic gap bounds in the heavily loaded case for the $(1+\beta)$-process and graphical balanced allocation (apart from $\beta=1$ or the graph being a clique, both equivalent to \TwoChoice).

\textbf{Further Related Work.}
Our model for $k = 1$ is equivalent to the $d$-\textsc{Thinning} process for $d=2$, where for each ball, a random bin is ``suggested'' and based on the bin's load, the ball is either allocated there or it is allocated to a second bin chosen uniformly and independently. 
Generalizing the results of \cite{FG18} for $d=2$, Feldheim and Li~\cite{FL20} also analyzed an extension of $2$-\textsc{Thinning}, called $d$-\textsc{Thinning}. For $m = \Oh(n)$, they proved tight lower and upper bounds, resulting into an achievable gap of $(d+o(1)) \cdot \left( d \log n / \log \log n \right)^{1/d}$. %
Iwama and Kawachi~\cite{IK04} analyzed a special case of the threshold process for $m = n$ and for $k$ equally-spaced thresholds, proving a gap of 
$
(k + \Oh(1)) \sqrt[k + 1]{(k+1) \frac{\log{n}}{\log \left((k + 1) \log{n}\right)}}.
$
Mitzenmacher~\cite[Section 5]{M96} coined the term \textit{weak threshold process} for the two threshold process in a queuing setting, where a customer chooses two queues uniformly at random and enters the first one iff it is shorter than $T$. This and previous work~\cite{ELZ86, KRZ13, Z88} analyze the case of a fixed threshold for queues and they do not directly imply results for the heavily loaded case.

In another related work, Alon et al.~\cite{AGL10} established for the case $m = \Theta(n)$ a trade-off between the number of bits used for the representation of the load and the number of $d$ bin choices. This is a more restricted case of having a fixed number of non-adaptive queries. For $d=2$, Benjamini and Makarychev~\cite{IM12} obtained tight results for the gap, using a process very similar to the threshold process, but considering the case $m=\Theta(n)$ only. %

Czumaj and Stemann~\cite{CS01} investigated general allocation processes, in which the decision whether to take a second (or further) sample depends on the load of the lightest sampled bin. They obtained strong and tight guarantees, but they assume the full information model and also $m=\Oh(n)$ (see~\cite{BKSS13} for some results for $m \geq n$).
Other processes with inaccurate (or outdated) information about the load of a bin have been studied in an asynchronous environment~\cite{ABK18} or a batch-based allocation~\cite{BCE12}. 
However, the obtained bounds on the gap are only $\Oh(\log n)$. Other protocols that study the communication between balls and bins in more detail are~\cite{LW11,LPY19,GM10, S96}, but they assume that a ball can sample more than two bins.

After an earlier version of this paper went online, Feldheim, Gurel-Gurevich and Li~\cite{FGL21} extended the lower bounds for $2$-\textsc{Thinning} when $m = \Oh(n \log^2 n)$ and also provided an adaptive thinning process that matches the $\Omega(\log n/\log \log n)$ lower bound proved in this paper. Also, Los, Sauerwald and Sylvester~\cite{LSS21} proved that $\Threshold(m/n)$ achieves \Whp~a $\Theta(\log n)$ gap.

\textbf{Organization.}
This paper is organized as follows. In \cref{sec:notation}, we introduce our model more formally in addition to some notation used in the analysis. In \cref{sec:lower}, we present our lower bounds on processes with one query. In \cref{sec:one_quantile}, we present the upper bound for the quantile process with one query.
In \cref{sec:new_upper_bound_for_k_queries},
we present a generalized upper bound for $k \geq 2$ queries. \cref{sec:applications_abstract} contains our applications to $(1+\beta)$-process and graphical balanced allocations. We close in \cref{sec:conclusion} by summarizing our main results and pointing to some open problems. 
We also briefly present some experimental results in \cref{sec:experiments}. In \cref{sec:relations}, we formally relate the new quantile (and threshold) processes to each other and to other processes studied before (see~\cref{fig:process_overview} for an overview). %

\section{Notation, Definitions and Preliminaries}~\label{sec:notation}
We sequentially allocate $m$ balls (jobs) into $n$ bins (servers).
The load vector at step $t$ is $x^{(t)}=(x_1^{(t)},x_2^{(t)},\ldots,x_n^{(t)})$ and in the beginning, $x_i^{(0)} = 0$ for $i \in [n]$.
Also $y^{(t)}=(y_1^{(t)},y_2^{(t)},\ldots,y_n^{(t)})$ will be the permuted load vector, sorted decreasingly in load.
This can be described by \textbf{ranks}, which form a permutation of $[n]$ that satisfies
$
  r = \Rank^{(t)}(i) \Rightarrow y_r^{(t)}=x_i^{(t)}.
$
Following previous work, we analyze allocation processes in terms of the 
\[
\Gap(t):= \max_{1 \leq i \leq n} x_i^{(t)} - \frac{t}{n} = y_1^{(t)} - \frac{t}{n},
\]
i.e., the difference between maximum and average load at time $t \geq 0$. It is well-known that even for \TwoChoice, the gap between maximum and minimum load is $\Omega(\log n)$ for large $m$ (e.g.~\cite{PTW15}). Here our focus is on sequential allocation processes based on binary queries. That is, at each step $t$:
\begin{enumerate}\itemsep-3pt
\item Sample {\em two} bins independently and uniformly at random (with replacement).
\item Send the same $k$ binary queries to each of the two bins about their load. 
\item Allocate the ball in the lesser loaded one of the two bins (based on the answers to the queries), breaking ties randomly.
\end{enumerate}

\begin{figure}[t]
  \centering\includegraphics[scale=0.36]{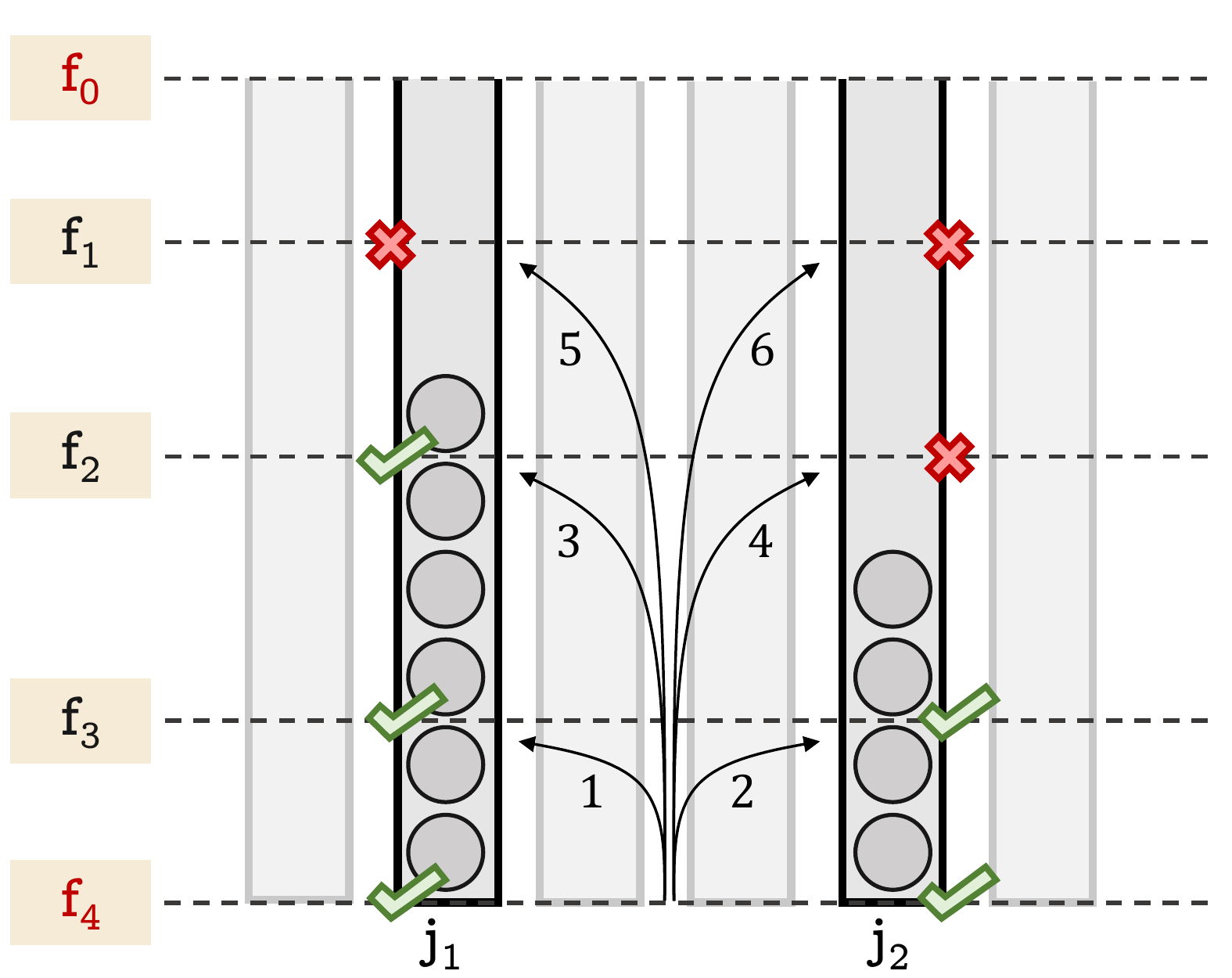}\quad\includegraphics[scale=0.36]{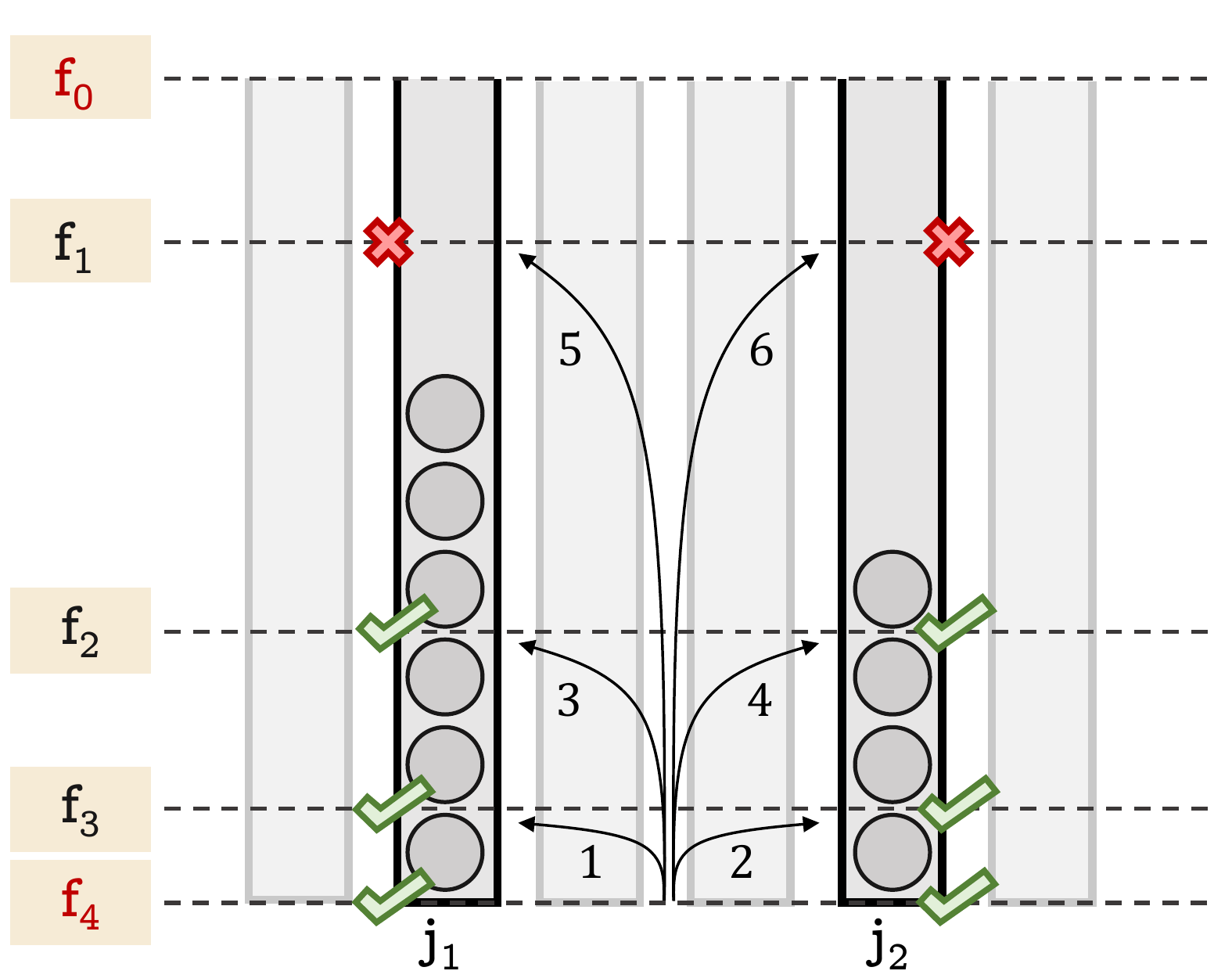}
  
  \caption{Example allocation using two $3$-threshold processes $(f_1,f_2,f_3)$. (\textbf{Left}): The ball is allocated in $j_2$, since $i_1 =2$ and $i_2 = 3$. (\textbf{Right}): For a different choice of thresholds, the process may not be able to differentiate the two loaded bins, so the ball will be allocated at random. }
  \label{fig:threshold_process_visualisation}
\end{figure}
We first describe \textbf{threshold-based processes}, where queries to each bin $j$ are of the type ``Is $x_j^{(t)} \geq f(t)$'' for some function $f$ that maps into $\mathbb{N}$. For example, we could ask whether the load of a bin is at least the average load. Formally, we denote such a process with two choices and $k$ queries by $\textsc{Threshold}(f_1,f_2,\ldots,f_k)$, where $f_1 > f_2 > \ldots > f_k$ are $k$ different load thresholds, that may depend on the time $t$, in which case we write $f_i(t)$. After sending all $k$ queries to a bin $j$, we receive the correct answers to all these queries and then we determine the $i$ ($0 \leq i \leq k$) for which,
\[
  x_j^{(t)} \in (f_{i+1}(t), f_{i}(t) ],
\]
where $f_{0}(t)= +\infty$ and $f_{k+1}(t) = -\infty$ (see \cref{fig:threshold_process_visualisation}). After having obtained two such numbers $i_1,i_2 \in \{0,1,\ldots,k\}$, one for each bin $j_1$ and $j_2$, we will allocate the ball ``greedily'', i.e., into $j_1$ if $i_1 < i_2$ and into $j_2$ if $i_1 > i_2$. If $i_1=i_2$, then we will break ties randomly. 

We proceed to define \textbf{quantile-based processes}. In this process, queries to a bin $j$ are of the type ``Is $x_j^{(t)} \geq y_{\delta(t) \cdot n}^{(t)}$?'', for some function $\delta$ that maps $t$ into $\{1/n,2/n,\ldots,1\}$. For example if $\delta=1/2$, we are querying whether the load of a bin is at most the median load. We denote such a process with two choices and $k$ queries by $\Quantile(\delta_1,\delta_2,\ldots,\delta_k)$, where $\delta_1 < \delta_2 < \ldots < \delta_k$ are $k$ different quantiles, which may depend on the time $t$. After sending all $k$ queries to a bin $j$ in step $t$, we receive the correct answers and then we determine the $i$ ($0 \leq i \leq k$) for which,
\[
   \Rank^{(t)}(j) \in ( \delta_{i}(t) \cdot n, \delta_{i+1}(t) \cdot n],
\]
where $\delta_0(t) = 0$ and $\delta_{k+1}(t) = 1$. As before, we allocate the ball to the bin with smaller $i$-value and break ties randomly.

\Quantile~and \Threshold~processes can be classified into oblivious processes and adaptive processes, depending on the type of queries. In an \textbf{oblivious process}, the queries $f_1,f_2,\ldots$ (or $\delta_1,\delta_2,\ldots$) may only depend on $t$ (as well as $n$) ---a special case is a \textbf{uniform process} where $\delta_1,\delta_2,\ldots$ are constants (independent of $t$), and the $f_i$'s are of the form $t/n + f_i(n)$.
In an \textbf{adaptive process}, queries in step $t$ may depend on the full history of the process, i.e., the load vector $x^{(t-1)}$, so each query $i$ involves a function $f_i( x^{(t-1)})$, but this must be specified before receiving any answers. In the adaptive setting, a $k$-quantile process can simulate any $k$-threshold process, by setting the quantile to the largest $\delta_i(t)$ such that $y_{\delta_i(t) \cdot n} \leq f_i(t)$ (\cref{lem:any_threshold_process_can_be_simulated_by_adaptive_quantile}).

The \textbf{$d$-\textsc{Thinning} process} \cite{FG18} works as follows. For each ball to be allocated, an overseer can inspect up to $d$ randomly sampled bins in an online fashion, and based on all previous history, can accept or reject each bin (however, one of the $d$ proposed bins must be accepted).

The \textbf{\DChoice process}~\cite{ABK99} (sometimes also called $\textsc{Greedy}[d]$) is the process where, for each ball, $d$ bins are chosen uniformly at random and the ball is placed in the least loaded bin. We will refer to the special case $d = 1$ as the \textbf{\OneChoice process}, and $d=2$ as the \textbf{\TwoChoice process}. The \textbf{$(1 + \beta)$-process}~\cite{PTW15} is the process where each ball is placed with probability $\beta$ according to \TwoChoice and with probability $1-\beta$ according to \OneChoice. 

Finally, in \textbf{graphical balanced allocation} \cite{KP06,PTW15}, we are given an undirected graph $G$ with $n$ vertices corresponding to $n$ bins. For each ball to be allocated, we select an edge $\{u,v\} \in E(G)$  uniformly at random, and place the ball in the lesser loaded bin among $\{u,v\}$. As pointed out in \cite{PTW15}, this setting can be generalized even to hypergraphs.

Following~\cite{PTW15} and generalizing the processes above, \textbf{an allocation process} can be described by a \textbf{probability vector} $p^{(t)} =(p_1^{(t)},p_2^{(t)},\ldots,p_n^{(t)})$ for step $t$, where $p_i^{(t)}$ is the probability for incrementing the load of the $i$-th most loaded bin. Following the idea of \textbf{majorization} (see Lemma~\ref{lem:old_majorisation}), if two processes with (time-invariant) probability vectors $p$ and $q$, for all $i \in [n]$ satisfy $\sum_{j \leq i} p_j \leq \sum_{j \leq i} q_j$, then there is a coupling between the allocation processes with sorted load vectors $y(p)$ and $y(q)$ such that $\sum_{j \leq i} y_i^{(t)}(p) \leq \sum_{j \leq i} y_i^{(t)}(q)$ for all $i \in [n]$ (\textbf{$q$ majorizes $p$}). 

Finally, we define the \textbf{height} of a ball as $i\geq 1$ if it is the $i$\textsuperscript{th} ball added to the bin.

Many statements in this work hold only for sufficiently large $n$, and several constants are chosen generously with the intention of making it easier to verify some technical inequalities.

\section{Basic Relations between Allocation Processes}\label{sec:relations}

In this section we collect several basic relations between allocation processes, following the notion of majorization~\cite{PTW15}. \cref{fig:process_overview} gives a high-level overview of some of these  relations, along the with the derived and implied gap bounds.

\begin{figure}[H]
\begin{center}
\makebox[\textwidth][c]{
\scalebox{0.73}{
\begin{tikzpicture}
[scale=1,
edge/.style={thick,-{Stealth[scale=1.3]}},
rec/.style={rounded corners=6pt,thick,draw=black},
rrec/.style={rounded corners=6pt,thick,fill=red!20!,draw=red}]
\node[rec,anchor=north,thick,fill=black!5] (1) at (5,25)
{
\begin{minipage}{0.3\textwidth}
\begin{center}
\OneChoice \\
$=$ \\
$(1+\beta)$ ($\beta=0$) \\
$=$ \\
\Quantile($\delta_1=1$) 
\end{center}
~\vspace{-0.75em}
\[
 \Gap(m)=\Theta \left( \sqrt{ \frac{m}{n} \log(n) } \right)
\]
\end{minipage}
};

\node[rec,anchor=north,thick,fill=black!5] (2) at (5,3)
{
\begin{minipage}{0.3\textwidth}
\begin{center}
\TwoChoice \\
$=$ \\
$(1+\beta)$ ($\beta=1$) \\
$=$ \\
\Quantile($\frac{1}{n},\frac{2}{n},\ldots,\frac{n-1}{n}$) 
\end{center}
~\vspace{-0.75em}
\[
 \Gap(m)= \log \log n  + \Theta(1)
\]
\end{minipage}
};

\node[rec,anchor=north] (b1) at (-4,20)
{
\begin{minipage}{0.25\textwidth}
\begin{center}
$(1+\beta)$ ($\beta =o(1)$)
\end{center}
\[
 \Gap(m)=\Theta \left( \frac{\log n}{\beta}  \right)
\]
\end{minipage}
};

\node[rec,anchor=north] (b2) at (-4,17)
{
\begin{minipage}{0.25\textwidth}
\begin{center}
$(1+\beta)$ ($\beta \in (0,1)$)
\end{center}
\[
 \Gap(m)=\Theta \left( \log n \right)
\]
\end{minipage}
};

\node[rec,anchor=north] (b3) at (-4,13)
{
\begin{minipage}{0.35\textwidth}
\begin{center}
$(1+\beta)$ ($\beta = 1 - 2^{-0.5 (\log n)}$)
\end{center}
\[
 \Gap(m)=\Oh\left( \sqrt{\log n} \right)
\]
\end{minipage}
};

\node[rec,anchor=north] (b4) at (-4,10)
{
\begin{minipage}{0.35\textwidth}
\begin{center}
$(1+\beta)$ ($\beta = 1 - 2^{-0.5 (\log n)^{2/3}}$)
\end{center}
\[
 \Gap(m)=\Oh\left( (\log n)^{1/3} \right)
\]
\end{minipage}
};

\node[rec,anchor=north] (b5) at (-4,7)
{
\begin{minipage}{0.35\textwidth}
\begin{center}
$(1+\beta)$ ($\beta = 1-\frac{1}{\operatorname{poly}(n)}$)
\end{center}
\[
 \Gap(m)=\Theta(\log \log n)
\]
\end{minipage}
};

\node[rec,anchor=north] (q2) at (14,18)
{
\begin{minipage}{0.3\textwidth}
\begin{center}
\Quantile($\delta_1$), $\delta_1 \in (0,1)$) \\
$= $ \\
$2$-\textsc{Thinning} 
\end{center}
\vspace{-0.75em}
\begin{align*}
 \Gap(m) &=\Oh \left( \log n \right) \\
 \Gap(m) &=\Omega \left( \frac{\log n}{\log \log n} \right) 
\end{align*}
~\vspace{-0.75em}~
\end{minipage}
};

\node[rec,anchor=north] (q3) at (14,13)
{
\begin{minipage}{0.32\textwidth}
\begin{center}
\Quantile($2^{-0.5 \sqrt{\log n}},2^{-1}$)
\end{center}
\[
 \Gap(m)=\Oh( \sqrt{\log n})
\]
\end{minipage}
};

\node[rec,anchor=north] (q4) at (14,10)
{
\begin{minipage}{0.48\textwidth}
\begin{center}
\Quantile($2^{-0.5(\log n)^{2/3}},2^{-0.5(\log n)^{1/3}},2^{-1}$)
\end{center}
\[
 \Gap(m)=\Oh( (\log n)^{1/3} )
\]
\end{minipage}
};
\node[rec,anchor=north] (q5) at (14,7)
{
\begin{minipage}{0.34\textwidth}
\begin{center}
\Quantile($\delta_1,\ldots,\delta_{\Theta(\log \log n)}$)
\end{center}
\[
 \Gap(m)=\Theta( \log \log n)
\]
\end{minipage}
};

\draw[edge] (b1) to (b2); 
\draw[edge] (b2) to (b3); 
\draw[edge] (b3) to (b4); 
\draw[edge,dotted] (b4) to (b5); 
\draw[edge] (b5.south) to (2.west);

\draw[edge] (q2)  to (q3); 
\draw[edge] (q3)  to (q4); 
\draw[edge,dotted] (q4)  to (q5); 
\draw[edge] (q5.south) to (2.east); 

\draw[-stealth,blue,line width=3pt] (1) to node[pos=0.5,below=7pt,thick,sloped]{\huge{Power of Two Choices}} (2);

\draw[-stealth,red,line width=3pt,in=-180,out=180] (q2.165) to node[pos=0.5,left=0.5cm,below=0.6cm,thick,rotate=-90]{\huge{Power of Two Queries}} (q3.175);

\draw[-{Stealth[scale=1.3]}, thick] (b2.5) to [bend right=-10] node[pos=0.96,above=2pt,rotate=-4]{
\begin{minipage}{0.3\textwidth}
majorizes
\end{minipage}
}
 (q2.150);
 
 \draw[{Stealth[scale=1.3]}-,  thick] (b3.5) to [bend right=-10] node[pos=0.32,above=2pt,rotate=5]{
\begin{minipage}{0.3\textwidth}
majorizes
\end{minipage}
}
 (q3.185);
 
  \draw[{Stealth[scale=1.3]}-, thick] (b4.5) to [bend right=-10] node[pos=0.35,above=2pt,rotate=5]{
\begin{minipage}{0.3\textwidth}
majorizes
\end{minipage}
}
 (q4.180);
 
  \draw[{Stealth[scale=1.3]}-, thick] (b5.5) to [bend right=-10] node[pos=0.32,above=2pt,rotate=5]{
\begin{minipage}{0.3\textwidth}
majorizes
\end{minipage}
}
 (q5.180);

\draw[edge, text width=5cm, align=center] (1.200) to node[pos=0.5,above,sloped]{
increase probability \\ of having a second choice
} (b1.north);
\draw[edge] (1.-20) to node[pos=0.5,above,sloped]{
\begin{minipage}{0.29\textwidth}
increase number of binary queries to estimate bin load
\end{minipage}
} (q2.north);

\end{tikzpicture}
}
}
\end{center}

\caption{Overview of bounds on $\Gap(m)$ for various allocation processes that interpolate between \OneChoice and \TwoChoice. All stated upper bounds are valid for any $m \geq 1$, while lower bounds may only hold for certain ranges of $m$. Some of the majorization results in the figure only hold for a suitable \RelaxedQuantile process.}\label{fig:process_overview}
\end{figure}

Recall that the \TwoChoice probability vector is, for $i \in [n]$:
\[
p_i = \frac{2i - 1}{n^2}.
\]
The $(1+\beta)$ probability vector~\cite{PTW15} interpolates between those of \OneChoice and \TwoChoice, so for any $i \in [n]$,
\[
p_i = (1 - \beta) \cdot \frac{1}{n} + \beta \cdot \frac{2i - 1}{n^2}.
\]
For the process $\Quantile{(\delta_1,\ldots,\delta_k)}$, it is straightforward to verify that the probability vector satisfies for any $i \in [n]$:
\begin{align}
p_i = \begin{cases}
\frac{\delta_1}{n} & 1 \leq i \leq \delta_1 \cdot n, \\
\frac{\delta_1 + \delta_2}{n} & \delta_1 \cdot n < i \leq \delta_2 \cdot n, \\
\ \ \ \vdots & \\
\frac{\delta_{k-1} + \delta_k}{n} & \delta_{k-1} \cdot n < i \leq \delta_k \cdot n, \\
\frac{1 + \delta_k}{n} & \delta_k \cdot n < i.
\end{cases} \label{eq:def_quantile}
\end{align}

The next lemma shows that we can always execute the $\Quantile(\delta)$ in the same way as $2$-\textsc{Thinning}:
\begin{lem}\label{lem:adaptive_quantile_equivalence}
Consider a quantile process $\Quantile(\delta)$ with one query. This process can be always transformed into an equivalent instance of $2$-\textsc{Thinning}: Sample a bin, if its rank is greater than $n \cdot \delta(t)$, then place the ball there; otherwise, place the ball in a randomly chosen bin.
\end{lem}
\begin{proof}
Let $p$ be the probability vector of the $\Quantile(\delta)$ process and let $q$ be the probability vector of the $2$-\textsc{Thinning} process described in the statement of the lemma. We will show that $p = q$. Let $R$ denote the set of bins with rank $> n \cdot \delta(t)$. Let $B_1$ and $B_2$ be the two bin choices at some time step. We consider two cases, based on the rank of bin $i \in [n]$.

\textbf{Case 1 (light bin):} $i$ has rank $> n \cdot \delta(t)$, then
\begin{align*}
p_i & = \frac{1}{2}\Pro{B_1 = i, B_2 \in R} + \frac{1}{2}\Pro{B_1 \in R, B_2 = i} \\ & \quad + \Pro{B_1 = i, B_2 \notin R} + \Pro{B_1 \notin R, B_2 = i} \\
 & = \left(1 - \delta(t)\right) \cdot \frac{1}{n} \cdot \frac{1}{2} \cdot 2 + 2 \cdot \frac{1}{n} \cdot \delta(t) = \frac{\delta(t) + 1}{n},
\end{align*}
\begin{align*}
q_i & = \Pro{B_1 = i} + \Pro{B_1 \notin R, B_2 = i} = \frac{1}{n} + \delta(t) \cdot \frac{1}{n} = \frac{\delta(t) + 1}{n}.
\end{align*}

\textbf{Case 2 (heavy bin):} $i$ has rank $\leq n \cdot \delta(t)$, then
\begin{align*}
p_i = \frac{1}{2}\Pro{B_1 = i, B_2 \notin R} + \frac{1}{2}\Pro{B_1 \notin R, B_2 = i} %
  = \frac{1}{2} \cdot 2 \cdot \frac{1}{n} \cdot \delta(t) = \frac{\delta(t)}{n},
\end{align*}
\[
q_i = \Pro{B_1 \in R, B_2 = i} = \delta(t) \cdot \frac{1}{n}.
\]\end{proof}

Similarly, for $\Threshold{(f)}$ we have:
\begin{lem}\label{lem:adaptive_threshold_equivalence}
Consider a threshold process $\textsc{Threshold}(f)$ with one query. This process can be always transformed into the following equivalent process: For the first sampled bin $i$, if its load is smaller than $f(t)$, place the ball; otherwise, place the ball in another randomly chosen bin $j$.
\end{lem}
\begin{proof}
The proof is similar to the one of \cref{lem:adaptive_quantile_equivalence}, setting $R$ to be the set of bins with load less than $f(t)$.
\end{proof}

\begin{obs} \label{obs:two_choice_is_a_quantile_process}
For any $n \geq 0$, the $\Quantile{(\frac{1}{n}, \frac{2}{n}\ldots, \frac{n-1}{n})}$ process is equivalent to the \TwoChoice process.
\end{obs}
\begin{proof}
The probability vector of the $\Quantile{(\frac{1}{n}, \frac{2}{n}\ldots, \frac{n-1}{n})}$ process is equal to that of \TwoChoice, since
\[
p_i = \frac{\delta_{i - 1} + \delta_i}{n} = \frac{i - 1 + i}{n^2} = \frac{2i - 1}{n^2},
\]
where we have used $\delta_0 = 0$ and $\delta_n = 1$ for convenience.
\end{proof}

\begin{obs} \label{obs:adding_one_quantile_leads_to_better}
For $k < n - 1$, for any $\delta', \delta_1, \ldots , \delta_k$ quantiles, the $\Quantile{(\delta_1, \ldots , \delta_k)}$ process majorizes $\Quantile{(\delta_1 , \ldots , \delta_i , \delta',\allowbreak  \delta_{i + 1}, \ldots, \delta_k)}$.
\end{obs}
\begin{proof}
Consider the $\Quantile{(\delta_1 , \ldots , \delta_i , \delta', \delta_{i + 1}, \ldots, \delta_k)}$ process. The additional quantile $\delta'$ allows us to distinguish between pairs of ranks in $(\delta_i \cdot n, \delta_{i+1} \cdot n]$, that were not distinguishable by $\Quantile{(\delta_1, \ldots , \delta_k)}$. So, the probability vector of the new process is obtained from the old one by moving probability mass from the lower part of the probability vector to the higher part. Hence, majorization follows.
\end{proof}%

By combining \cref{obs:two_choice_is_a_quantile_process} and \cref{obs:adding_one_quantile_leads_to_better}, we get:
\begin{cor} \label{cor:any_quantile_process_majorises_two_choice}
Any $\Quantile{(\delta_1, \ldots , \delta_k)}$ process majorizes \TwoChoice.
\end{cor}
\begin{proof}
Given any $\Quantile{(\delta_1, \ldots , \delta_k)}$, by incrementally adding the $n - 1 - k$ missing quantiles of the form $j/n$ for $j \in [n]$, we obtain a sequence of quantile processes where each process majorizes the next, by \cref{obs:adding_one_quantile_leads_to_better}. The last process is $\Quantile{(\frac{1}{n}, \ldots, \frac{n-1}{n})}$ which is \TwoChoice, by \cref{obs:two_choice_is_a_quantile_process}.
\end{proof}

\begin{lem}\label{lem:thomas}
For any $\delta \in (0,1)$ and any $\beta \in (0,1)$ with $ \beta \leq \delta \leq 1-\beta$,
the process \Quantile$(\delta)$ is majorized by a $(1+\beta)$-process. In particular, the gap of the quantile process is stochastically smaller than that of the $(1+\beta)$-process.
\end{lem}

Note that for any given $\delta \in (0,1)$, $\beta:=\min \{ \delta, 1-\delta \}$ always satisfies the precondition of the lemma. Conversely, for any given $\beta \leq 1/2$, we have $ \beta \leq 1/2 \leq (1-\beta)$, and thus we can set $\delta:=1/2$. 
\begin{proof}
Let $p$ be the probability vector for the $\Quantile{(\delta)}$ and $q$ for the $(1+\beta)$-process. Recall that $p_i = \frac{\delta}{n}$ for $1 \leq i \leq n \cdot \delta$ and $p_i = \frac{1+\delta}{n}$ for $n \cdot \delta < i \leq n$. The claim will follow immediately once we establish that: (i) For any $1 \leq i \leq n \cdot \delta$, $p_{i} \leq q_{i}$, (ii) For any $n \cdot \delta < i \leq n$, $p_{i} \geq q_i$.

For the first inequality, note that using $\delta \leq 1-\beta$,
\[
 q_i \geq (1-\beta) \cdot \frac{1}{n} \geq \delta \cdot \frac{1}{n} = p_i.
\]
For the second inequality, we have, using $ \beta \leq \delta$,
\[
 q_i = (1-\beta) \cdot \frac{1}{n} + \beta \cdot \frac{2(i-1)}{n^2} \leq (1-\beta) \frac{1}{n} + \beta \cdot \frac{2}{n} = \frac{1}{n} + \frac{\beta}{n} \leq p_i.
\]\end{proof}

\begin{figure}
\begin{tikzpicture}[scale=1]

\begin{axis}[domain=0:10,
  samples=40,
  grid=both,xmin=0.5,xmax=10.5,ymin=0,ymax=0.3,
 legend pos=north west]
\addplot[domain=1:10,opacity=0.6,color=green,draw=green, mark=*] plot coordinates
    {
    (1,0.01)
    (2,0.03)
    (3,0.05)
    (4,0.07)
    (5,0.09)
    (6,0.11)
    (7,0.13)
    (8,0.15)
    (9,0.17)
    (10,0.19)
    };
     \addlegendentry{\TwoChoice};
     \addplot[domain=1:10,opacity=0.6,color=red,draw=red, mark=*] plot coordinates
    {
    (1,0.064)
    (2,0.072)
    (3,0.08)
    (4,0.088)
    (5,0.096)
    (6,0.104)
    (7,0.112)
    (8,0.12)
    (9,0.128)
    (10,0.136)
    };
     \addlegendentry{$(1+\beta)$, $\beta=0.4$};
     
    \addplot[domain=1:10,opacity=0.6,color=blue,draw=blue, mark=*] plot coordinates
    {
    (1,0.06)
    (2,0.06)
    (3,0.06)
    (4,0.06)
    (5,0.06)
    (6,0.06)
    (7,0.16)
    (8,0.16)
    (9,0.16)
    (10,0.16)
    };
     \addlegendentry{$\Quantile(0.6)$}; 
     
\end{axis}

\end{tikzpicture}%
~\hspace{0.2cm}~
\begin{tikzpicture}[scale=1]

\begin{axis}[domain=0:10,
  samples=40,
  grid=both,xmin=0.5,xmax=10.5,ymin=0,ymax=1.1, legend pos=north west]
\addplot[domain=1:10,opacity=0.6,color=green,draw=green, mark=*] plot coordinates
    {
    (1,0.01)
    (2,0.04)
    (3,0.09)
    (4,0.16)
    (5,0.25)
    (6,0.36)
    (7,0.49)
    (8,0.64)
    (9,0.81)
    (10,1)
    };
     \addlegendentry{\TwoChoice};
     \addplot[domain=1:10,opacity=0.6,color=red,draw=red, mark=*] plot coordinates
    {
    (1,0.064)
    (2,0.136)
    (3,0.216)
    (4,0.304)
    (5,0.4)
    (6,0.504)
    (7,0.616)
    (8,0.736)
    (9,0.864)
    (10,1)
    };
     \addlegendentry{$(1+\beta)$, $\beta=0.4$};
     
    \addplot[domain=1:10,opacity=0.6,color=blue,draw=blue, mark=*] plot coordinates
    {
    (1,0.06)
    (2,0.12)
    (3,0.18)
    (4,0.24)
    (5,0.3)
    (6,0.36)
    (7,0.52)
    (8,0.68)
    (9,0.84)
    (10,1)
    };
     \addlegendentry{$\Quantile(0.6)$}; 
     
\end{axis}

\end{tikzpicture}
\caption{Illustration of the probability vector $(p_1,p_2,\ldots,p_{10})$ and cumulative probability distribution of \TwoChoice, $(1+\beta)$ with $\beta=0.4$ and \Quantile$(0.6)$, which is sandwiched by the other two processes.}\label{fig:domination}
\end{figure}
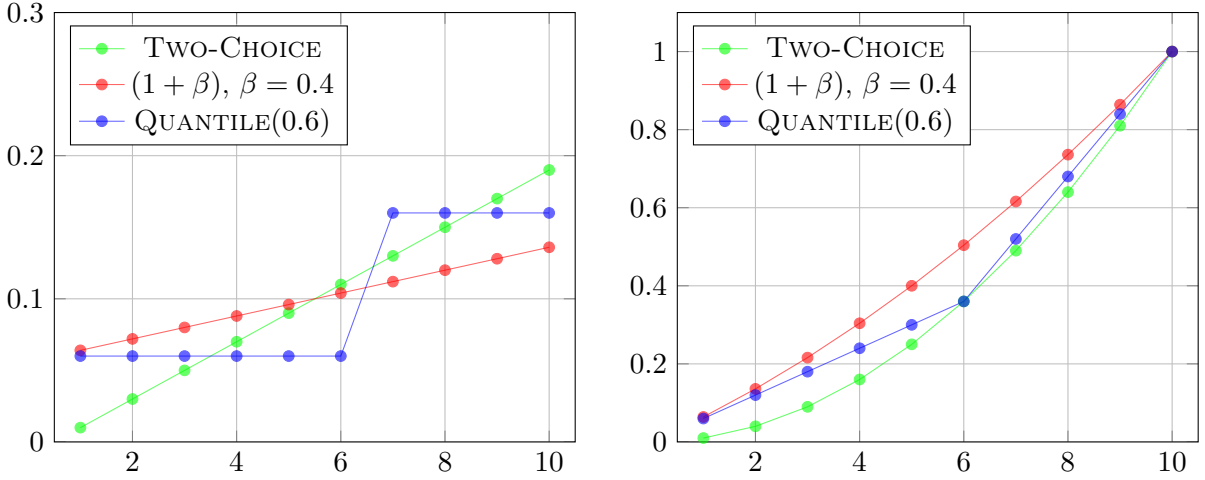

The majorization results in \cref{cor:any_quantile_process_majorises_two_choice} and \cref{lem:thomas} are illustrated in \cref{fig:domination} for $n=10$.

\begin{lem} \label{lem:any_threshold_process_can_be_simulated_by_adaptive_quantile}
Any $\Threshold(f_1, \ldots, f_k)$ process can be simulated by an adaptive quantile process with $k$ queries.
\end{lem}
\begin{proof}
Consider an arbitrary time step $t \geq 0$. Since the process is adaptive, we are allowed to determine the value of $\delta_j(t)$ by looking at the load distribution $x^{(t)}$. We want to choose $\delta_j(t)$, such that comparing the rank $i \leq \delta_j(t) \cdot n$ gives the same answer as $f_j(t) \leq x_i^{(t)}$ for every $i \in [n]$. This can be achieved by choosing $\delta_j(t)$ to be the largest possible quantile such that $y_{\delta_j(t) \cdot n} \leq f_j(t)$. This way any $i \leq \delta_j(t) \cdot n$ will have $x_i^{(t)}\leq \delta_j(t)$ and these will be the only such $i$'s by construction. Hence, at each time step the probability vectors of $\Quantile{(\delta_1, \ldots , \delta_k)}$ and $\Threshold{(f_1, \ldots , f_k)}$ will be the same.
\end{proof}

\begin{lem}\label{lem:quantile_can_be_simulated_by_adaptive_threshold}
Any step $t$ of a $\Quantile{(\delta_1, \ldots, \delta_k)}$ process can be simulated by first choosing $f_1(t),f_2(t),\ldots,f_k(t)$ randomly (from a suitable distribution depending on $x^{(t)}$ and $\delta_1(t),\ldots,\delta_k(t)$) and then running $\Threshold(f_1,f_2,\ldots,f_k)$.
\end{lem}
In other words, there is a reduction from $\Quantile$ to adaptive $\Threshold$, but the $\Threshold$ process must have the ability to randomize between different instances of $\Threshold$.
\begin{proof}

\begin{figure}[t]
  \centering
  \includegraphics[scale=0.6]{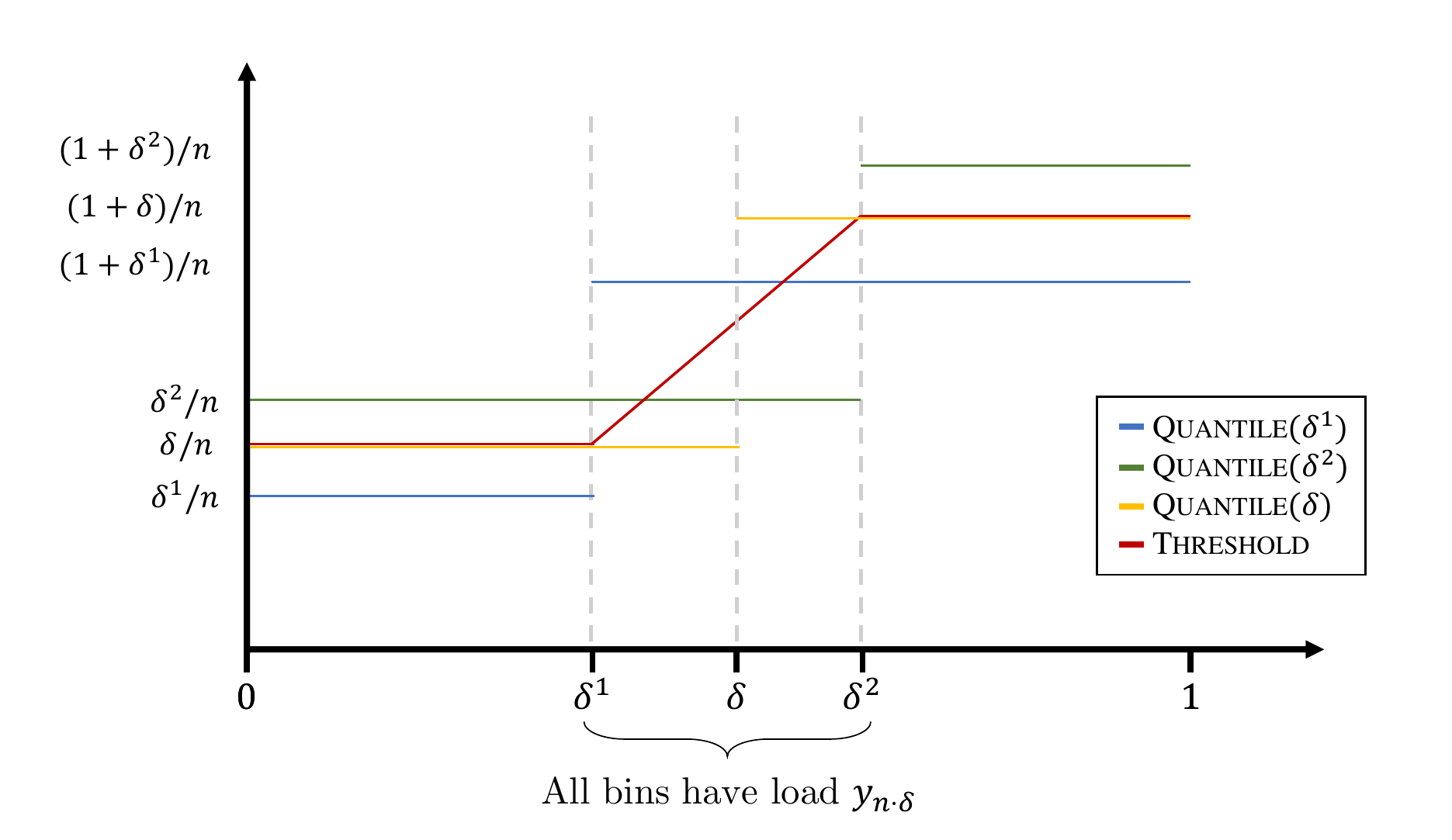}
  \caption{The threshold process which uses a threshold of $y_{n \cdot \delta}$ probability $\alpha$ and $y_{n \cdot \delta} + 1$ with probability $1 - \alpha$, corresponds to mixing the probability vectors of $\Quantile{(\delta^1)}$ and $\Quantile{(\delta^2)}$. The resulting probability vector differs from $\Quantile{(\delta)}$ only in the region $(n \cdot \delta_1, n \cdot \delta_2]$, where by design all bins have load $y_{n \cdot \delta}$. Hence, the effect of the two processes is indistinguishable.}
  \label{fig:threshold_can_simulate_quantile}
\end{figure}
Let us first prove the claim for $k=1$, that is,
$\Quantile{(\delta)}$ can be simulated by an adaptive randomized threshold process with one threshold. Since we only analyze one time-step $t$, we will for simplicity omit this dependency and write $\delta=\delta(t)$.

Let $\delta^1$ be the quantile where the values equal to $y_{n \cdot \delta}$ start and $\delta^2$, where they end (so $\delta^1 \leq \delta \leq \delta^2$). Sampling between a threshold of $y_{n \cdot \delta}$ and $y_{n \cdot \delta} + 1$ with probability $\alpha \in [0, 1]$ interpolates between the $\Quantile{(\delta^1)}$ and $\Quantile{(\delta^2)}$. Let $p^1$ and $p^2$ be the probability vectors for $\Quantile{(\delta^1)}$ and $\Quantile{(\delta^2)}$, then the probability vector $q$ for this adaptive randomized threshold process is given by,
\[
q_i = \alpha \cdot p_{i}^1 + (1-\alpha) \cdot p_{i}^2.
\]
At $i \leq n \cdot \delta^1 \leq n \cdot \delta^2$, we have,
\[
q_i = \alpha \cdot \frac{\delta^1}{n} + (1-\alpha) \cdot \frac{\delta^2}{n}.
\]
We pick $\alpha=\frac{\delta^2-\delta}{\delta^2-\delta^1} \in [0,1]$ so that $q_i = \frac{\delta}{n}$ for $i \leq n \cdot \delta^1$. Then for $i \geq n \cdot \delta^2 \geq n \cdot \delta^1$, we get
\[
q_i = \alpha \cdot \frac{1 + \delta^1}{n} + (1-\alpha) \cdot \frac{1 + \delta^2}{n} = 
\frac{\alpha + (1-\alpha)}{n} +  \frac{\alpha \cdot \delta^1 + (1-\alpha) \cdot \delta^2}{n} = \frac{1 + \delta}{n},
\]
by the choice of $\alpha$. So at the indices $i \in [n] \setminus (n\delta^1, n\delta^2]$ agree with $\Quantile{(\delta)}$.

At the indices $n \cdot \delta^1 < i \leq n \cdot \delta^2$, the probability is shared between bins with the same load, so the effect is indistinguishable (see \cref{fig:threshold_can_simulate_quantile}), in terms of the resulting load vectors.

We will extend this idea to $k > 1$ quantiles, by replacing each quantile $\delta_j$ with a mixture of two thresholds $y_{n \cdot \delta_j}$ and $y_{n \cdot \delta_j} + 1$ with probability $\alpha_j$. For this, we define $\delta_j^1$ and $\delta_j^2$ with $\delta_j^2 \geq \delta_j \geq \delta_j^1$ to be the left and right quantiles for the values of $y_{n \cdot \delta_j}$.

To argue that there exist coefficients $\alpha_j$ such that the two processes are equivalent, we start with the probability vector $q$ of the $\Quantile{(\delta_1, \ldots, \delta_k)}$ process. For each $j \in [k]$, construct the probability vector $q^j$ which agrees with $q$ at all $i \leq n \cdot \delta_j$, except possibly for values equal to $y_{n \cdot \delta_j}$. For these values at $i \leq \delta_j \cdot n$, we will ensure that the processes have the same aggregate probability, so the effect on these bins will be indistinguishable.

In each step we create probability vectors $p^{1j}$ and $p^{2j}$, by adding quantiles $\delta_j^1$ and $\delta_j^2$ respectively to $q^{j-1}$. These affect only the values of the entries in $(n \cdot \delta_{j - 1}, n \cdot \delta_{j+1}]$. As in the one query case, we choose $\alpha_j := \frac{\delta_j^2 - \delta_j}{\delta_j^2 - \delta_j^1}$ such that, for $i \in (n \cdot \delta_{j-1}, n \cdot \delta_j^1]$
\[
q_i^j = \alpha_j \cdot \left( \frac{\delta_{j-1} + \delta_j^1}{n} \right) + (1-\alpha_j) \cdot \left( \frac{\delta_{j-1} + \delta_j^2}{n} \right) = \frac{\delta_{j-1}}{n} + \alpha_j \cdot \frac{\delta_j^1}{n} + (1-\alpha_j) \cdot \frac{\delta_j^2}{n} = \frac{\delta_{j-1} + \delta_j}{n} = q_i,
\]
and for $i \in (n \cdot \delta_j^2, n \cdot \delta_{j+1}]$,
\[
q_i^j = \alpha_j \cdot \left( \frac{\delta_j^1 + \delta_{j+1}}{n} \right) + (1-\alpha_j) \cdot \left( \frac{\delta_j^2 + \delta_{j+1}}{n} \right) = \frac{\delta_{j+1}}{n} + \frac{\alpha_j \delta_j^1 + (1-\alpha_j)\delta_j^2}{n} = \frac{\delta_{j-1} + \delta_j}{n} = q_i.
\]
The linear weighting preserves the following property: Let $B$ be a set of bins, then if $\sum_{b \in B} p_b^{1j} = \sum_{b \in B} p_b^{2j}$ then $\sum_{b \in B} q_b^j = \sum_{b \in B} p_b^{1j} = \sum_{b \in B} p_b^{2j}$. This implies that:
\begin{enumerate} \itemsep0pt
  \item If $p_i^{1j} = p_i^{2j}$, then $q_i^j = p_i^{1j} = p_i^{2j}$.
  \item Let $B_x$ be the set of bins in $[1, \delta_{j-1} \cdot n]$ with equal load $x$. By the inductive argument, in $q^j$ the probability of allocating a ball to $x$ will be the same as in that of $q$.
\end{enumerate}
Hence, this ensures that each step extends the agreement of probability vector $q^j$ and $q$ to each bin $i \in [1, \delta_{j+1} \cdot n]$. The only possible exceptions are bins with equal load, where the probability mass is just rearranged among them. Hence, $q^{k}$ will be equivalent to $q$ for the given load vector.\end{proof}

\begin{lem}\label{lem:threshold_thinning}
For any $k \geq 1$, a \Quantile$(\delta_1,\ldots,\delta_k)$ process can be simulated by an adaptive (and randomized) $(2k)$-\textsc{Thinning} process.
\end{lem}
\begin{proof}
We may assume that \Quantile$(\delta_1,\ldots,\delta_k)$ will process $2k$ queries one by one, and alternate between the two bins. First, send the largest quantile to bin $i_1$, then send the largest to bin $i_2$, then send the second largest to bin $i_1$, etc. and stop as soon as you receive a negative answer. Therefore, for ease of notation, let us set $\gamma_{i}:=\delta_{k-i}$ for $i \in [k]$.

Further, let $i_1$ and $i_2$ be two chosen bins, and $\tilde{i}$ be the bin where the ball is finally placed. 
Note that
\[
 \Pro{ \Rank^{(t)}(\tilde{i}) \leq n \cdot \gamma_{j} } =   \gamma_j \cdot \gamma_j.
\]
since $\tilde{i} \in \{i_1,i_2\}$ will be of rank at least $n \cdot \gamma_j$ if and only if both bins $i_1$ and $i_2$  satisfy $\Rank^{(t)}(i_1) \leq n \cdot \gamma_j$ and $\Rank^{(t)}(i_2) \leq n \cdot \gamma_j$; and those bins are chosen independently.

On the other hand, consider now an adaptive $(2k)$-\textsc{Thinning} process with increasing load thresholds $f_1 \leq f_2 \leq \ldots \leq f_{2k}$ and $2k$ bin choices $i_1,i_2,\ldots,i_{2k}$, which are chosen uniformly and independently at random. Each load threshold $f_{j}$ applied to bin $i_{j}$ will be randomized so that it simulates a \Quantile$(\gamma_{\lfloor (j-1)/2 \rfloor})$ see (\cref{lem:quantile_can_be_simulated_by_adaptive_threshold}). Further, let $\overline{i}$ be the final bin of this allocation process. 

First, the bin $i_{\ell}$ in iteration $\ell$ will not be accepted with probability
\[
  \Pro{ \Rank^{(t)}(i_{\ell}) \leq  n \cdot \gamma_{1+\lfloor \ell/2 \rfloor}  } = \gamma_{1+\lfloor \ell/2 \rfloor},
\]
and using the independence of the first $2j$ sampled different bins, we obtain 
\begin{align*}
  \Pro{ \Rank^{(t)}(\overline{i}) \leq n \cdot \gamma_{j} } 
  &= \prod_{\ell=1}^{2j}  \Pro{ \Rank^{(t)}(i_{\ell}) \leq n \cdot \gamma_{1+\lfloor \ell/2 \rfloor} } \\
  &= \gamma_{1} \cdot \gamma_{1} \cdot \gamma_{2} \cdot \gamma_{2} \cdot \ldots \cdot \gamma_{j} \cdot \gamma_{j} \leq \gamma_{j} \cdot \gamma_{j}
  \leq  \Pro{ \Rank^{(t)}(\tilde{i}) \leq n \cdot \gamma_{j} }.
\end{align*}\end{proof}

\section{Lower Bounds for One Quantile and One Threshold} \label{sec:lower}

In the lightly loaded case (i.e., $m=n$), \cite{FG18} proved an upper bound of $(2+o(1)) \cdot (\sqrt{ 2 \log n/ \log \log n})$ on the maximum load for a uniform \textsc{Threshold}$(f)$-process with $f=\sqrt{ 2 \log n/ \log \log n}$ (\cite{FL20} extended this to $d>2$). They also proved that this strategy is asymptotically optimal. In~\cite[Problem~1.3]{FG18}, the authors suggest that the $\Oh(\sqrt{ \log n/ \log \log n})$ bound on the gap extends to the heavily loaded case. Here we will disprove this, establishing a slightly larger lower bound of $\Omega(\sqrt{\log n})$ (\cref{thm:new_adaptive_quantile_lower_bound}). We also derive additional lower bounds (\cref{thm:adaptive_quantile_lower_bound} and \cref{cor:large_gap_in_nlog2n_interval}) that demonstrate that any \Quantile~or \textsc{Threshold} process will ``frequently'' attain a gap which is even as large as $\Omega(\log n/\log \log n)$.

Let us describe the intuition behind this bound in case of uniform quantiles, neglecting some technicalities. Consider $\Quantile(\delta)$ and the equivalent $2$-\textsc{Thinning} instance where a ball is placed in the first bin if its load is among the $(1-\delta) \cdot n$ lightest bins, and otherwise it is placed in a new (second) bin chosen uniformly at random (Lemma~\ref{lem:adaptive_quantile_equivalence}). We have two cases:
\begin{enumerate}\itemsep-1pt
\item[] \textbf{Case 1:} We choose most times a ``large'' $\delta$. Then we allocate approximately $m \cdot \delta$ balls to their second bin choice which is uniform over all $n$ bins. This will lead to a behavior close to \OneChoice (\cref{lem:two_case_quantile}).
\item[] \textbf{Case 2:} We choose most times a ``small'' $\delta$. Then we allocate approximately $m \cdot (1-\delta)$ balls with the first bin choice, which is a \OneChoice process over the $n \cdot (1-\delta)$ lightest bins. As we establish in \cref{lem:analysis_of_p2}, for small $\delta$ there are simply ``too many'' light bins that will reach a high load level, so the process is again close to \OneChoice.
\end{enumerate}

\begin{restatable}{thm}{adaptivelowerbound} \label{thm:adaptive_quantile_lower_bound}
For any adaptive $\Quantile(\delta)$ (or \textsc{Threshold}$(f)$) process, 
\[
\Pro{\max_{t \in [1, n \log^2 n]} \Gap( t )  \geq \frac{1}{8} \cdot \frac{\log n}{\log \log n  }} \geq 1-o(n^{-2}).
\]
\end{restatable}

In fact, as shown in~\cref{cor:large_gap_in_nlog2n_interval}, this lower bound holds for a significant proportion of time-steps. 
We also show a lower bound for fixed $m$, which is derived in a similar way
as \cref{thm:adaptive_quantile_lower_bound}, but with a different parameterization of ``large'' and ``small'' quantiles:

\begin{restatable}{thm}{newadaptivequantilelowerbound} \label{thm:new_adaptive_quantile_lower_bound}
For any adaptive $\Quantile(\delta)$ (or \textsc{Threshold}$(f)$) process, with $m= K \cdot n \sqrt{\log n}$ balls for $K = 1/10$, it holds that
\[
\Pro{ \Gap( m ) \geq \frac{1}{20} \sqrt{\log n} } \geq 1-o(n^{-2}).
\]
\end{restatable}

\subsection{Preliminaries for Lower Bounds}
\label{sec:lower_bounds_prelims}

Let us first formalize the intuition of the lower bound. Recall that we will analyze the adaptive case, which means that the quantiles at each step $t$ may depend on the full history of the process, or, equivalently, on the load vector $(x_1^{(t-1)},x_2^{(t-1)},\ldots,x_n^{(t-1)})$. We also remind the reader that any adaptive \textsc{Threshold}$(f)$ process can be simulated by $\Quantile(\delta)$ (\cref{lem:any_threshold_process_can_be_simulated_by_adaptive_quantile}), which is why we will do the analysis below for $\Quantile(\delta)$ only.

The next lemma proves that if within $n$ consecutive allocations a large quantile is used too often, then  $\Quantile(\delta)$ restricted to the heavily loaded bins generates a high maximum load, similar to \OneChoice.
\begin{lem}\label{lem:two_case_quantile}
Consider any adaptive $\Quantile(\delta)$ process during the time-interval $[t,t+n)$. If $\Quantile(\delta)$ allocates at least $n/(\log n)^2$ balls with a quantile larger than $(\log n)^{-2}$ in $[t, t + n)$, then
\[
\Pro{\Gap(t + n) \geq \frac{1}{8} \frac{\log n}{\log \log n}} \geq 1 - o(n^{-4}).
\]
\end{lem}
\begin{proof}
Assume there are at least $n/(\log n)^2$ allocations with quantile larger than $(\log n)^{-2}$. Then, using \cref{lem:multiplicative_factor_chernoff_for_binomial}, \wp~at least $1 - o(n^{-4})$, at least $\frac{1}{e} \frac{n}{\log^2 n} \cdot \frac{1}{\log^2{n}} \geq \frac{n}{\log^5 n}$ balls are thrown using \OneChoice.

Consider now the load configuration before the \emph{batch}, i.e. the next $n$ balls are allocated. If $\Gap(t) \geq \log n$, then $\Gap(t + n) \geq \frac{1}{8} \log n / \log \log n$, as a load can decrease by at most $1$ in $n$ steps. So we can assume $\Gap(t) < \log n$. Let $B$ be the set of bins whose load is at least the average load at time $t$, then $|B| \geq n/ \log n$. Using \cref{lem:multiplicative_factor_chernoff_for_binomial}, \wp~at least $1-o(n^{-4})$ the batch will allocate at least $n/(\log n)^{6}$ balls to the bins of $B$. Hence, by \cref{lem:lower_bound_max_load_whp}, at least one bin in $B$ will increase its load by an additive factor of $\frac{1}{7}\log n/\log \log n$ \wp~at least $1 - o(n^{-4})$. Since the average load only increases by one during the batch, there will be a gap of $\frac{1}{8}\log n/\log \log n$ w.h.p., and our claim is established.
\end{proof}

The next lemma implies that if for most allocations the largest quantile is too small, then the allocations on the lightest bins follows that of \OneChoice, and we end up with a high maximum load. 
\begin{lem}\label{lem:analysis_of_p2}
Consider any adaptive $\Quantile(\delta)$ process with $m = n \log^2 n$ balls that allocates at most $n$ balls with a quantile larger than $(\log n)^{-2}$. Then,
\[
\Pro{\Gap(m) \geq 0.2 \log n} \geq 1 - o(n^{-2}).
\]
\end{lem}
The proof of this Lemma is similar to \cref{lem:two_case_quantile}, but a bit more complex. We define a coupling between the $\Quantile(\delta)$ process and the \OneChoice process. We couple the allocation of balls whose first sample is among the $(1-\delta(t)) \cdot n$-lightest bins with a \OneChoice process.
The balls whose first sample is among the $\delta(t) \cdot n$-heaviest bins are allocated differently, and cause our process to diverge from an original \OneChoice process. However, we prove that the number of different allocations is too small to change the order of the gap.

\begin{proof}
We will use the following coupling between the allocations of $\Quantile{(\delta)}$ and \OneChoice. At each step $t \in [1,n \log^2 n]$, we first sample a bin index $j \in [n]$ uniformly at random. In the \OneChoice process, we place the ball in the $j$-th most loaded bin. In the \Quantile{} process:
\begin{enumerate}
  \item If $j > \delta(t) \cdot n$, we place the ball in the $j$-th most loaded bin (of \Quantile{}), and we say that the processes agree.
  \item If $j \leq \delta(t) \cdot n$, we sample another bin index $\tilde{j} \in [n]$ uniformly at random and place the ball in the $\tilde{j}$-th most loaded bin (of \Quantile{}).
\end{enumerate}
Let $y^{(s)}$ and $z^{(s)}$ be the sorted load vectors of \OneChoice and the \Quantile{} process respectively at step $s \geq 0$. Further, let $L(s) :=  d_{\ell_1}(y^{(s)}, z^{(s)})$ be the $\ell_1$-distance between these vectors. Note that $L(0)=0$. If in a step both processes place a ball in the $j$-th most loaded bin, using a simple coupling argument (see~ \cref{lem:increasing_kth_distance_same} below for details) it follows that
\[
  L(t+1) \leq L(t).
\]
Otherwise, if in a step the processes place a ball in a different bin, since only two positions in the load vectors can increase by one, then
\[
 L(t+1) \leq L(t) + 2.
\]
Hence by induction over $s$, if $k$ is the number of steps for which the processes disagree, then
\[
 L(n \log^2 n ) \leq 2\cdot k.
\]

We will next show an upper bound on $k$, which in turn implies an upper bound on $L(n \log^2 n )$. 
First, for each of the at most $n$ steps $t \in [1,n \log^2 n]$ for which $\delta(t) \geq (\log n)^{-2}$, we (pessimistically) assume that the two processes always disagree.
Secondly, for the at most $n \log^2 n$ steps $t \in [1,n \log^2 n]$ with $\delta(t) \leq (\log n)^{-2}$, using a Chernoff bound (\cref{lem:multiplicative_factor_chernoff_for_binomial}), we have \wp\ $1-o(n^{-2})$ in at most $(n \log^2 n )\cdot (\log n)^{-2} \cdot e = ne$ of these steps $s$, the case that $j \leq \delta(s) \cdot n$, i.e., the two processes disagree. Now if this event occurs,
\[
 k \leq n \cdot 1 + n \cdot e \leq 2 n \cdot e \qquad \Rightarrow \qquad 
 L(n \log^2 n) \leq 4 n \cdot e.
\]

By \cref{lem:one_choice_close_to_max_load}, there are constants $a = 0.4, c = 0.25$ such that with probability $1-o(n^{-2})$, the \OneChoice load vector $y^{(n \log^2 n)}$ has at least
$c n \log n$ balls with height at least $\frac{a}{2} \log n$. However, any load vector which has no balls at height $\frac{a}{2} \log n$ must have a $\ell_1$-distance of at least $c n \log n$ to $y^{(n \log^2 n)}$, and thus we conclude by the union bound that $\Gap(n \log^2 n)\geq \frac{a}{2} \log n$ holds with probability $1-2 o(n^{-2})$.\end{proof}

\begin{lem} \label{lem:increasing_kth_distance_same}
Let $y$ and $z$ be two decreasingly sorted load vectors. Consider the sorted vectors $y+\mathbf{e}_i$ and $z+\mathbf{e}_i$ after incrementing the value at index $i$. Then, $d_{\ell_1}(y, z) \geq d_{\ell_1}(y+\mathbf{e}_i, z +\mathbf{e}_i)$.
\end{lem}
\begin{proof}
If the items being updated end up both in the same indices (after sorting), then their $\ell_1$ distance remains unchanged.

Let $u := y_i$ and $v := z_i$ for the updated index $i$ in the (old) sorted load vector. To obtain the new sorted load vector, we have to search in both $y$ and $z$ from right to left for the leftmost entry being equal to $u$ and being equal to $v$, respectively, and then increment these values. Then, there are the following three cases to consider (in bold is the value to be updated): 

\textbf{Case 1} $u < v$: Let $v < w_1 \leq \ldots \leq w_k$, where $w_k$ is the matching value for $u+1$ in $z$, then $w_k > v \Rightarrow w_k \geq u + 2$
\[
\begin{array}[t]{c|cccccccc}
y & \ldots & u   & \ldots & u   & u & \ldots & \mathbf{u} & \ldots \\ \hline
z & \ldots & w_k & \ldots & w_1 & v & \ldots & \mathbf{v} & \ldots
\end{array}
\ \  \raisebox{-0.5\normalbaselineskip}{$\to$} \ \ 
\begin{array}[t]{c|cccccccc}
y+\mathbf{e}_i & \ldots & u+1   & \ldots & u   & u & \ldots & u  & \ldots\\ \hline
z+\mathbf{e}_i & \ldots & w_k & \ldots & w_1 & v+1 & \ldots & v & \ldots \\
\multicolumn{1}{c}{} &  & 
\multicolumn{1}{l@{}}{%
  \raisebox{.5\normalbaselineskip}{%
  \rlap{$\underbrace{\hphantom{\mbox{$u+1$}}}_{-1}$}}%
    }  & &  & 
\multicolumn{1}{l@{}}{%
  \raisebox{.5\normalbaselineskip}{%
  \rlap{$\underbrace{\hphantom{\mbox{$v+1$}}}_{+1}$}}%
    } &  & & 
\end{array}
\]

\textbf{Case 2} $u < v$: Let $u < w_1 \leq \ldots \leq w_k$, where $w_k$ is the matching value for $v+1$ in $y$
\[
\begin{array}[t]{c|cccccccc}
y & \ldots & w_k & \ldots & w_1 & u & \ldots & \mathbf{u} & \ldots \\ \hline
z & \ldots & v   & \ldots & v   & v & \ldots & \mathbf{v} & \ldots
\end{array} \ \  \raisebox{-0.5\normalbaselineskip}{$\to$} \ \ 
\begin{array}[t]{c|cccccccc}
y+\mathbf{e}_i & \ldots & w_k & \ldots & w_1 & u+1 & \ldots & u & \ldots \\ \hline
z+\mathbf{e}_i & \ldots & v+1 & \ldots & v   & v & \ldots & v & \ldots \\
\multicolumn{1}{c}{} &  & 
\multicolumn{1}{l@{}}{%
  \raisebox{.5\normalbaselineskip}{%
  \rlap{$\underbrace{\hphantom{\mbox{$v+1$}}}_{\leq 1}$}}%
    }  & &  & 
\multicolumn{1}{l@{}}{%
  \raisebox{.5\normalbaselineskip}{%
  \rlap{$\underbrace{\hphantom{\mbox{$u+1$}}}_{-1}$}}%
    } &  & & 
\end{array}
\]

\textbf{Case 3} $u = v$: Let $u < w_1 \leq \ldots \leq w_k$, where $w_k$ is the matching value for $u+1$ in $z$
\[
\begin{array}[t]{c|cccccccc}
y & \ldots & w_k & \ldots & w_1 & u & \ldots & \mathbf{u} & \ldots \\ \hline
z & \ldots & u   & \ldots & u   & u & \ldots & \mathbf{u} & \ldots
\end{array}
\ \  \raisebox{-0.5\normalbaselineskip}{$\to$} \ \ 
\begin{array}[t]{c|cccccccc}
y+\mathbf{e}_i & \ldots & w_k   & \ldots & w_1 & u+1 & \ldots & u & \ldots \\ \hline
z+\mathbf{e}_i & \ldots & u+1 & \ldots & u   & u & \ldots & u & \ldots \\
\multicolumn{1}{c}{} &  & 
\multicolumn{1}{l@{}}{%
  \raisebox{.5\normalbaselineskip}{%
  \rlap{$\underbrace{\hphantom{\mbox{$u+1$}}}_{-1}$}}%
    }  & &  & 
\multicolumn{1}{l@{}}{%
  \raisebox{.5\normalbaselineskip}{%
  \rlap{$\underbrace{\hphantom{\mbox{$u+1$}}}_{+1}$}}%
    } &  & & 
\end{array}
\]
\end{proof}

\subsection{Lower Bound for a Range of Values (Theorem~\ref{thm:adaptive_quantile_lower_bound})} \label{sec:lower_bounds_for_adaptive_quantiles}

With \cref{lem:two_case_quantile} and \cref{lem:analysis_of_p2} proven in the previous subsection, we can now derive a lower bound for any adaptive $\Quantile(\delta)$ (or \textsc{Threshold}$(f)$) process, establishing \cref{thm:adaptive_quantile_lower_bound}. After the proof, we also state two simple consequences that follow immediately from this result.

\adaptivelowerbound*

\begin{proof}
Since any adaptive \textsc{Threshold}$(f)$ can be simulated by an adaptive $\Quantile(\delta)$ process  (see \cref{lem:any_threshold_process_can_be_simulated_by_adaptive_quantile}), it suffices to prove the claim for adaptive 
$\Quantile(\delta)$ processes.
We will allow the adversary to run two processes, and then choose one that achieves a gap of $< \frac{1}{8}\log{n} / \log{\log{n}}$ (if such exists): 
\begin{itemize}
  \item \textbf{Process $P_1$.} The adversary has to allocate $m=n \log^2 n$ balls into $n$ bins. The adversary wins if for all steps $t \in [m]$, $\Gap(t) < \frac{1}{8}\log n/\log \log n$, and, Condition $C_1$, at least $n$ out of the $m$ quantiles are larger than $(\log n)^{-2}$.
  \item \textbf{Process $P_2$.} The adversary has to allocate $m=n\log^2{n}$ balls into $n$ bins. The adversary wins if $\Gap(m)< \frac{1}{8}\log n/\log \log n$ and, Condition $C_2$, at least $m-n=n \log^2n -n$ out of the $m$ quantiles are at most $(\log n)^{-2}$.
\end{itemize}

Note that the conditions $C_1$ and $C_2$ form a disjoint partition. We will prove that the adversary cannot win any of the two games with probability greater than $n^{-2}$.
Now recall the original process, the one we would like to analyze:
\begin{itemize}
\item \textbf{Process $P_3$ (adaptive} $\Quantile(\delta)$\textbf{)}. The adversary has to allocate $m=n\log^2n$ balls into bins at each step. The adversary wins
if $\Gap(t)< \frac{1}{8}\log n/\log \log n$ for all $t \in [m]$.
\end{itemize}

We will show below that $\Pro{ \mbox{adversary wins $P_1$} } = o(n^{-2})$ and $\Pro{ \mbox{adversary wins $P_2$} } = o(n^{-2})$, and these bounds hold for the best possible strategies an adversary can use in each game, respectively. 
Assuming that these bounds hold and by noticing that exactly one of $C_1$ and $C_2$ must hold for $P_3$,
\[
\Pro{P_3 \text{ wins}} = \Pro{P_3 \text{ wins}, C_1} + \Pro{P_3 \text{ wins}, C_2} \leq \Pro{P_1 \text{ wins}} + \Pro{P_2 \text{ wins}} \leq o(n^{-2}).
\]

\textbf{Analysis of Process 1:} 
Let $\mathcal{E}_t$ be the event that (i) \Quantile{} allocates at least $n/(\log n)^2$ balls with a quantile larger than $(\log n)^{-2}$ in the interval $[t,t+n)$, and (ii) $\Gap(t+n) < \frac{1}{8}\log n/\log \log n$. Note that this is the negation of \cref{lem:two_case_quantile}, so by union bound over $1 \leq t \leq m - n$,
\[
 \Pro{ \bigcup_{t=1}^{m-n} \mathcal{E}_t } \leq n \log^2 n \cdot o(n^{-4}) = o(n^{-2}).
\]
Note that if none of the $\mathcal{E}_t$ for $1 \leq t \leq m-n$ occur, then the adversary allocates at most $n/(\log n)^{2} \cdot (\log n)^2 \geq n$ out of the $m$ balls with a quantile at least $(\log n)^{-2}$. Therefore,
\[
\Pro{\mbox{adversary wins $P_1$}} \leq o(n^{-2}).
\]

\textbf{Analysis of Process 2:} The analysis of $P_2$ follows directly by \cref{lem:analysis_of_p2}.\end{proof}

Let us also observe a slightly stronger statement which follows directly from Theorem~\ref{thm:adaptive_quantile_lower_bound}:

\begin{cor} \label{cor:large_gap_in_nlog2n_interval}
Any adaptive process $\Quantile(\delta)$ satisfies:
\[
  \Pro{ \bigcup_{t \in [1, n \log^2 n]} \left\lbrace \min_{s \in \left[ t,t+\frac{1}{16} n \frac{\log n }{ \log \log n}\right)} \Gap(s) \geq \frac{1}{16} \cdot \frac{\log n}{\log \log n} \right\rbrace} \geq 1-n^{-2}.
\]
\end{cor}
\begin{proof}[Proof of \cref{cor:large_gap_in_nlog2n_interval}]
If there is a step $t$ for which $\Gap(t) \geq \frac{1}{8} \cdot \log n / \log \log n $, then for any $s$ with $t \leq s \leq t+\frac{1}{16} \cdot \log n/\log \log n$, $\Gap(s) \geq \Gap(t) - (s-t)/n \geq \frac{1}{16} \cdot \log n/ \log \log n$. Hence the statement follows from \cref{thm:adaptive_quantile_lower_bound}.
\end{proof}

In other words, the corollary states that for at least $\Omega(n \log n/\log \log n)$ (consecutive) steps in $[1,\Theta(n \log ^2 n)]$, the gap is $\Omega(\log n/\log \log n)$. This is in  contrast to the behavior of the process $\Quantile(\delta_1,\delta_2)$, for which our result in \cref{sec:new_upper_bound_for_k_queries} implies that with high probability the gap is {\em always} below $\Oh(\sqrt{\log n})$ during any time-interval of the same length.

For uniform $\Quantile(\delta)$, we are always running either process $P_1$ or $P_2$, so the following strengthened version of \cref{thm:adaptive_quantile_lower_bound} holds:
\begin{cor}\label{cor:uniform_quantile_lower_bound}
For any uniform $\Quantile(\delta)$ process for $m = n \log^2 n$ balls,
\[
\Pro{\Gap(m) \geq \frac{1}{8} \cdot \frac{\log n}{\log \log n}} \geq 1 - o(n^{-2}).
\]
\end{cor}
\begin{proof}
Since $\delta$ is fixed, in the proof of \cref{thm:adaptive_quantile_lower_bound}, we are always running either process $P_1$ or $P_2$. For process $P_1$, $\mathcal{E}_{m-n}$ holds w.p.\ $1 - o(n^{-4})$, so there is an $\Omega(\log n / \log \log n)$ gap at $m$. For process $P_2$, there is an $\Omega(\log n / \log \log n)$ gap at $m$ w.p.\ $1 - o(n^{-2})$. Hence, in both cases the gap at step $m$ is $\Omega(\log n / \log \log n)$ w.p. $1-o(n^{-2})$.
\end{proof}

\subsection{Lower Bound for Fixed \texorpdfstring{$m = \Theta(n \sqrt{\log n})$ (\cref{thm:new_adaptive_quantile_lower_bound})}{m = Theta(n sqrt(log n))}} \label{sec:lower_bound_for_fixed_m}

We now prove a version of \cref{thm:adaptive_quantile_lower_bound} that establishes a lower bound of $\Omega(\sqrt{\log n})$ on the gap for a {\em fixed} value $m$. It follows the same proof as \cref{thm:adaptive_quantile_lower_bound} except that the parameters are different: (i) $m = \Theta(n \sqrt{\log n})$ and (ii) Condition $C_1$ is defined as having at least $m \cdot e^{-\sqrt{\log n}}$ out of the $m$ quantiles being at least $e^{-\sqrt{\log n}}$. \cref{lem:new_two_case_quantile} is the modified \cref{lem:two_case_quantile} and \cref{lem:new_analysis_of_p2} is the modified \cref{lem:analysis_of_p2}.

\begin{lem} \label{lem:new_two_case_quantile}
Consider any adaptive $\Quantile(\delta)$ process during the time-interval $[t,t+n)$. If $\Quantile(\delta)$ allocates at least $n / e^{\sqrt{\log {n}}}$ balls with a quantile larger than $e^{-\sqrt{\log {n}}}$ in $[t, t + n)$, then
\[
\Pro{\Gap(t + n) \geq \frac{1}{5} \sqrt{\log n}} \geq 1 - o(n^{-4}).
\]
\end{lem}

\begin{proof}

Assume there are at least $n / e^{\sqrt{\log {n}}}$ allocations with quantile larger than $e^{-\sqrt{\log {n}}}$. Then, using \cref{lem:multiplicative_factor_chernoff_for_binomial}, \wp~at least $1 - o(n^{-4})$, at least $\frac{1}{e} \frac{n}{ e^{\sqrt{\log {n}}}} \cdot \frac{1}{e^{\sqrt{\log {n}}}} \geq \frac{n}{e^{3\sqrt{\log {n}}}}$ balls are thrown using \OneChoice.

Consider now the load configuration before the batch is allocated. If $\Gap(t) \geq \frac{1}{4} \sqrt{\log n}$, then $\Gap(t + n) \geq \frac{1}{5} \sqrt{\log n}$, as a load can decrease by at most $1$ in $n$ steps. So we can assume $\Gap(t) < \frac{1}{4} \sqrt{\log n}$. Let $B$ be the set of bins whose load is at least the average load at time $t$, then $|B| \geq n / (\frac{1}{4} \sqrt{\log n})$. Using \cref{lem:multiplicative_factor_chernoff_for_binomial}, \wp~at least $1-o(n^{-4})$ the batch will allocate at least $n/(e \cdot e^{3\sqrt{\log {n}}} \cdot (\frac{1}{4} \sqrt{\log n})) \geq n/e^{4\sqrt{\log {n}}}$ balls to the bins of $B$. Hence, using \cref{lem:new_lower_bound_max_load_whp} with $c = 1/2$, $u = 4$ and $k = \frac{2}{9}$ at least one bin in $B$ will increase its load by an additive factor of $\frac{2}{9} \sqrt{\log{n}}$ \wp\ at least $1 - o(n^{-4})$. Since the average load only increases by one during the batch, we have created a gap of $\frac{2}{9} \sqrt{\log{n}} - 1 > \frac{1}{5} \sqrt{\log{n}}$, and our claim is established.
\end{proof}

\begin{lem}\label{lem:new_analysis_of_p2}
Consider any adaptive $\Quantile(\delta)$ process with $m = K n \sqrt{\log n}$ balls that allocates at most $n$ balls with a quantile larger than $e^{-\sqrt{\log n}}$, then
\[
\Pro{\Gap(m) \geq \frac{1}{20} \sqrt{\log n}} \geq 1 - o(n^{-2}),
\]
where $K = 1/10$.
\end{lem}
\begin{proof}
Let $C = 1/20$. We will use the same coupling as in the proof of \cref{lem:analysis_of_p2}. We now obtain an upper bound on $k$, which in turn implies an upper bound on $L(K n \sqrt{\log n})$. 
First, for each of the at most $n$ steps $t \in [1,K n \sqrt{\log n}]$ for which $\delta(t) \geq e^{-\sqrt{\log n}}$, we (pessimistically) assume that the two processes always disagree.
Secondly, for the at most $K n \sqrt{\log n}$ steps $t \in [1,K n \sqrt{\log n}]$ with $\delta(t) \leq e^{-\sqrt{\log n}}$, using a Chernoff bound (\cref{lem:multiplicative_factor_chernoff_for_binomial}), we have \wp\ $1-o(n^{-2})$ in at most $e \cdot (K n \sqrt{\log n} )\cdot e^{-\sqrt{\log n}}$ of these steps $s$ the case that $j \leq \delta(s) \cdot n$, i.e., the two processes disagree. Now if this event occurs,
\[
 k \leq n \cdot 1 + n \cdot e \leq 2 n \cdot e \qquad \Rightarrow \qquad 
 L(K n \sqrt{\log n}) \leq 4 \cdot e\cdot (K n \sqrt{\log n} )\cdot e^{-\sqrt{\log n}}.
\]

By \cref{lem:new_one_choice_close_to_max_load}, with probability $1-o(n^{-2})$, the \OneChoice load vector $y^{(K n \sqrt{\log n})}$ has at least
$e^{-0.21 \sqrt{\log n}} \cdot C n\sqrt{\log{n}}$ balls with at least $(K + C)\cdot \sqrt{\log n}$ height. However, any load vector which has no balls at height $(K + C)\cdot \sqrt{\log n}$ must have a $\ell_1$-distance of at least $e^{-0.21 \sqrt{\log n}} \cdot C n\sqrt{\log{n}} \cdot K \cdot \sqrt{\log n} > L(K n \sqrt{\log n})$ to $y^{(K n \sqrt{\log n})}$, and thus we conclude by the union bound that $\Gap(K n \sqrt{\log n}) \geq C \sqrt{\log n}$ holds with probability $1-2 o(n^{-2})$. \end{proof}

\newadaptivequantilelowerbound*

\begin{proof}
Since any adaptive \textsc{Threshold}$(f)$ can be simulated by an adaptive $\Quantile(\delta)$ process (see \cref{sec:notation}), it suffices to prove the claim for adaptive $\Quantile(\delta)$ processes.
We will allow the adversary to run two processes, and then choose one that achieves a gap smaller than $C \sqrt{\log n}$ (if such exists): 
\begin{itemize}
  \item \textbf{Process $P_1$.} The adversary has to allocate $m=K n \sqrt{\log n}$ balls into $n$ bins. The adversary wins if for step $m$, $\Gap(m) < C \sqrt{\log n}$, and, Condition $C_1$, at least $(K n \sqrt{\log n}) \cdot e^{-\sqrt{\log n}}$ out of the $m$ quantiles are larger than $e^{-\sqrt{\log n}}$.%
  \item \textbf{Process $P_2$.} The adversary has to allocate $m=K n \sqrt{\log n}$ balls into $n$ bins. The adversary wins if for step $m$, $\Gap(m) < C \sqrt{\log n}$ and, Condition $C_2$, at least $m-(K n \sqrt{\log n}) \cdot e^{-\sqrt{\log n}}$ out of the $m$ quantiles are at most $e^{-\sqrt{\log n}}$.
\end{itemize}

We will prove that the adversary cannot win any of the two games with probability greater than $n^{-2}$.
Now recall the original process, the one we would like to analyze:
\begin{itemize}
\item \textbf{Process $P_3$ (adaptive} $\Quantile(\delta)$\textbf{)}. The adversary has to allocate $m=K n \sqrt{\log n}$ balls into bins using one adaptive query at each step. The adversary wins if $\Gap(m) < C \sqrt{\log n}$.
\end{itemize}
Again, we will show below that $\Pro{ \mbox{adversary wins $P_1$} } = o(n^{-2})$ and $\Pro{ \mbox{adversary wins $P_2$} } = o(n^{-2})$, and these bounds imply that $\Pro{ P_3 \text{ wins}} = o(n^{-2})$. We now turn to the analysis of $P_1$ and $P_2$:

\textbf{Analysis of Process 1:} 
Let $\mathcal{E}_t$ be the event that (i) \Quantile{} allocates at least $n \cdot e^{-\sqrt{\log n}}$ balls with a quantile at least $e^{-\sqrt{\log n}}$ in the interval $[t,t+n]$, and (ii) $\Gap(t+n) \leq \frac{1}{5} \sqrt{\log n}$. Note that this is the negation of \cref{lem:new_two_case_quantile}, so by union bound over $1 \leq t \leq m - n$,
\[
 \Pro{ \bigcup_{t=1}^{m-n} \mathcal{E}_t } \leq K \cdot n \sqrt{\log n} \cdot o(n^{-4}) = o(n^{-2}).
\]
Note that if none of the $\mathcal{E}_t$ for $1 \leq t \leq m-n$ occur, then we either have $\Gap(t) \geq \frac{1}{5} \sqrt{\log {n}}$ at some time $t \leq m$ (implying $\Gap(m) \geq (\frac{1}{5} - \frac{1}{10}) \sqrt{\log n} \geq \frac{1}{20} \sqrt{\log n}$), or the adversary allocates less than $\frac{n}{e^{\sqrt{\log n}}} \cdot K \sqrt{\log n}$ out of the $m$ balls with a quantile at least $e^{-\sqrt{\log n}}$. Therefore,
\[
\Pro{\mbox{adversary wins $P_1$}} = o(n^{-2}).
\]

\textbf{Analysis of Process 2:} The analysis of $P_2$ follows directly by \cref{lem:new_analysis_of_p2}.
\end{proof}

\section{Upper Bounds for One Quantile} \label{sec:one_quantile}

In this section we study the $\Quantile(\delta)$ process for constant $\delta \in (0, 1)$. %
This analysis will also serve as the basis for the $k$-quantile case with $k > 1$ in Section~\ref{sec:new_upper_bound_for_k_queries}. First, we define the following exponential potential function (similarly to~\cite{PTW15}): 
For any time-step $s \geq 0$, 
\begin{align*}
 \Phi_{0}^{(s)} := \sum_{i=1}^n \exp \left( \alpha_2 \cdot (x_i^{(s)} - \frac{s}{n} )^{+} \right),
\end{align*}
where $z^{+}=\max(z,0)$ and $\alpha_2 > 0$ to be specified later. We first remark that with the results in~\cite{PTW15}, a bound on the expected value of $\Phi_{0}$ can be easily derived:
\begin{thm}[cf.~Theorem~2.10 in~\cite{PTW15}]\label{thm:one_plus_beta_theorem}
Consider any allocation process with probability vector $p$ that is (i) non-decreasing in $i$, $p_i \leq p_{i+1}$ and (ii) for some $0 < \epsilon < 1/4$,  
\[
p_{n/3} \leq \frac{1-4\epsilon}{n} \quad \text{ and } \quad p_{2n/3} \geq \frac{1 + 4\epsilon}{n}.
\]
Then, for $0 < \alpha_2 < \epsilon / 6$, we have for any $s \geq 0$, $\Ex{ \Phi_{0}^{(s)} } \leq cn$, where $c = \frac{40 \cdot 128^3}{\epsilon^5}$.
\end{thm}

In particular, by verifying the condition on the probability vector and applying Markov's inequality, we immediately obtain an upper bound of $\Oh(\log n)$ on the gap.
\begin{restatable}{thm}{medianquantilegap} \label{thm:median_quantile_logn_gap_whp}
For the quantile process $\Quantile(\delta)$ with $\delta \in [1/3,2/3]$ and any $m \geq 1$,
\[
\Pro{\Gap(m) \leq 300 \log n} \geq 1 - \Oh(n^{-2}).
\]
\end{restatable}
\begin{proof}
We will show that the $\Quantile(\delta)$ process for $\delta \in [1/3, 2/3]$ satisfies the preconditions of \cref{thm:one_plus_beta_theorem}. For the potential $\Phi_0$, we pick $\alpha_2:=0.01$. The probability vector of $\Quantile(\delta)$ is clearly non-decreasing in $i$, and choosing $\epsilon:=\frac{1}{12}$ it also satisfies 
\[
p_{n/3} = \frac{\delta}{n} \leq \frac{1 - 4 \cdot \frac{1}{12}}{n} = \frac{3}{4} \cdot \frac{1}{n} \quad \text{and} \quad p_{2n/3} = \frac{1 + \delta}{n} \geq \frac{1 + 4 \cdot \frac{1}{12}}{n} = \frac{1 + \frac{1}{3}}{n}.
\]
Hence \cref{thm:one_plus_beta_theorem} yields $\ex{\Phi_{0}^{(m)}} \leq cn$. Using Markov's inequality, $\Pro{\Phi_{0}^{(m)} \leq n^3} \geq 1 - \Oh(n^{-2})$ for sufficiently large $n$. Assume the gap is $\Gap(m) > 300 \cdot\log n$, then
\[
\Phi_{0}^{(m)} \geq \exp{ (\alpha_2  \cdot \Gap(m))} > \exp{(0.01 \cdot 300 \cdot \log n)} = n^{3},
\]
which is a contradiction. Hence, $\Gap(m) \leq 300 \cdot \log n$ \wp~at least $1 - \Oh(n^{-2})$.
\end{proof}

We note that an alternative way of proving a gap bound of $\Oh(\log n)$ is to use the fact that any $\Quantile(\delta)$ process with $\delta \in (0,1)$ is majorized by a $(1+\beta)$-process with $\beta=\min\{\delta,1-\delta\}$ (see Lemma~\ref{lem:thomas}). However, this proof is less direct and leads to a slightly worse constant, which is why we  apply~\cref{thm:one_plus_beta_theorem} to $\Quantile(\delta)$ instead.

However, to analyze the process with more than one quantile in the next section, we will need a tighter analysis. We prove the following refined version of \cref{thm:one_plus_beta_theorem}:
\begin{restatable}{thm}{highprobbetapotential} 
\label{thm:high_prob_beta_potential}
Consider any probability vector $p$ that is $(i)$ non-decreasing in $i$, i.e., $p_{i} \leq p_{i+1}$ and $(ii)$ for $\epsilon = 1/12$, 
\[
p_{n/3} \leq \frac{1-4\epsilon}{n} \quad \text{ and } \quad p_{2n/3} \geq \frac{1 + 4\epsilon}{n}.
\]
Then, for any $t \geq 0$ and $\alpha_{2}:=0.0002$, $c := c_{\epsilon, \alpha_2} := 2 \cdot  40 \cdot 128^3 \cdot \epsilon^{-7} \cdot 4 \cdot \alpha_2^{-1}$, 
\[
\Pro{ \bigcap_{s \in [t, t + n \log^5 n]} \{ \Phi_0^{(s)} \leq 2cn \} } \geq 1 - n^{-3}.
\]
\end{restatable}

Note that \cref{thm:high_prob_beta_potential} not only implies a gap of $\Oh(\log n)$ using Markov's inequality (as \cref{thm:one_plus_beta_theorem}), but also that for any fixed time $s$, the number of bins  with load at least $s/n + \lambda$  is at most $2cn / \exp(\alpha_2 \cdot \lambda)$ for any $\lambda \geq 0$. In particular, for any $\lambda=\Theta(\log n)$, only a polynomially small fraction of all bins have load at least $s/n + \lambda$.

\begin{figure}[H]
\begin{tikzpicture}[
  txt/.style={anchor=west,inner sep=0pt},
  Dot/.style={circle,fill,inner sep=1.25pt},
  implies/.style={-latex,dashed},yscale=0.95]

\def\betaO{0}
\def\betaA{1}
\def\betaB{4}
\def\betaC{7}
\def\End{13.1}
\def\yO{-4.4}

\draw[dashed] (\betaA, 0) -- (\betaA, \yO);
\node[anchor=north] at (\betaA, \yO) {$t$};
\draw[dashed] (\betaB, -2) -- (\betaB, \yO);
\node[anchor=north] at (\betaB, \yO) {$t + n \log^2 n$};
\draw[dashed] (\End, 0) -- (\End, \yO);
\node[anchor=north] at (\End, \yO) {$t + n \log^5 n$};

\node (indstep) at (\betaA,0.5) {};
\node[txt] at (indstep) {$\Gamma_1^{(s)} \leq cn^9$ for all $s \in [t, t + n \log^5 n]$ (\cref{lem:gamma_potential_linear}~$(ii)$+Markov's Inequality)};
\draw[|-|, thick] (\betaA, 0) -- (\End, 0) ;

\node (PsiSmall) at (\betaA+0.1,-0.5) {};
\node[txt] at (PsiSmall) {$\Gamma_2^{(t)} \leq n^{4/3}$};
\node[Dot] at (\betaA, -0.75){};

\node (ExistsPsiLinear) at (\betaA+0.5,-1.5) {};
\node[txt] at (ExistsPsiLinear) {$\exists s_0 : \Gamma_2^{(s_0)} \leq cn$ \Whp};
\node[Dot] at (\betaA+1, -2){};
\draw[|-|, thick] (\betaA, -2) -- (\betaB, -2);

\node (ExPsiLinear) at (\betaA + 1, -3) {};
\node[txt] at (\betaA + 1, -3) {$\Gamma_2^{(s)} \leq cn$ for all $s \in [s_0, t + n \log^5 n]$};
\draw[|-|, thick] (\betaA+1, -3.5) -- (\End, -3.5);

\draw[-latex, thick] (\betaO - 0.5, \yO) -- (\End + .8, \yO);

\draw[implies] (indstep) edge[bend right=90] (PsiSmall);
\node[anchor=east, black!50!blue] at (\betaA - 0.45, -0) {\small \cref{lem:gamma_1_poly_implies}~$(ii)$};

\draw[implies] (PsiSmall) edge[bend right=70] (ExistsPsiLinear);
\node[anchor=east, black!50!blue] at (\betaA - 0.2, -1) {\small \cref{lem:gamma_1_poly_n_implies_gamma_2_linear_whp}};

\draw[implies] (ExistsPsiLinear) edge[bend right=90] (ExPsiLinear);
\node[anchor=east, black!60!green] at (\betaA + 0.1, -2.7) {\small Starting point};

\draw[implies,black!60!green] (\betaC - 1.2, 0.2) edge[bend left=30] (\betaC + 1.1, -3.2);
\node[text width=4cm, anchor=east,black!60!green] at (\betaC + 3.5, -2) {\small Bounded difference (\cref{lem:gamma_1_poly_implies}~$(iii)$)};

\draw[implies,black!60!green] (\betaC + 4.2, 0.2) edge[bend left=50] (\betaC + 1.1, -3.2);
\node[anchor=east,black!60!green, text width=3cm] at (\betaC + 6.5, -2) {\small Expectation drop \\ (\cref{lem:gamma_conditional_inequality}~$(i)$)};

\node[black!60!green] at (\betaC - 0.5, -4) {\small %
Completion of the Proof of \cref{thm:high_prob_beta_potential} (\cref{sec:completing_high_prob_beta_potential_proof})};

\end{tikzpicture}
\caption{Outline for the proof of \cref{thm:high_prob_beta_potential}. Results in green are used in the application of the concentration inequality (\cref{thm:chung_lu_theorem_8_5})~in \cref{thm:high_prob_beta_potential}.} 
\label{fig:base_case_proof_outline}
\end{figure}
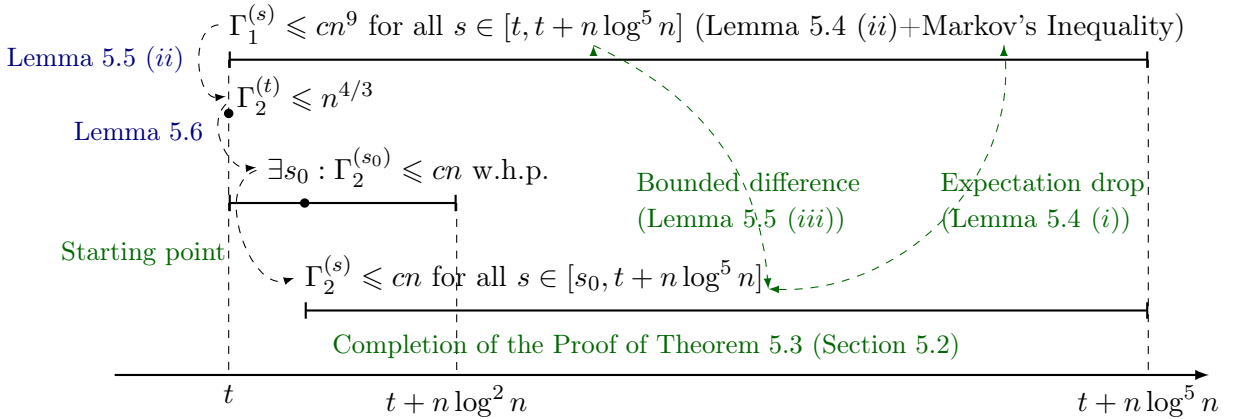

\textbf{Proof Outline of Theorem~\ref{thm:high_prob_beta_potential}.} In order to prove that $\Phi_0$ is small, we will reduce it to the potential function $\Gamma$ used in~\cite{PTW15}:
\[
\Gamma^{(s)} := \sum_{i=1}^n \left( \exp \bigl(\alpha (x_i^{(s)} - s/n) \bigr) + \exp \bigl(-\alpha (x_i^{(s)} - s/n) \bigr) \right),
\]
for some constant $0 < \alpha < 1/(6 \cdot 12)$.
Note that if $\alpha = \alpha_2$, then $\Phi_0^{(s)} \leq \Gamma^{(s)}$, so it suffices to upper bound $\Gamma^{(s)}$. It is crucial that this potential includes both the $\exp(\alpha(\cdot))$ and $\exp(-\alpha(\cdot))$ terms, as otherwise the potential may not decrease, even if it is large~(see~\cite[Appendix]{PTW15}). %
\begin{restatable}[Theorem 2.9 and 2.10 in~\cite{PTW15}]{lem}{gammapotentiallinear} \label{lem:gamma_conditional_inequality}\label{lem:gamma_potential_linear}
For any process satisfying the conditions of \cref{thm:high_prob_beta_potential}, $(i)$ for any $t \geq 0$,
\[
\Ex{\Gamma^{(t+1)} \mid \Gamma^{(t)}} \leq \Big(1 - \frac{\epsilon'_{\alpha}}{n}\Big) \cdot \Gamma^{(t)} + c',
\]
where $\epsilon'_{\alpha} := \frac{\alpha \epsilon}{4}$ and $c' := \frac{40 \cdot 128^3}{\epsilon^5}$.
Furthermore, $(ii)$ for any $t \geq 0$, $\Ex{\Gamma^{(t)}} \leq cn$.
\end{restatable}
To obtain the stronger statement that $\Gamma^{(t)}=\Oh(n)$ \Whp, we will be using two instances of the potential function: $\Gamma_1$ with $\alpha_1 = 0.01$ and $\Gamma_2$ with $\alpha_2 = 0.0002$; so $\Gamma_1 \geq \Gamma_2$. The interplay between these two potentials is shown in \cref{fig:base_case_proof_outline}. We pick $\alpha_1$ such that $12.1 \cdot \frac{\alpha_1}{\alpha_2} < \frac{1}{3}$ and hence the additive change of $\Gamma_2$ (given $\Gamma_1$ is small) is at most $n^{1/3}$:

\begin{restatable}{lem}{gammaonepolyimplications} \label{lem:gamma_1_poly_implies}
For any $t \geq 0$, if $\Gamma_1^{(t)} \leq cn^9$, then, $(i)$
$
\bigl|x_i^{(t)} - \frac{t}{n} \bigr| \leq \frac{9.1}{\alpha_1} \log n$ for all $i \in [n]$, $(ii)$
$\Gamma_2^{(t)} \leq n^{4/3},
$
and, $(iii)$
$
 | \Gamma_2^{(t+1)} - \Gamma_2^{(t)} | \leq n^{1/3}.
$
\end{restatable}

The precondition of \cref{lem:gamma_1_poly_implies} is easy to satisfy thanks to \cref{lem:gamma_conditional_inequality} and Markov's inequality. The next lemma proves a weaker version of~\cref{thm:high_prob_beta_potential}, in the sense that the potential $\Gamma_2^{(s)}$ is small in \emph{at least} one step. Note that due to the choice of $\alpha_1$ and $\alpha_2$, we have $c > \frac{2c'}{\epsilon'_{\alpha_2}}$.  
\begin{restatable}{lem}{gammaonepolynimpliesgammatwolinearwhp} \label{lem:gamma_1_poly_n_implies_gamma_2_linear_whp}
For any $t \geq n \log^2 n$, for constants $c'>0, \epsilon'_{\alpha_2} >0$ defined as above,
\[
\Pro{\bigcup_{s \in [t - n \log^2 n, t]} \{ \Gamma_2^{(s)} \leq \frac{2c'}{\epsilon'_{\alpha_2}} \cdot n \}} \geq 1 - 2cn^{-8}.
\]
\end{restatable}

To prove the strong version that $\Gamma_2^{(s)}$ is small at \emph{all} time-steps in $[t, t + n \log^5 n]$, we will use \cref{lem:gamma_1_poly_n_implies_gamma_2_linear_whp} to obtain a starting point $s_0$. For the following time-steps, we bound the expected value of $\Gamma_2^{(s)}$ for $s \geq s_0$, using \cref{lem:gamma_potential_linear}. Then we apply a concentration inequality for supermartingales~(Theorem~\ref{thm:chung_lu_theorem_8_5}), and use the bounded difference $| \Gamma_2^{(s+1)} - \Gamma_2^{(s)} | \leq n^{1/3}$ for all $s \geq t$ (\cref{lem:gamma_1_poly_implies}). %

\subsection{Tools for Theorem~\ref{thm:high_prob_beta_potential}}

\gammaonepolyimplications*

\begin{proof}%
\emph{First Statement.} For any bin $i \in [n]$,
\[
\Gamma_1^{(t)} \leq cn^9 \Rightarrow e^{\alpha_1\cdot(x_i^{(t)} - \frac{t}{n})} + e^{-\alpha_1\cdot(x_i^{(t)} - \frac{t}{n})} \leq cn^9 \Rightarrow 
x_i^{(t)} - \frac{t}{n} \leq \frac{9.1}{\alpha_1} \log n \, \wedge \,
 \frac{t}{n} - x_i^{(t)} \leq \frac{9.1}{\alpha_1} \log n,
\]
where in the second implication we used $\log c + \frac{9}{\alpha_1} \log n \leq \frac{9.1}{\alpha_1} \log n$, for sufficiently large $n$.

\emph{Second Statement.} By the definition of $\Gamma_2^{(t)}$ and the bound on each bin load,
\[
 \Gamma_2^{(t)} < 2 \cdot \sum_{i=1}^n \exp\Big( \alpha_2 \cdot \frac{9.1}{\alpha_1} \cdot \log n \Big)
 \leq 2n \cdot n^{1/4} < n^{4/3}.
\]

\emph{Third Statement.} Consider $\Gamma_2^{(t+1)}$ as a sum over $2n$ exponentials, which is obtained from $\Gamma_2^{(t)}$ by slightly changing the values of the $2n$ exponents. The total $\ell_1$-change in the exponents is upper bounded by $4$, as we will increment one entry in the load vector $x^{(t)}$ (and this entry appears twice), and we will also increment the average load by $\frac{1}{n}$ in all $2n$ exponents. Since $\exp(.)$ is convex, the largest change is upper bounded by the (hypothetical) scenario in which the largest exponent increases by $4$ and all others remain the same,
\begin{align*}
 \left| \Gamma_2^{(t+1)} - \Gamma_2^{(t)} \right| 
 & \leq \exp \Big( \alpha_2 \cdot \max\{ x_{\max}^{(t)} + 4 - t/n, t/n - x_{\min}^{(t)} - 4 \} \Big) \\
 & \leq e^{4 \alpha_2} \cdot \exp\Big( \alpha_2 \cdot \frac{9.1}{\alpha_1} \cdot \log n \Big) = e^{4 \alpha_2} \cdot \exp \Big( 0.0002 \cdot \frac{9.1}{0.01} \cdot \log n \Big) \leq n^{1/3}.
\end{align*}
\end{proof}

\begin{clm}
\label{clm:large_gamma_exponential_drop}
For any step $t\geq 0$, $\ex{\Gamma_2^{(t+1)} \,\mid\, \Gamma_2^{(t)}, \Gamma_2^{(t)} \geq \frac{2c'}{\epsilon'_{\alpha_2}} \cdot n} \leq 
(1-\frac{\epsilon'_{\alpha_2}}{2n}) \cdot \Gamma_2^{(t)}$.
\end{clm}
\begin{proof}
If  $\Gamma_2^{(t)} \geq \frac{2c'}{\epsilon'_{\alpha_2}} \cdot n$, then the inequality of \cref{lem:gamma_conditional_inequality} yields,
\begin{align*}
\ex{\Gamma_2^{(t+1)} \,\mid\, \Gamma_2^{(t)}, \Gamma_2^{(t)} \geq \frac{2c'}{\epsilon'_{\alpha_2}} \cdot n} & \leq \Gamma_2^{(t)} - \frac{\epsilon'_{\alpha_2}}{n} \cdot \Gamma_2^{(t)} + c' \\
 & \leq
\Gamma_2^{(t)} - \frac{\epsilon'_{\alpha_2}}{2n} \cdot \Gamma_2^{(t)} + \Big(c'- \frac{\epsilon'_{\alpha_2}}{2n} \cdot \Gamma_2^{(t)}\Big)
 \leq 
\Big(1-\frac{\epsilon'_{\alpha_2}}{2n}\Big) \cdot \Gamma_2^{(t)}.
\end{align*}
\end{proof}

\gammaonepolynimpliesgammatwolinearwhp*
The proof of this lemma relies on the two statements in \cref{lem:gamma_potential_linear}.

\begin{proof}
By \cref{lem:gamma_potential_linear}, using Markov's inequality at time $t-n\log^2 n$, we have
\[
\Pro{\Gamma_1^{(t-n\log^2n)} \leq cn^9} \geq 1 - cn^{-8}. 
\]
Assuming $\Gamma_1^{(t-n\log^2n)} \leq cn^9$, then
the second statement of \cref{lem:gamma_1_poly_implies}
implies
$\Gamma_2^{(t- n\log^2n)} \leq n^{4/3}$. By \cref{clm:large_gamma_exponential_drop} if at some step $\Gamma_2^{(r)} > \frac{2c'}{\epsilon'_{\alpha_2}} \cdot n$, then \[
\Ex{\Gamma_2^{(r+1)} \,\mid\, \Gamma_2^{(r)}} \leq 
(1-\frac{\epsilon'_{\alpha_2}}{2n}) \cdot \Gamma_2^{(r)}.
\]
For any $r \in [t - n \log^2 n, t]$, we define the ``killed'' potential function,
\[
\tilde{\Gamma}_2^{(r)} := \Gamma_2^{(r)} \cdot \mathbf{1}_{\bigcap_{ \tilde{r} \in [t - n\log^2 n, r]} \{ \Gamma_2^{(\tilde{r})} > \frac{2c'}{\epsilon'_{\alpha_2}} \cdot n\} }.
\]
This satisfies the inequality of \cref{lem:gamma_conditional_inequality}  without any constraint on the value of $\tilde{\Gamma}_2^{(r)}$, that is,
\[
\Ex{\tilde{\Gamma}_2^{(r+1)} \,\mid\, \tilde{\Gamma}_2^{(r)}} \leq (1-\frac{\epsilon'_{\alpha_2}}{2n}) \cdot \tilde{\Gamma}_2^{(r)}.
\]
Inductively applying this for $n \log^2 n$ steps, and noting that $\epsilon'_{\alpha_2} = \frac{\alpha_2 \epsilon}{4} < 1$, %
\[
\Ex{\tilde{\Gamma}_2^{(t)} \mid \tilde{\Gamma}_2^{(t - n \log^2 n)}} \leq e^{- \frac{n \log^2 n}{n}} \cdot n^{4/3} \leq n^{-7}. 
\]
So, by Markov's inequality, %
\[
 \Pro{\tilde{\Gamma}_2^{(t)} \geq n  \, \mid\, \Gamma_1^{(t-n\log^2n)} \leq cn^9} \leq n^{-{8}} \Rightarrow \Pro{\tilde{\Gamma}_2^{(t)} \geq n} < 2cn^{-8}.
\]
Due to the definition of $\tilde{\Gamma}_2$
we conclude that~w.p.~at least $1 - 2cn^{-8}$, there must be at least one time step $s \in [t - n\log^2 n, t]$, $\Gamma_2^{(s)} \leq \frac{2c'}{\epsilon'_{\alpha_2}} \cdot n$.
\end{proof}

\subsection{Completing the proof of Theorem~\ref{thm:high_prob_beta_potential}} \label{sec:completing_high_prob_beta_potential_proof}

\highprobbetapotential*
\begin{proof}%

The proof will be concerned with time-steps $\in [t - n \log^2 n,t+n \log^5 n]$. First, by applying
 \cref{lem:gamma_1_poly_n_implies_gamma_2_linear_whp}, 
 \begin{align*}
\Pro{\bigcup_{s \in [t - n \log^2 n, t]} \{ \Gamma_2^{(s)} \leq \frac{2c'}{\epsilon'_{\alpha_2}} \cdot n \} } &\geq 1 - 2cn^{-8}.
 \end{align*}
Assuming that such a time $s \in [t - n \log^2n,t]$ indeed exists, we partition the time-steps $r \in [s,t+n \log^5 n]$ into red and green phases (see \cref{fig:gamma_2_small_large_regions}):
\begin{enumerate}\itemsep-4pt
    \item \textbf{Red Phase:} The process at step $r$ is in a red phase if $\Gamma_2^{(r)} > \frac{2c'}{\epsilon'_{\alpha_2}} \cdot n =: T$.
    \item \textbf{Green Phase:} Otherwise, the process is in a green phase.
\end{enumerate}

\begin{figure}[t]
    \centering
    \includegraphics[height=5cm]{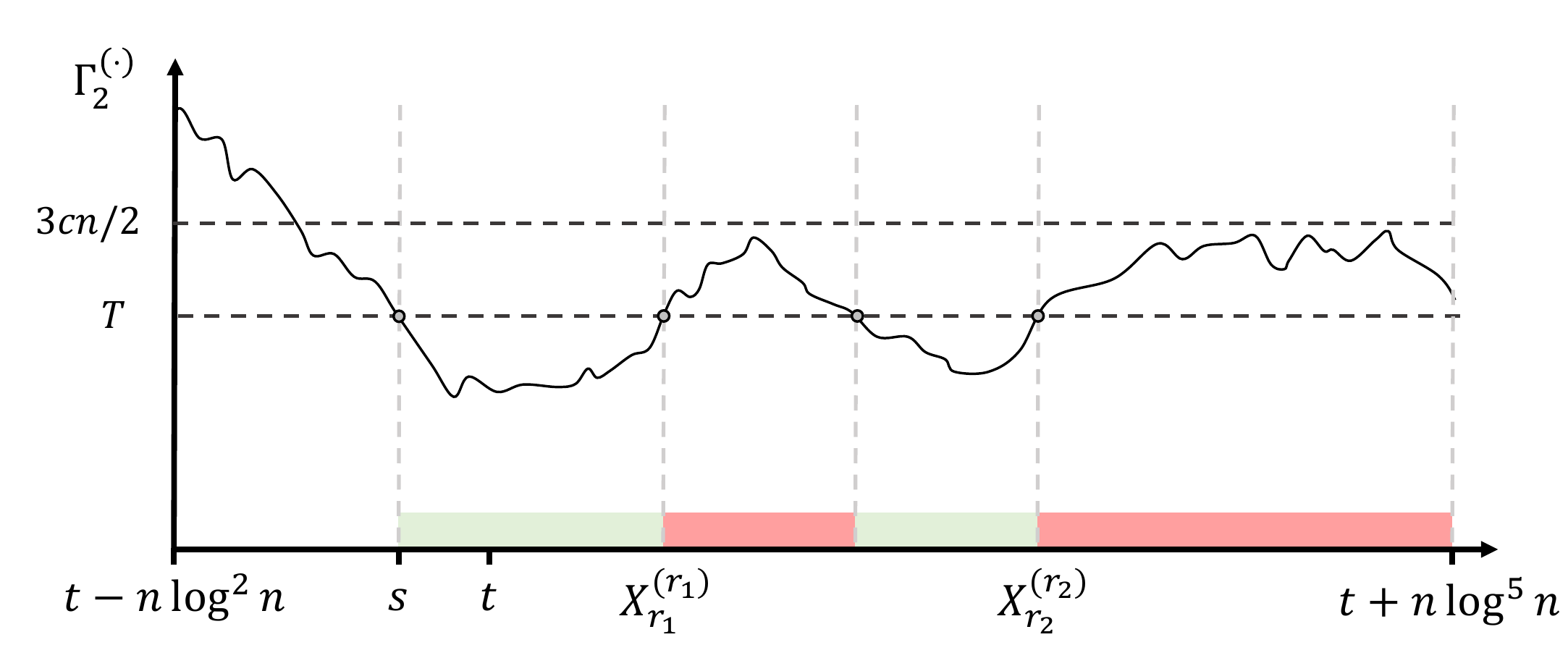}
    \caption{Green phases indicate time-steps where $\Gamma_2^{(r)}$ is small and red phases indicate time-steps for which the potential is large and drops (in expectation). The main objective is to prove a $<3cn/2$ guarantee at every point within a red phase using a concentration inequality.} %
    \label{fig:gamma_2_small_large_regions}
\end{figure}

Note that by the choice of $s$, the process is at a green phase at time $s$, which means that every red phase is preceded by a green phase. Obviously, for steps in a green phase, we have $\Gamma_2^{(r)} \leq T$. Hence for $r$ being a (possible) first step of a red phase after a green phase, it follows that $\Gamma_{2}^{(r+1)} \leq e^{\alpha_2} \cdot \Gamma_{2}^{(r)} \leq 2 \cdot \Gamma_{2}^{(r)}$, and therefore
\begin{align}
\Gamma_{2}^{(r+1)} \leq 2 \cdot T. \label{eq:twoT}
\end{align}

The remaining part of the proof is to analyze $\Gamma_2^{(r)}$ during the time steps of a red phase, and to establish that $\Gamma_{2}^{(r)}=\Oh(n)$ holds for all $r \in [s, t + n \log^5 n]$ (see \cref{fig:gamma_2_small_large_regions}).

The idea of this partitioning is that  within a red phase, i.e., $\Gamma_2^{(r)} \geq \frac{2c'n}{\epsilon'_{\alpha_2}}$, so by \cref{clm:large_gamma_exponential_drop}, \begin{align}
\Ex{\Gamma_2^{(r+1)} \mid \Gamma_2^{(r)} \geq \frac{2c'n}{\epsilon'_{\alpha_2}}} \leq \Gamma_2^{(r)}. \label{eq:keypoint}
\end{align}
In order to analyze the behavior of $\Gamma_2$ till the end of a red phase, we define for every $r \in [s,t+n \log^5 n]$ being the (potential) beginning of a red phase, a stopping time $\tau(r) := \min \{ u \geq r \colon \Gamma_2^{(u)} \leq \frac{2c'n}{\epsilon'_{\alpha_2}} \}$. Further, define
\[
  X_{r}^{(u)} := \Gamma_{2}^{(u \wedge \tau_{r})}.
\]

Our goal is to apply a concentration inequality for supermartingales (\cref{thm:chung_lu_theorem_8_5}) to $X_{r}^{(u)}$. As a preparation, we will first derive some basic bounds for $\Gamma_1^{(r)}$.
For any time step $r \in [t - n \log^2 n, t+n \log^5 n]$, 
applying \cref{lem:gamma_potential_linear} and using Markov's inequality, we have that $\Pro{\Gamma_1^{(r)} \leq c n^{9}} \geq 1 - cn^{-8}$. Hence, by the union bound over $[t - n \log^2 n, t+n \log^5 n]$, 
\[
\Pro{ \bigcap_{u \in [t-n \log^2 n, t+ n \log^ 5 n]} \{ \Gamma_1^{(u)} \leq c n^{9} \} } \geq 1 - \frac{2\log^5 n}{n^7}.
\]
Following the notation in~\cref{thm:chung_lu_theorem_8_5}, we set
$\overline{B^{(r)}} := \bigcap_{u \in [r, t + n \log^5 n]} \{ \Gamma_1^{(u)} \leq c n^9 \}$. As for the three preconditions in ~\cref{thm:chung_lu_theorem_8_5}, we obtain by choice of $\overline{B^{(r)}}$: 
\begin{enumerate}\itemsep-2pt
  \item $\ex{X_r^{(u)} \mid \mathcal{F}_r^{(u-1)}} \leq X_r^{(u - 1)}$. This holds, since if event $\overline{B^{(r)}}$ occurs, we can apply \cref{clm:large_gamma_exponential_drop} to deduce that~\cref{eq:keypoint} holds.
  \item %
  \begin{align*}
  \var{X_i^{(u)} \mid \mathcal{F}_i^{(u-1)}} & \leq \frac{1}{4} \big| \bigl( \max X_{i}^{(u)} - \min X_i^{(u)}  \bigr) \, \mid \, \mathcal{F}_i^{(u-1)} \bigr|^2 \\
  & \leq \frac{1}{4} \big| \bigl( \max X_{i}^{(u)} - \min X_i^{(u)}  \bigr) \, \mid \, \mathcal{F}_i^{(u-1)} \bigr|^2 \leq n^{2/3},
  \end{align*}
  where the first inequality follows by Popovicius' inequality, the second and third by the triangle inequality and \cref{lem:gamma_1_poly_implies} (third statement).
  \item $X_r^{(u)} - \ex{X_r^{(u)} \mid \mathcal{F}_r^{(u-1)}} \leq \bigl| \Gamma_2^{(u)} - \Gamma_2^{(u-1)} \bigr| \mid \mathcal{F}_r^{(u-1)} \leq 2n^{1/3}$ by \cref{lem:gamma_1_poly_implies} (third statement).
\end{enumerate}

Since each possible red phase will end before $t+2n \log^5 n$, applying \cref{thm:chung_lu_theorem_8_5} for any $u \in [r,t+2n \log ^5 n]$ gives
\[
\Pro{X_r^{(u)} \geq X_r^{(r)} + \frac{cn}{2}} \leq \exp\left( - \frac{c^2n^2/4}{2 \cdot (2n \log^5 n) \cdot (2 n^{1/3}) } \right) + \frac{2\log^5 n}{n^7} \leq  \frac{3 \log^5 n}{n^7}.
\]
Also recall that $X_{r}^{(r)} \leq 2 T \leq cn$ by \cref{eq:twoT}, so if a red phase starts at time $r$, then with probability $1-\frac{3 \log^5 n}{n^7}$, $\Gamma_{2}^{(u)}$ will always be $\leq \frac{3cn}{2}$. Now taking the union bound over $u \in [r, t + 2 n \log^5 n]$ and %
taking a union bound over all possible starting points of a red phase yields:
\[
\Pro{\bigcup_{r \in [s, t + n \log^5 n]}  \bigcup_{u \in [r, t + n \log^5 n]} \{ X_{r}^{(u)} > \frac{3cn}{2} \} } \leq 3 \cdot \frac{\log^5 n}{n^7} \cdot (4 n^2 \log^{10} n) \leq \frac{1}{2} n^{-4}.
\]
Hence with probability $1-\frac{1}{2}n^{-4}-2cn^{-8} \geq 1-n^{-4}$, it holds that $\Gamma_2^{(r)} \leq \frac{3cn}{2}$ for all time-steps $r$ which are within a red phase in $[s,t+n \log^ 5 n] \subseteq [t,t+n \log^5 n]$. 
Since $\Gamma_2^{(r)} \leq T \leq cn$ holds (deterministically) by definition for all time-steps $r$ within a green phase, the theorem follows.\end{proof}

\section{Upper Bounds for More Than One Quantile} \label{sec:new_upper_bound_for_k_queries}

\subsection{Upper Bounds on the Original Quantile Process and Consequences}

We now generalize the analysis from \cref{sec:one_quantile} for one quantile to $2 \leq k \leq  \kappa \cdot \log \log n$ quantiles, where $\kappa:=1/\log(10^4)$. %
We emphasize that our chosen quantiles are oblivious and even uniform, i.e., independent of $t$ (but dependent on $n$). Specifically, we define 
\[
\tilde{\delta}_{i} = \begin{cases}
  \frac{1}{2} &  \quad \text{ for } i = k, \\
  2^{-0.5 (\log n)^{(k-i)/k}} &  \quad \text{ for $1 \leq i < k$ },
  \end{cases}
\]
and let each $\delta_i$ be $\tilde{\delta}_i$ rounded up to the nearest multiple of $\frac{1}{n}$. The intuition is that the largest quantile $\delta_k=\frac{1}{2}$ ensures that the load distribution is at least ``coarsely'' balanced, analogous to the $(1+\beta)$-process. 
All smaller quantiles $\delta_1,\delta_2,\ldots,\delta_{k-1}$ almost always returns a negative answer, but they gradually reduce the probability of allocating to a heavily loaded bin. 

\begin{thm}[simplified version of Theorem \ref{thm:multiple_quantiles_relaxed}]\label{thm:new_multiple_quantiles}
For any integer $2 \leq k \leq \kappa \log \log n$, consider the $\Quantile(\delta_1,\delta_2,\ldots,\delta_k)$ process with the $\delta_i$'s defined above. Then for any $m \geq 1$,
\[
 \Pro{ \Gap(m) \leq 1000 \cdot k \cdot (\log n)^{1/k} } \geq 1 - n^{-3}. %
\]
\end{thm}

For $k=2$ and $k=3$, \cref{thm:new_multiple_quantiles} directly implies the following corollary:
\begin{cor}
For $k=2$, the process $\Quantile(2^{-0.5 \sqrt{\log n}},\frac{1}{2})$ satisfies for any $m \geq 1$,
\[
 \Pro{ \Gap(m) \leq 2000 \cdot \sqrt{ \log n} } \geq 1 - n^{-3}. %
\]
Similarly, for $k=3$ the process $\Quantile(2^{-0.5 (\log n)^{2/3}},2^{-0.5 (\log n)^{1/3}},\frac{1}{2})$ satisfies for any $m \geq 1$,
\[
 \Pro{ \Gap(m) \leq 3000 \cdot (\log n)^{1/3} } \geq 1 - n^{-3}. %
\] %
\end{cor}
Using the fact that any allocation process with $k$ quantiles majorizes a suitable adaptive (and randomized) $2k$-\textsc{Thinning} process~(\cref{lem:threshold_thinning}), we also obtain:
\begin{cor}
For any even $d \leq \frac{2}{\kappa} \log \log n$, there is an (adaptive and randomized) $d$-\textsc{Thinning}  process, satisfying for any $m \geq 1$,
$
 \Pro{ \Gap(m) \leq 2000 \cdot d \cdot (\log n)^{ (2/d) } } \geq 1 - n^{-3}. %
$
\end{cor}

In particular for the $d$-\textsc{Thinning} process where the $j$-th decision is given by the quantile $\delta_{j}$, we get an extension of~\cite[Theorem 1.1]{FL20} to the heavily-loaded case

\begin{lem}
Consider the $d$-\textsc{Thinning} process for $d < \kappa \log \log n$ induced by the quantiles $\delta_1, \ldots , \delta_{d-1}$, given by $\delta_j = 2^{-0.5(\log n)^{j/(d-1)}}$. Then there is a constant $c > 0$ such that for every $m \geq 1$,
\[
\Pro{\Gap(m) \leq c \cdot (d-1) \cdot (\log n)^{1/(d-1)}} \geq 1 - n^{-2}.
\]
\end{lem}
\begin{proof}
We will show that this $d$-\textsc{Thinning} process is majorized by the $\Quantile(\delta_1, \ldots, \delta_{d-1})$ process. We consider two cases for bins $i \in [n]$:
\begin{itemize}
 \item \textbf{Case 1 [$i \leq n \cdot \delta_1$]:} We allocate to  $i \leq n \cdot \delta_1$ if the first $d$ samples are heavy and the last one is equal to $i$. So,
 \[
 p_i = \delta_1 \cdot \delta_2 \cdot \ldots \cdot \delta_{d} \cdot \frac{1}{n} \leq \frac{\delta_1}{n}.
 \]
 \item \textbf{Case 2 [$j < d - 1$, $n \cdot \delta_{j} < i \leq n \cdot \delta_{j+1}$]:} Let $j_1, \ldots, j_{d}$ be the sampled bins. The probability of allocating to $i$ is if the first $\ell_1 \geq j$ samples where heavy and then we picked $i$:
 \begin{align*}
 p_i & = \frac{1}{n} \cdot \sum_{\ell_1 = j}^k \prod_{\ell_2 = 1}^{\ell_1} \Pro{j_{\ell_2} \leq n \cdot \delta_{\ell_2} } \\
 & = \frac{1}{n} \cdot \delta_1 \cdot \ldots \cdot \delta_j \cdot (1 + \delta_{j+1} + \delta_{j+1} \cdot \delta_{j+2} + \ldots + \delta_{j+1} \cdot \ldots \cdot \delta_k) \\
 & \leq \frac{1}{n} \cdot \delta_1 \cdot \ldots \cdot \delta_j \cdot \Big(1 + \delta_{j+1} + \delta_{j+1} \cdot \frac{1}{2} + \delta_{j+1} \cdot \frac{1}{2^2} + \ldots \Big) \\
 & \leq \frac{1}{n} \cdot \delta_{j} \cdot (1 + 2 \delta_{j+1}) \\
 & \leq \frac{1}{n} \cdot (\delta_{j}  + 2 \cdot \delta_{j}\cdot \delta_{j+1}) \\
 & \leq \frac{1}{n} \cdot (\delta_{j}  + \delta_{j+1}),
 \end{align*}
 using that $\delta_j \leq 1/2$ for any $j$.
\end{itemize}
Hence, by \cref{thm:new_multiple_quantiles}  majorization the $d$-\textsc{Thinning} process \Whp has an $\Oh((d-1) \cdot (\log n)^{1/(d-1)})$ gap.
\end{proof}

Finally, for $k=\Theta(\log \log n)$, the bound on the gap in \cref{thm:new_multiple_quantiles} is $C \cdot \log \log n$ for some (large) constant $C>0$. Surprisingly, this matches the gap of the full information setting (\TwoChoice process), even though the \Quantile{} process behaves quite differently. For instance, \Quantile{} cannot discriminate
among the $n/2$ most lightly loaded bins.
Also since any \Quantile{} process majorizes \TwoChoice (see~\cref{cor:any_quantile_process_majorises_two_choice}), we deduce:
\begin{cor}
For \TwoChoice, there is a constant $C > 0$ such that for any $m \geq 1$,
$
 \Pro{ \Gap(m) \leq C \log \log n } \geq 1 - n^{-3}.
$
\end{cor}

This result originally shown in~\cite{BCSV06} proved the tighter bound $\Gap(m) = \log_2 \log n \pm \Oh(1)$, \Whp~However, their analysis combines sophisticated tools from Markov chain theory and computer-aided calculations. The simpler analysis in \cite{TW14} derives the same gap bound up to an additive $\Oh(\log \log \log n)$ term, but the error probability is much larger, i.e., $\Theta((\log \log n)^{-4})$. 
In comparison to their bound, our result achieves a much smaller error probability of $\Oh(n^{-3})$, but it comes at the cost of a multiplicative constant in the gap bound.

\subsection{Relaxed Quantile Process and Outline of the Inductive Step}

We now define a class of processes $\RelaxedQuantile_{\gamma}(\delta_1,\ldots,\delta_k)$, which relaxes the definition of $\Quantile{(\delta_1,\ldots,\delta_k)}$, with $1 \leq k \leq \kappa \log \log n$ and a relaxation factor $\gamma \geq 1$. The probability vector $p$ of such a process satisfies four conditions: $(i)$, for each $i \in [n]$,
\[
p_i \leq  \begin{cases}
\gamma \cdot \frac{\delta_1}{n} & \text{for }1 \leq i \leq \delta_1 \cdot n, \\
\gamma \cdot \frac{\delta_1 + \delta_2}{n} & \text{for }\delta_1 \cdot n < i \leq \delta_2 \cdot n, \\
\ \ \ \vdots & \\
\gamma \cdot \frac{\delta_{k-1} + \delta_k}{n} & \text{for }\delta_{k-1} \cdot n< i \leq \delta_k \cdot n, \\
\end{cases}
\]
$(ii)$ the probability vector $p$ is non-decreasing in $i$, $(iii)$ $p_{n/3} \leq \frac{1-4\epsilon}{n}$ and, $(iv)$ $p_{2n/3} \geq \frac{1+4\epsilon}{n}$ for some $0 < \epsilon < 1/4$. Note that the process $\Quantile(\delta_1,\delta_2,\ldots,\delta_k)$ with the $\delta_i$'s as defined above falls into this class with $\gamma=1$ and $\eps = 1/12$ (cf.~\cref{eq:def_quantile}).

\begin{thm}[generalization of Theorem~\ref{thm:new_multiple_quantiles}]
\label{thm:multiple_quantiles_relaxed}
Consider a $\RelaxedQuantile_{\gamma}(\delta_1,\delta_2,\ldots,\delta_k)$ process with the $\delta_i$'s defined above. Let $2 \leq k \leq \kappa \log \log n$ and $1 \leq \gamma \leq 6$. Then for any $m \geq 1$,
\[
 \Pro{ \Gap(m) \leq 1000 \cdot k \cdot (\log n)^{1/k} } \geq 1 - n^{-3}. %
\]
\end{thm}

\paragraph{Reduction of~\cref{thm:multiple_quantiles_relaxed} to \cref{lem:new_inductive_step}.}

The proof of \cref{thm:multiple_quantiles_relaxed} employs some type of layered induction over $k$ different, super-exponential potential functions. Generalizing the definition of $\Phi_{0}^{(s)}$ from Section~\ref{sec:one_quantile}, for any $0 \leq j \leq k-1$:
\[
 \Phi_{j}^{(s)} := \sum_{i=1}^n \exp \Big( \alpha_2 \cdot (\log n)^{j/k} \cdot \Big(x_i^{(s)} - \frac{s}{n} - \frac{2}{\alpha_2} j (\log n)^{1/k} \Big)^{+} \Big),
\]
where $\alpha_2 = 0.0002$ (recall $z^{+}=\max\{z,0\}$). 
We will then employ this series of potential functions $j=0,1,\ldots,k-1$ to analyze the process over the time-interval $s \in [m - n \log^5 n, m]$.

The next lemma (\cref{lem:new_inductive_step}) formalizes this inductive argument. It shows that if for all steps $s$ within some suitable time-interval, the number of balls of height at least $\frac{s}{n} + \frac{2}{\alpha_2}j (\log n)^{1/k}$ is small, then the number of balls of height at least $\frac{s}{n} + \frac{2}{\alpha_2}(j+1) (\log n)^{1/k}$ is even smaller. This ``even smaller'' is encapsulated by the (non-constant) base of $\Phi_j$, which increases in $j$; however, this comes at the cost of reducing the time-interval slightly by a $\Theta(n \log^3 n)$ term. Finally, for $j=k-1$, we can conclude that at step $s=m$, there are no balls of height $\frac{s}{n} + \frac{2}{\alpha_2}k (\log n)^{1/k}$. Hence we can infer that the gap is $\Oh( k \cdot (\log n)^{1/k})$, and \cref{thm:new_multiple_quantiles} follows (see \cref{sec:proof_of_main_appendix} for the complete proof of this step).

\begin{lem}[\textbf{Inductive Step}]\label{lem:new_inductive_step}
Assume that for some $1 \leq j \leq k \leq \frac{1}{\log(10^4)} \log \log n$, the process $\RelaxedQuantile_{\gamma}(\delta_1,\ldots,\delta_k)$ with the $\delta_i$'s as defined before, and $\gamma \leq 6$ and $t \geq 0$ satisfies:
\[
  \Pro{\bigcap_{s \in [\beta_{j-1},t+n \log^5 n]} \{ \Phi_{j-1}^{(s)} \leq 2cn\} } \geq 1 - \frac{(\log n)^{8(j-1)}}{n^4},
\]
where $\beta_{j} := t + 2jn \log^3 n$ and $c = c_{1/12, \alpha_2}$ (see~\cref{thm:high_prob_beta_potential}). Then,
it
also satisfies:
\[
  \Pro{ \bigcap_{s \in [\beta_j, t+n \log^5 n]} \{ \Phi_{j}^{(s)} \leq 2cn \} } \geq 1 - \frac{(\log n)^{8j}}{n^4}.
\]
\end{lem}

As in \cref{sec:one_quantile}, we will also use a second version of the potential function to extend an expected bound on the potential into a \Whp~bound. Intuitively, we exploit the property that potential functions will have linear expectations for a range of coefficients. %
With this in mind, we define the following potential function for any $0 \leq j \leq k - 1$,
\begin{align*}
 \Psi_{j}^{(s)} := \sum_{i=1}^n \exp \Big( \alpha_1 \cdot (\log n)^{j/k} \cdot \Big(x_i^{(s)} - \frac{s}{n} - \frac{2}{\alpha_2} j (\log n)^{1/k} \Big)^{+} \Big),
\end{align*}
where $\alpha_1 = 0.01$. Note that $\Psi_j$ is defined in the same way as $\Phi_j$ with the only difference that $\alpha_1$ is significantly larger $\alpha_2$. The interplay between $\Psi_{j}$ and $\Phi_{j}$ is similar to the interplay between $\Gamma_1$ and $\Gamma_2$ in the proof of \cref{thm:high_prob_beta_potential}, but some extra care is needed. In particular, while underloaded bins with load of $m/n-\Theta(\log n)$ contribute heavily to $\Gamma_1$ (or $\Gamma_2$), their contribution has to be eliminated here in order to derive a gap bound better than $\Oh(\log n)$.

\subsection{Proof Outline of Lemma~\ref{lem:new_inductive_step}.}
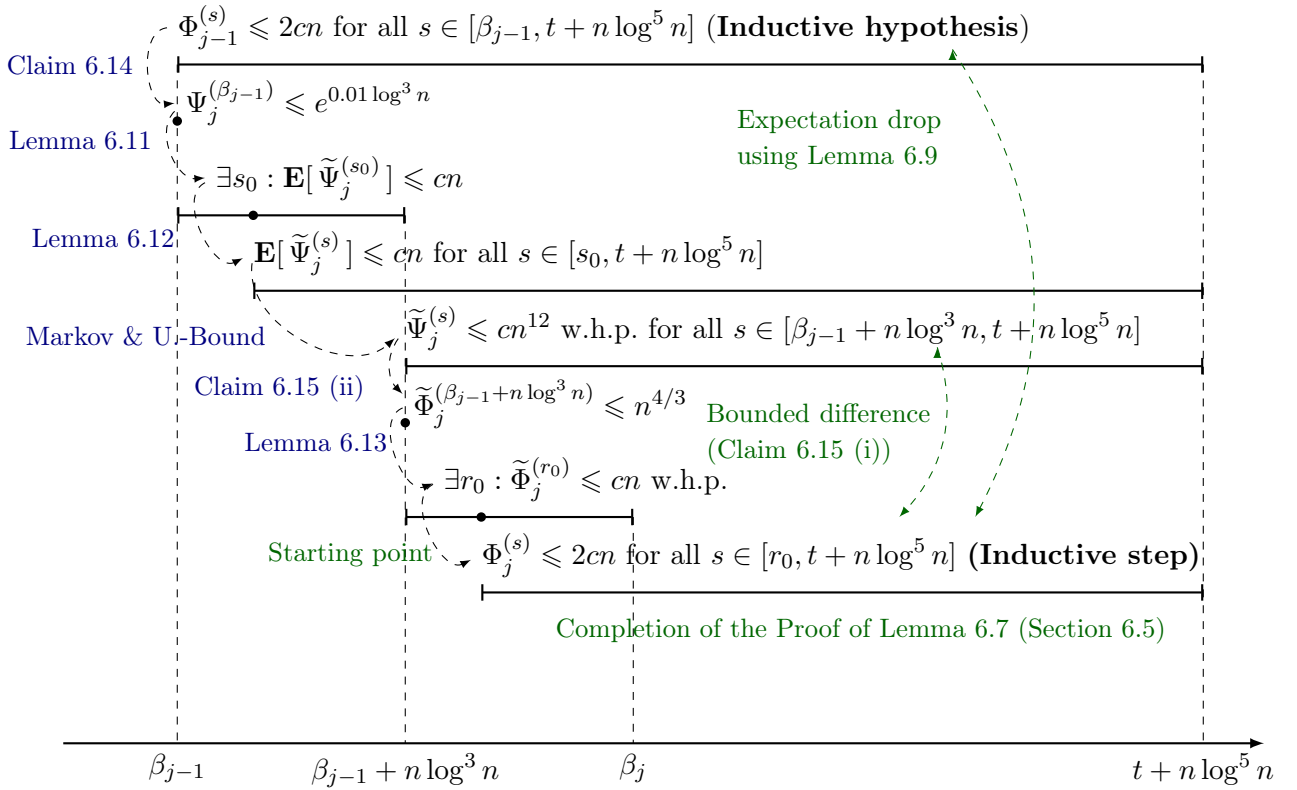
\begin{figure}[H]
\begin{tikzpicture}[
  txt/.style={anchor=west,inner sep=0pt},
  Dot/.style={circle,fill,inner sep=1.25pt},
  implies/.style={-latex,dashed}]

\def\betaO{0}
\def\betaA{1}
\def\betaB{4}
\def\betaC{7}
\def\End{14.5}
\def\yO{-9}

\draw[dashed] (\betaA, 0) -- (\betaA, \yO);
\node[anchor=north] at (\betaA, \yO) {$\beta_{j-1}$};
\draw[dashed] (\betaB, -2) -- (\betaB, \yO);
\node[anchor=north] at (\betaB, \yO) {$\beta_{j-1} + n \log^3 n$};
\draw[dashed] (\betaC, -6) -- (\betaC, \yO);
\node[anchor=north] at (\betaC, \yO) {$\beta_j$};
\draw[dashed] (\End, 0) -- (\End, \yO);
\node[anchor=north] at (\End, \yO) {$t + n \log^5 n$};

\node (indstep) at (\betaA,0.5) {};
\node[txt] at (indstep) {$\Phi_{j-1}^{(s)} \leq 2cn$ for all $s \in [\beta_{j-1}, t + n \log^5 n]$ (\textbf{Inductive hypothesis})};
\draw[|-|, thick] (\betaA, 0) -- (\End, 0) ;

\node (PsiSmall) at (\betaA+0.1,-0.5) {};
\node[txt] at (PsiSmall) {$\Psi_{j}^{(\beta_{j-1})} \leq e^{0.01 \log^3 n}$};
\node[Dot] at (\betaA, -0.75){};

\node (ExistsPsiLinear) at (\betaA+0.5,-1.5) {};
\node[txt] at (ExistsPsiLinear) {$\exists s_0 : \ex{\tilde{\Psi}_{j}^{(s_0)}} \leq cn$};
\node[Dot] at (\betaA+1, -2){};
\draw[|-|, thick] (\betaA, -2) -- (\betaB, -2);

\node (ExPsiLinear) at (\betaA + 1, -2.5) {};
\node[txt] at (\betaA + 1, -2.5) {$\ex{\tilde{\Psi}_j^{(s)}} \leq cn$ for all $s \in [s_0, t + n \log^5 n]$};
\draw[|-|, thick] (\betaA+1, -3) -- (\End, -3);

\node (PsiPoly) at (\betaB, -3.5) {};
\node[txt] at (PsiPoly) {$\tilde{\Psi}_j^{(s)} \leq cn^{12}$ \Whp~for all $s \in [\beta_{j-1}+n \log^3n, t + n \log^5 n]$};
\draw[|-|, thick] (\betaB, -4) -- (\End, -4);

\node (PhiPoly) at (\betaB+0.1,-4.5) {};
\node[txt] at (\betaB+0.1,-4.5) {$\tilde{\Phi}_{j}^{(\beta_{j-1} + n \log^3 n)} \leq n^{4/3}$};
\node[Dot] at (\betaB, -4.75){};

\node (ExistsPhiLinear) at (\betaB+0.5,-5.5) {};
\node[txt] at (ExistsPhiLinear) {$\exists r_0 : \tilde{\Phi}_{j}^{(r_0)} \leq cn$ \Whp};
\node[Dot] at (\betaB+1, -6){};
\draw[|-|, thick] (\betaB, -6) -- (\betaC, -6);

\node (PhiLinear) at (\betaB+1,-6.5) {};
\node[txt] at (\betaB+1,-6.5) {$\Phi_{j}^{(s)} \leq 2cn$ for all $s \in [r_0, t + n \log^5 n]$ \textbf{(Inductive step)}};
\draw[|-|, thick] (\betaB+1, -7) -- (\End, -7);

\draw[-latex, thick] (\betaO - 0.5, \yO) -- (\End + .8, \yO);

\draw[implies] (indstep) edge[bend right=90] (PsiSmall);
\node[anchor=east, black!50!blue] at (\betaA - 0.45, -0) {\small \cref{clm:phi_small_implies_psi_plus_one_small}};

\draw[implies] (PsiSmall) edge[bend right=70] (ExistsPsiLinear);
\node[anchor=east, black!50!blue] at (\betaA - 0.2, -1) {\small \cref{lem:there_exists_linear_ex_psi_plus_one}};

\draw[implies] (ExistsPsiLinear) edge[bend right=90] (ExPsiLinear);
\node[anchor=east, black!50!blue] at (\betaA + 0.1, -2.3) {\small \cref{lem:exists_s_st_ex_psi_linear}};

\draw[implies] (ExPsiLinear) edge[bend right=70] (PsiPoly);
\node[anchor=east, black!50!blue] at (\betaA + 1.3, -3.6) {\small Markov \& U.-Bound};

\draw[implies] (PsiPoly) edge[bend right=40] (PhiPoly);
\node[anchor=east, black!50!blue] at (\betaB - 0.4, -4.3) {\small \cref{clm:psi_potential_poly_implies}~(ii)};

\draw[implies] (PhiPoly) edge[bend right=90] (ExistsPhiLinear);
\node[anchor=east, black!50!blue] at (\betaB - 0.1, -5) {\small \cref{lem:exists_s_st_phi_linear_whp}};

\draw[implies] (ExistsPhiLinear) edge[bend right=90] (PhiLinear);
\node[anchor=east, black!60!green] at (\betaB +0.5, -6.5) {\small Starting point};

\draw[implies,black!60!green] (\betaC + 4, -3.75) edge[bend left=30] (\betaC + 3.5, -6);
\node[text width=4cm, anchor=east,black!60!green] at (\betaC + 5.1, -4.9) {\small Bounded difference (\cref{clm:psi_potential_poly_implies} (i))};

\draw[implies,black!60!green] (\betaC + 4.2, 0.2) edge[bend left=30] (\betaC + 4.5, -6);
\node[anchor=east,black!60!green, text width=3cm] at (\betaC + 4.5, -1) {\small Expectation drop \\ using \cref{lem:rec_inequality_phi_psi}};

\node[black!60!green] at (\betaC + 3, -7.5) {\small %
Completion of the Proof of \cref{lem:new_inductive_step} (\cref{sec:proof_of_key})};

\end{tikzpicture}
\caption{Outline for the proof of \cref{lem:new_inductive_step}. Results in blue are given in \cref{sec:auxiliary}, while results in green are used in the application of the concentration inequality (\cref{thm:chung_lu_theorem_8_5})~in \cref{sec:proof_of_key}. }\label{fig:beautiful} 
\label{fig:proof_outline}
\end{figure}

We will now give a summary of the main technical steps in the proof of \cref{lem:new_inductive_step} (an illustration of the key steps is shown in \cref{fig:beautiful}). On a high level, the proof mirrors the proof of~\cref{thm:high_prob_beta_potential}; however, there are some differences, especially in the final part of the proof.

First, fix any $1 \leq j \leq k - 1$. Then the inductive hypothesis ensures that $\Phi_{j-1}^{(r)}$ is small for $r \in [\beta_{j-1}, t + n \log^5 n]$. From that, it follows by a simple estimate that  $\Psi_j^{(\beta_{j-1})} \leq e^{0.01 \log^3 n}$ (\cref{clm:phi_small_implies_psi_plus_one_small}). Using a multiplicative drop (\cref{lem:rec_inequality_phi_psi}) repeatedly, it follows that there exists $u \in [\beta_{j-1}, \beta_{j-1} + n\log^3 n]$, $\ex{\Psi_j^{(u)}} \leq cn$ (\cref{lem:there_exists_linear_ex_psi_plus_one}). Then by \cref{lem:exists_s_st_ex_psi_linear}, this statement is extended to the time-interval $[\beta_{j-1} + n \log^3 n, t + n \log^5 n]$.
By simply using Markov's inequality and a union bound, we can deduce that $\Psi_j^{(r)} \leq cn^{12}$ for all $r \in [\beta_{j-1} + n \log^3 n, t + n \log^5 n]$. By a simple relation between two potentials, this implies $\Phi_j^{(r)} \leq n^{4/3}$ (\cref{clm:psi_potential_poly_implies}~(ii)). Now using a multiplicative drop (\cref{lem:rec_inequality_phi_psi}) guarantees that this becomes $\Phi_j^{(r)} \leq cn$ \Whp~for \emph{a single} time-step $r \in [\beta_{j-1}, \beta_j]$ (\cref{lem:exists_s_st_phi_linear_whp}). 

To obtain the stronger statement which holds for all time-steps $r \in [\beta_{i-1},\beta_j]$, we will use a concentration inequality. The key point is that whenever $\Psi_j^{(s)} \leq cn^{12}$, then the absolute difference $|\Phi_j^{(s+1)} - \Phi_j^{(s)}|$ is at most $n^{1/3}$, because $12.1 \frac{\alpha_2}{\alpha_1} < 1/3$ (\cref{clm:psi_potential_poly_implies}~(ii)). This is crucial so that applying the supermartingale concentration bound~\cref{thm:chung_lu_theorem_8_5}~from~\cite{CL06} to $\Phi_j$ yields an $\Oh(n)$ guarantee for the entire time interval.

In \cref{sec:auxiliary} we collect and prove all lemmas and claims mentioned above. After that, in \cref{sec:proof_of_key} we use these lemmas to complete the Proof of \cref{lem:new_inductive_step}.

\subsection{Auxiliary Definitions and Claims for the proof of Lemma~\ref{lem:new_inductive_step}}\label{sec:auxiliary}

In the following, we will always implicitly assume that $1 \leq j \leq k - 1$, as the base case $j=0$ has been done. We define the following event, which will be used frequently in the proof: 
\[\mathcal{E}_{j-1}^{(s)} := \left\{ \Phi_{j-1}^{(s)} \leq 2cn \right\}.
\] 
Recall that the induction hypothesis asserts that $\mathcal{E}_{j-1}^{(s)}$ holds for all steps $s \in [\beta_{j-1},t+n \log^5 n]$.
In the following arguments we will be working frequently with the ``killed'' versions of the potentials, i.e., we condition on $ \mathcal{E}_{j-1}^{(s)}$ holding on all time steps:
\[
\tilde{\Phi}_{j}^{(s)} := \Phi_{j}^{(s)} \cdot \mathbf{1}_{ \cap_{r\in [\beta_{j-1}, s]} \mathcal{E}_{j-1}^{(r)} } \text{ and } \tilde{\Psi}_{j}^{(s)} := \Psi_{j}^{(s)} \cdot \mathbf{1}_{ \cap_{r\in [\beta_{j-1}, s]} \mathcal{E}_{j-1}^{(r)} }. 
\]

As the proof of \cref{lem:new_inductive_step} requires several claims and lemmas, the remainder of this section is divided further in:
\begin{enumerate}
\item Analysis of the (expected) drop of the potentials $\Phi_j$ and $\Psi_j$. (\cref{sec:expected_drop})
\item Auxiliary (probabilistic) lemmas based on these drop results. (\cref{sec:probabilistic_drop})
\item (Deterministic) inequalities that involve one or two potentials. (\cref{sec:deterministic_claims})
\end{enumerate}
After that, we proceed to complete the proof of \cref{lem:new_inductive_step} in \cref{sec:proof_of_key}.

\subsubsection{Analysis of the Drop of the Potentials \texorpdfstring{$\Phi_j$ and $\Psi_j$}{Phi\_j and Psi\_j}} \label{sec:expected_drop}

We define $\alpha_j^{(s)} := \frac{s}{n}+ \frac{2}{\alpha_2} \cdot j (\log n)^{1/k}$, so that when $\mathcal{E}_{j-1}^{(s)}$ holds, then $y_{n \cdot \delta_{k-j}}^{(s)} \leq  \alpha_j^{(s)} - 1$; this will be established in the next lemma below.

\begin{lem} \label{lem:quantile_bound}
For any step $s \geq 1$, if $\mathcal{E}_{j-1}^{(s)}$ holds then $y_{n \cdot \delta_{k-j}}^{(s)} \leq  \alpha_j^{(s)} - 1$.
\end{lem}
\begin{proof}
Assuming the opposite $y_{n \cdot \delta_{k-j}}^{(s)} >  \alpha_j^{(s)} - 1$, we conclude
\begin{align*}
\Phi_{j-1}^{(s)} & \geq \sum_{i=1}^{n \cdot \delta_{k-j}}
\exp\Big( \alpha_2 (\log n)^{(j-1)/k} \cdot \Big(\alpha_j^{(s)} - \frac{s}{n} - \frac{2}{\alpha_2} \cdot (j-1) (\log n)^{1/k} \Big)^+ \Big) 
\\
&\geq n \cdot 2^{-0.5(\log n)^{j/k}} \cdot e^{\alpha_2 \cdot (\log{n})^{(j-1)/k} \cdot \frac{2}{\alpha_2} (\log n)^{1/k} } \\
&\geq n \cdot 2^{-0.5 (\log n)^{j/k}} \cdot e^{2 \cdot (\log{n})^{j/k}} \\
& > 2 c n,
\end{align*}
since $(e^2 \cdot 2^{-0.5})^{2 \cdot (\log{n})^{j/k}} > (e^2 \cdot 2^{-0.5})^{10^4} > 2c$ for sufficiently large $n$, which contradicts $\mathcal{E}_{j-1}^{(s)}$.
\end{proof}

\begin{lem} \label{lem:rec_inequality_phi_psi} For any step $s \geq \beta_{j-1} = t+2jn \log^3 n$,
\[
\Ex{\Phi_j^{(s+1)} \,\big|\, \mathcal{E}_{j-1}^{(s)}, \Phi_{j}^{(s)}  } \leq \Big(1 - \frac{1}{n} \Big) \cdot \Phi_j^{(s)} + 2,
\]
and
\[
\Ex{\Psi_j^{(s+1)} \,\big|\, \mathcal{E}_{j-1}^{(s)}, \Psi_{j}^{(s)}  } \leq \Big(1 - \frac{1}{n} \Big) \cdot \Psi_j^{(s)} + 2.
\]
\end{lem}

\begin{proof} We will prove the statement for the potential function $\Psi_j$. The same proof holds for $\Phi_j$, since the only steps dependent on the coefficients ($\alpha_1$ vs.~$\alpha_2$) are \cref{lem:quantile_bound} to obtain the bound $y_{n \cdot \delta_{k - j}} < \alpha_j^{(t)} - 1$ and the facts that $2^{0.5} > e^{0.01} > e^{\alpha_2}$ and $0.5 \cdot \alpha_1 \cdot (\log n)^{j/k} > 0.5 \cdot \alpha_2 \cdot (\log n)^{j/k} > 1.2$ (which also hold for $\alpha_2$).

In the following part of the proof, we will break down $\Psi_{j}^{(s+1)}$ (and, similarly, $\Psi_{j}^{(s)}$) as follows:
\begin{align*}
 \Psi_{j}^{(s+1)} = \sum_{i=1}^n \Psi_{j, i}^{(s+1)} &= \sum_{i=1}^n \exp \left( 0.01 (\log n)^{j/k} \cdot ( x_i^{(s+1)} - \alpha_j^{(s+1)} )^+ \right).
\end{align*}
Then we will split this sum into bins that have load at least (or less than) $\alpha_j^{(s)}$, i.e.,
\begin{align*}
 \Psi_{j}^{(s+1)} &= \sum_{i \colon x_i^{(s)} \geq \alpha_j^{(s)} } \Psi_{j, i}^{(s+1)} + \sum_{i \colon x_i^{(s)} < \alpha_j^{(s)} } \Psi_{j, i}^{(s+1)}. 
 \end{align*}
After that we will apply linearity of expectation to bound the expected value of the potential at step $s+1$.

\textbf{Case 1}: First, consider the contribution of a bin $i$ with $x_i^{(s)} \geq \alpha_j^{(s)}$ to $\Psi_j^{(s+1)}$,
\[
\Psi_{j,i}^{(s+1)} = \begin{cases} 
\exp{\left(0.01(\log{n})^{j/k}(1 - \frac{1}{n}) \right)}\Psi_{j,i}^{(s)} & \quad \text{if } x_i^{(s+1)} = x_i^{(s)} + 1, \\
\exp{\left(0.01(\log{n})^{j/k}\left( - \frac{1}{n} \right) \right)}\Psi_{j,i}^{(s)} & \quad \text{otherwise.} \end{cases}
\]
Define $u_i := \Pro{x_i^{(s+ 1)} = x_i^{(s)} + 1 \, \big| \, \mathcal{E}_{j-1}^{(s)}, \Psi_{j}^{(s)}}$. We have:
\begin{align*}
\lefteqn{\Ex{  \Psi_{j,i}^{(s+1)}  \, \big| \, \mathcal{E}_{j-1}^{(s)}, \Psi_{j}^{(s)}} 
 = e^{0.01(\log{n})^{j/k}(1 - \frac{1}{n}) }\cdot\Psi_{j,i}^{(s)} \cdot u_i + e^{0.01(\log{n})^{j/k}(- \frac{1}{n})}\cdot\Psi_{j,i}^{(s)} \cdot (1 - u_i)} \\
 & \quad = \left(  e^{0.01(\log{n})^{j/k}(1 - \frac{1}{n})} - e^{0.01(\log{n})^{j/k}(- \frac{1}{n})} \right) \cdot \Psi_{j,i}^{(s)} \cdot u_i 
  + e^{0.01(\log{n})^{j/k}(- \frac{1}{n})} \cdot\Psi_{j,i}^{(s)}.
\end{align*}
Note that since $y_{n \cdot \delta_{k-j}}^{(s)} \leq \alpha_j^{(s)} - 1$ (see \cref{lem:quantile_bound}), bin $i$ must be among the $(\delta_{k-j})$-th heaviest bins. To increment the load of bin $i$, $i$ has to be one of the two randomly chosen bins, and the other choice must be a bin whose load is at most $y_{n \cdot \delta_{k-j}}^{(s)}$, hence
$u_i \leq \frac{2}{n} \delta_{k-j} \leq \frac{3 \gamma}{n} \tilde{\delta}_{k-j} \leq \frac{18}{n} \tilde{\delta}_{k-j}$ (see \cref{clm:choice_independent_of_potential} below this lemma for details), which yields
\begin{align*}
 \Ex{  \Psi_{j,i}^{(s+1)}  \, \big| \, \mathcal{E}_{j-1}^{(s)}, \Psi_{j}^{(s)}} & \quad  \leq e^{0.01(\log{n})^{j/k}(1 - \frac{1}{n})} \cdot \Psi_{j,i}^{(s)} \cdot \frac{18}{n} \tilde{\delta}_{k-j} 
  + e^{0.01(\log{n})^{j/k}(- \frac{1}{n})} \cdot \Psi_{j,i}^{(s)} \\
 &  \quad = e^{-0.01(\log{n})^{j/k} \frac{1}{n}} \cdot \Psi_{j,i}^{(s)} \cdot \left( 1 + \frac{18}{n} \cdot \frac{e^{0.01(\log{n})^{j/k}}}{2^{0.5(\log{n})^{j/k}}} \right) \\
 & \quad = \Psi_{j,i}^{(s)} \cdot \left( 1 - 0.6 \cdot \frac{0.01}{n} \cdot (\log n)^{j/k} \right) \cdot \left( 1 + \frac{18}{n} \cdot \left( \frac{e^{0.01}}{2^{0.5}} \right)^{(\log{n})^{j/k}}\right) \\
 &\quad < \Psi_{j,i}^{(s)} \cdot \left(1 - \frac{1.2}{n} \right) \cdot \left( 1 + \frac{0.06}{n} \right) \\
 & \quad < \Psi_{j,i}^{(s)} \cdot \left(1 - \frac{1}{n} \right),
\end{align*}
where we have used that $e^x < 1 + 0.6 \cdot x$ for any $-1.1 \leq x < 0$ and $(\log n)^{j/k} > \exp(\frac{1}{k} \log \log n) > \exp(\frac{\log(10^4)}{\log \log n} \cdot \log \log n) = 10^4$ for sufficiently large $n$. In conclusion, we have shown that for any bin $i$ with load $x_i^{(s)} \geq \alpha_j^{(s)}$,
\[
\Ex{  \Psi_{j,i}^{(s+1)} \, \big| \, \mathcal{E}_{j-1}^{(s)}, \Psi_{j}^{(s)} }
\leq \Psi_{j,i}^{(s)}  \cdot \left(1 - \frac{1}{n} \right).
\]

\textbf{Case 2}: Let us now consider the contributions of a bin $i$ with $x_i^{(s)} < \alpha_j^{(s)}$ to $\Psi_j^{(s+1)}$. Note that out of those bins, only bins $i$ with $x_i^{(s)} \in [ \alpha_j^{(s)} - 1, \alpha_j^{(s)})$ can change the potential.
Hence,
\[
\Psi_{j,i}^{(s+1)} \leq \begin{cases} 
\exp{\left(0.01(\log{n})^{j/k}(1 - \frac{1}{n}) \right)} & \quad \text{if } x_i^{(s+1)} = x_i^{(s)} + 1, \\
1 & \quad \text{otherwise.} \end{cases}
\]
Since $y_{n \cdot \delta_{k-j}} \leq  \alpha_j^{(s)} - 1$, we can conclude as in the previous case that such a bin $i$ is incremented with probability at most $u_i \leq \frac{18}{n} \delta_{k-j}$, so
\begin{align*}
 \Ex{  \Psi_{j,i}^{(s+1)} \, \big| \, \mathcal{E}_{j-1}^{(s)}, \Psi_{j}^{(s)} }
&\leq 
(1-u_i) \cdot
\Psi_{j,i}^{(s)} 
+ u_i  \cdot \exp \left( 0.01 (\log n)^{j/k} ( 1-\frac{1}{n}) \right) \\
&\leq 1 \cdot 1 + \frac{18}{n} \tilde{\delta}_{k-j} \cdot \exp \left( 0.01 (\log n)^{j/k} ( 1-\frac{1}{n}) \right).
\end{align*}

Combining the two cases, we find that

\begin{align*}
  \Ex{ \Psi_j^{(s+1)}  \, \big| \, \mathcal{E}_{j-1}^{(s)}, \Psi_{j}^{(s)}  } 
 & = \sum_{i \colon x_i^{(s)} \geq \alpha_j^{(s)}} \Ex{  \Psi_{j,i}^{(s+1)} \, \big| \, \mathcal{E}_{j-1}^{(s)}, \Psi_{j}^{(s)} 
 }
 +  \sum_{i \colon x_i^{(s)} < \alpha_j^{(s)}} \Ex{  \Psi_{j,i}^{(s+1)} \, \big| \, \mathcal{E}_{j-1}^{(s)}, \Psi_{j}^{(s)} 
 } \\
 &\leq \sum_{i \colon x_i^{(s)} \geq \alpha_j^{(s)}} \Psi_{j,i}^{(s)} \cdot \left( 1 - \frac{1}{n} \right) \\
 & \quad \quad \quad + 
  \sum_{i \colon x_i^{(s)} < \alpha_j^{(s)}} \Psi_{j,i}^{(s)} +\frac{18}{n} \delta_{k-j}  \cdot e^{0.01(\log{n})^{j/k}(1 - \frac{1}{n})} \\
  &\leq \Psi_{j}^{(s)} \cdot \left( 1 - \frac{1}{n} \right) + n \cdot \frac{18}{n} \cdot 2^{-0.5(\log n)^{j/k}} e^{0.01(\log n)^{j/k}} + 1 \\
  &\leq \Psi_{j}^{(s)} \cdot \left( 1 - \frac{1}{n} \right) + 2,
\end{align*}
where the second inequality used the fact that $\Psi_{j,i}^{(s)} = 1$ for $x_i^{(j)} < \alpha_j^{(s)}$.
\end{proof}

\begin{clm}\label{clm:choice_independent_of_potential}
Let $\tilde{\Phi}_j^{(s)}$, $\mathcal{E}_{j-1}^{(s)}$ and $\alpha_j^{(s)}$ be defined as in \cref{lem:rec_inequality_phi_psi}. Then for any bin $i \in [n]$ with $x_i^{(s)} \geq \alpha_j^{(s)} $, we get
\[
\Pro{x_i^{(s+1)} = x_i^{(s)} + 1 \, \big| \,  \tilde{\Phi}_j^{(s)}, \mathcal{E}_{j-1}^{(s)}, x_i^{(s)} \geq \alpha_j^{(s)} } \leq  \frac{\gamma\delta}{n}.
\]
\end{clm}
\begin{proof}
By \cref{lem:quantile_bound} we get, $x_i^{(s)} \geq \alpha_j^{(s)} \geq y_{n \cdot \delta}^{(s)}$. By the definition of the process, incrementing bin $i$ depends only on $\Rank^{(s)}(i)$.
\[
\Pro{x_i^{(s+1)} = x_i^{(s)} + 1 \, \big| \,  \tilde{\Phi}_j^{(s)}, \mathcal{E}_{j-1}^{(s)}, \Rank^{(s)}(i) =r } = \Pro{x_i^{(s+1)} = x_i^{(s)} + 1 \, \big| \, \Rank^{(s)}(i) =r }
\]
So,
\begin{align*}
\lefteqn{\Pro{x_i^{(s+1)} = x_i^{(s)} + 1 \, \big| \,  \tilde{\Phi}_j^{(s)}, \mathcal{E}_{j-1}^{(s)}, x_i^{(s)} \geq \alpha_j^{(s)} }} \\
 & = \Pro{x_i^{(s+1)} = x_i^{(s)} + 1 \, \big| \,  \tilde{\Phi}_j^{(s)}, \mathcal{E}_{j-1}^{(s)}, x_i^{(s)} \geq \alpha_j^{(s)}, \Rank^{(s)}(i) \leq n \cdot \delta } \\
 & = \sum_{u} \Pro{x_i^{(s+1)} = x_i^{(s)} + 1 \, \big| \,  \Rank^{(s)}(i) = u, \tilde{\Phi}_j^{(s)}, \mathcal{E}_{j-1}^{(s)}, x_i^{(s)} \geq \alpha_j^{(s)}, \Rank^{(s)}(i) \leq n \cdot \delta }  \\
 & \quad \quad \quad \quad \cdot \Pro{\Rank^{(s)}(i) = u \, \big| \,  \tilde{\Phi}_j^{(s)}, \mathcal{E}_{j-1}^{(s)}, x_i^{(s)} \geq \alpha_j^{(s)}, \Rank^{(s)}(i) \leq n \cdot \delta } \\ 
 & = \sum_{u : y_u^{(s)} \geq \alpha_j^{(s)}} \Pro{x_i^{(s+1)} = x_i^{(s)} + 1 \, \big| \,  \Rank^{(s)}(i) = u} \\
 & \quad \quad \quad \quad \cdot \Pro{\Rank^{(s)}(i) = u \, \big| \,  \tilde{\Phi}_j^{(s)}, \mathcal{E}_{j-1}^{(s)}, x_i^{(s)} \geq \alpha_j^{(s)}, \Rank^{(s)}(i) \leq n \cdot \delta } \\
 & \leq \sum_{u : y_u^{(s)} \geq \alpha_j^{(s)}} \frac{\gamma\delta}{n} \cdot \Pro{\Rank^{(s)}(i) = u \, \big| \,  \tilde{\Phi}_j^{(s)}, \mathcal{E}_{j-1}^{(s)}, x_i^{(s)} \geq \alpha_j^{(s)}, \Rank^{(s)}(i) \leq n \cdot \delta } 
\intertext{The information in the previous potential functions, $\tilde{\Phi}_j^{(s)}$ and $\mathcal{E}_{j-1}^{(s)}$, does not allow to distinguish bins, so each one is equally likely to have any of the loads,}
 & = \sum_{u : y_u^{(s)} \geq \alpha_j^{(s)}} \frac{\gamma\delta}{n} \cdot \frac{1}{\left| u : y_u^{(s)} \geq \alpha_j^{(s)} \right|} = \frac{\gamma\delta}{n}. 
\end{align*}
\end{proof}

\subsubsection{Auxiliary Probabilistic Lemmas on the Potential Functions}\label{sec:probabilistic_drop}

The first lemma proves that $\tilde{\Psi}_{j}^{(s)}$ is small in expectation for at \emph{at least one} time-step. It relies on the multiplicative drop (\cref{lem:rec_inequality_phi_psi}), and the fact that precondition $\cap_{r\in [\beta_{j-1}, s]} \mathcal{E}_{j-1}^{(r)}$ holds due to the definition of the killed potential $\tilde{\Psi}_{j-1}$.
\begin{lem} \label{lem:there_exists_linear_ex_psi_plus_one}
There exists $s \in [\beta_{j-1}, \beta_{j-1} + n \log^3 n]$ such that $\ex{\tilde{\Psi}_{j}^{(s)}} \leq cn$.
\end{lem}
\begin{proof}
In this lemma, we analyze
$\tilde{\Psi}_{j}^{(s)}=\Psi_j^{(s)} \cdot \mathbf{1}_{\cap_{r\in [\beta_{j-1}, s]} \mathcal{E}_{j-1}^{(r)}  }$, so we will implicitly only deal with the case where $\mathcal{E}_{j-1}^{(r)}$ holds for all $r \in [\beta_{j-1}, \beta_{j-1} + n \log^3 n]$.

Since $\mathcal{E}_{j-1}^{(\beta_{j-1})}$ holds, $\Phi_{j-1}^{(\beta_{j-1})} \leq 2cn$ holds, so using \cref{clm:phi_small_implies_psi_plus_one_small}, we have $\Psi_{j}^{(\beta_{j-1})} \leq \exp(0.01(\log n)^3)$. Note that if at step $r \in [\beta_{j-1}, \beta_{j-1} + n \log^3 n]$, $\tilde{\Psi}_{j}^{(r)} \geq cn$, then the second inequality from \cref{lem:rec_inequality_phi_psi} implies,
\begin{equation} \label{eq:psi_i_plus_1_large_psi}
\Ex{\tilde{\Psi}_{j}^{(r+1)} \,\big|\, \tilde{\Psi}_{j}^{(r)}  } \leq \left(1 - \frac{1}{2n} \right) \cdot \tilde{\Psi}_{j}^{(r)},
\end{equation}
since whenever $\tilde{\Psi}_j^{(s)} = 0$, the inequality holds trivially. We define the killed potential function,
\[
\Lambda_{j}^{(r)} := \tilde{\Psi}_{j}^{(r)} \cdot \mathbf{1}_{\bigcap_{\tilde{r} \in [\beta_{j-1}, r]} \tilde{\Psi}_{j}^{(\tilde{r})} \geq cn},
\]
for $r \in [\beta_{j-1}, \beta_{j-1} + n \log^3 n]$. This satisfies the multiplicative drop in~(\ref{eq:psi_i_plus_1_large_psi}), but regardless of how large $\Lambda_j^{(r)}$ is. By inductively applying the inequality for $\Delta = n \log^3 n$ steps, we have
\[
\Ex{\Lambda_{j}^{(\beta_{j-1} +\Delta)}} = \Ex{\Ex{\Lambda_{j}^{(\beta_{j-1} +\Delta)} \,\big|\, \Lambda_{j}^{(\beta_{j-1})} } } \leq e^{- \frac{\Delta}{2n}} \cdot \Ex{\tilde{\Psi}_{j}^{(\beta_{j-1})}} \leq 1 \leq c.
\]
Hence, there exists $s \in [\beta_{j-1}, \beta_{j-1} + \Delta]$, such that $\Ex{\tilde{\Psi}_{j}^{(s)}} \leq cn$.
\end{proof}
Generalizing the previous lemma, and again exploiting the conditioning on $ \cap_{r\in [\beta_{j-1}, s]} \mathcal{E}_{j-1}^{(r)}$ of $\Psi_j^{(s)}$, we know prove that $\tilde{\Psi}_{j}^{(s)}$ is small in expectation for the entire time interval.
\begin{lem} \label{lem:exists_s_st_ex_psi_linear}
For all $s \in [\beta_{j-1}+n\log^3 n, t + n \log^5 n]$, $\ex{\tilde{\Psi}_j^{(s)}} \leq cn$.
\end{lem}
\begin{proof}
Again, note that this lemma analyzes
$\tilde{\Psi}_{j}^{(s)}=\Psi_j^{(s)} \cdot \mathbf{1}_{\cap_{r\in [\beta_{j-1}, s]} \mathcal{E}_{j-1}^{(r)}  }$, and thus we will implicitly only deal with the case where $\mathcal{E}_{j-1}^{(r)}$ holds for all $r \in [\beta_{j-1}, \beta_{j-1} + n \log^3 n]$.

By using \cref{lem:there_exists_linear_ex_psi_plus_one}, we get that $\ex{\tilde{\Psi}_{j}^{(r_0)}} \leq cn$ for some $r_0 \in [\beta_{j-1}, \beta_{j-1} +  n \log^3 n]$.
Further, by using the inequality of \cref{lem:rec_inequality_phi_psi} and the fact that $2 \leq c$, we have for any $r \in [\beta_{j-1}, t + n \log^5 n]$,
\[
\Ex{\tilde{\Psi}_j^{(r+1)} \,\big|\, \tilde{\Psi}_{j}^{(r)}  } \leq \left(1 - \frac{1}{n} \right) \cdot \tilde{\Psi}_j^{(r)} + c.
\]
For all $r \in [r_0, t + n \log^5 n]$ such that $\ex{\tilde{\Psi}_j^{(r)}} \leq cn$ holds, it follows that,
\[
\Ex{\tilde{\Psi}_j^{(r + 1)}} = \Ex{\Ex{\tilde{\Psi}_j^{(r+1)} \,\big|\, \tilde{\Psi}_{j}^{(r)}  } } \leq \left(1 - \frac{1}{n} \right) \cdot \Ex{\tilde{\Psi}_j^{(r)} } + c \leq \left(1 - \frac{1}{n} \right) cn + c = cn.
\]
Hence, starting from $\ex{\tilde{\Psi}_{j}^{(r_0)}} \leq cn$, it follows inductively that for any $s \in [r_0, t + n \log^5 n]$, $\ex{\tilde{\Psi}_j^{(s)}} \leq cn$. Since $r_0 \leq \beta_{j-1} + n \log^3 n$, the claim follows.
\end{proof}

We now switch to the other potential function $\tilde{\Phi}_j^{(s)}$, and prove that if it is polynomial in \emph{at least one step}, then it is also linear in \emph{at least one} step (not much later).
\begin{lem} \label{lem:exists_s_st_phi_linear_whp}
For all $1 \leq j < k$ it holds that,
\[
\Pro{\bigcup_{s \in [\beta_{j-1}, \beta_j]} \{ \tilde{\Phi}_j^{(s)} \leq cn \} ~\,\Big|\,~ \bigcup_{r \in [\beta_{j-1}, \beta_{j-1} + n \log^3 n]} \{ \tilde{\Phi}_j^{(r)} \leq n^{4/3} \} } \geq 1 - n^{-5}.
\]
\end{lem}
\begin{proof}

Note that if at step $\tilde{r}$, $\tilde{\Phi}_{j}^{(\tilde{r})} > cn$, then the first inequality from \cref{lem:rec_inequality_phi_psi} implies,
\begin{equation} \label{eq:phi_i_plus_1_large_phi}
\Ex{\tilde{\Phi}_{j}^{(\tilde{r}+1)} \,\big|\, \tilde{\Phi}_{j}^{(\tilde{r})}, \tilde{\Phi}_{j}^{(\tilde{r})} \geq cn } \leq \left(1 - \frac{1}{2n} \right) \cdot \tilde{\Phi}_{j}^{(\tilde{r})}.
\end{equation}
We define the killed potential function
\[
\Lambda_{j}^{(\tilde{r})} := \tilde{\Phi}_{j}^{(\tilde{r})} \cdot \mathbf{1}_{\bigcap_{\tilde{s} \in [r, \tilde{r})} \tilde{\Phi}_{j}^{(\tilde{s})} \geq cn},
\]
for $\tilde{r} \in [\beta_{j-1}, \beta_{j-1} + n \log^3 n]$. This  satisfies inequality~\ref{eq:phi_i_plus_1_large_phi} for all $\tilde{r} \in [\beta_{j-1}, \beta_{j-1} + n \log^3 n]$, regardless of the value of $\tilde{\Phi}_{j}^{(\tilde{r})}$.

Let $r \geq \beta_{j-1}$ be the smallest time-step such that $\tilde{\Phi}_j^{(r)} \leq n^{4/3}$ holds.
By inductively applying~\ref{eq:phi_i_plus_1_large_phi} for $\Delta = n \log^2 n$ steps, we have
\[
\Ex{\Lambda_{j}^{(r +\Delta)} \,\big|\, \Lambda_{j}^{(r)}, \tilde{\Phi}_j^{(r)} \leq n^{4/3} } \leq e^{- \frac{\Delta}{2n}} \cdot \tilde{\Phi}_{j}^{(r)} \leq e^{- \frac{1}{2} \log^2 n} \cdot n^{4/3} \leq n^{-4},
\]
for sufficiently large $n$. By Markov's inequality,
\[
\Pro{\Lambda_j^{(r + \Delta)} > c n \,\big|\, \tilde{\Phi}_j^{(r)} \leq n^{4/3}} \leq n^{-4}  \cdot \frac{1}{cn} \leq n^{-5}.
\]
Since, $\Lambda_j^{(r+\Delta)} \leq cn \Rightarrow \exists s \in [r,r+\Delta] \subseteq [\beta_{j-1}, \beta_j] \colon \, \tilde{\Phi}_j^{(s)} \leq cn$, we get the conclusion.
\end{proof}

\subsubsection{Deterministic Relations between the Potential Functions}\label{sec:deterministic_claims}

We collect several basic facts about the potential functions $\Phi_j^{(s)}$ and $\Psi_j^{(s)}$.

\begin{clm} \label{clm:phi_small_implies_psi_plus_one_small} For any $s \geq 0$,
\[
\Phi_j^{(s)} \leq 2cn \Rightarrow \Psi_{j+1}^{(s)} \leq \exp(0.01 \cdot \log^3 n).
\]
\end{clm}
\begin{proof}
Assuming $\Phi_j^{(s)} \leq 2cn$, implies that for any bin $i \in [n]$, 
\begin{align*}
\exp \left(\alpha_2\cdot (\log n)^{j/k} \cdot \left(x^{(s)}_{i} - \frac{s}{n} - \frac{2}{\alpha_2} j (\log n)^{1/k} \right) \right) \leq 2c n  \\ 
\Rightarrow ~~ x^{(s)}_{i} \leq \frac{s}{n} + \log(2c) + \frac{1}{\alpha_2} (\log n)^{\frac{k-j}{k}} + \frac{2}{\alpha_2} j (\log n)^{1/k} \leq \frac{s}{n} + 0.5 \cdot (\log n)^2,
\end{align*}
for sufficiently large $n$. Hence,
\[
\Psi_{j+1}^{(s)} \leq n \cdot \exp(0.01 \cdot (\log n)^{\frac{j + 1}{k}} \cdot 0.5 \cdot (\log n)^2) \leq \exp( 0.01 \cdot \log^3 n).
\]
\end{proof}

The next claim is crucial for applying the concentration inequality, since the third statement bounds the maximum additive change of $\Phi^{(s)}$ (assuming $\Psi^{(s)}$ is small enough):
\begin{clm} \label{clm:psi_potential_poly_implies}
For any $s \geq 0$, if $\Psi_j^{(s)} \leq c n^{12}$, then (i) $x_i^{(s)} \leq \frac{s}{n} + \frac{12.1}{\alpha_1} \cdot (\log n)^{\frac{k - j}{k}} + \frac{2}{\alpha_2} j (\log n)^{1/k}$ for all $i \in [n]$, (ii) $ \Phi_j^{(s)} \leq n^{4/3}$ and (iii) $| \Phi_j^{(s+1)} - \Phi_j^{(s)} | \leq n^{1/3}$.
\end{clm}
\begin{proof}
Let $s$ be some time-step with $\Psi_j^{(s)} \leq c n^{12}$. For $(i)$, assuming that $x_i^{(s)} > \frac{s}{n} + \frac{12.1}{\alpha_1} \cdot (\log n)^{\frac{k - j}{k}} + \frac{2}{\alpha_2} j (\log n)^{1/k}$, then we get $\Psi_j^{(s)} > \exp(\alpha_1 \cdot \frac{12.1}{\alpha_1} \cdot \log n ) = n^{12.1}$, which is a contradiction.
For $(ii)$, it suffices to prove $\Phi_{j,i}^{(s)} \leq n^{1/3}$, so,
\begin{align*}
     \Phi_{j,i}^{(s)}  
 &\leq 
 \exp \bigl( \alpha_2 \cdot (\log n)^{j/k} \cdot \bigl( x_i^{(s)}  - \frac{s}{n} - \frac{2}{\alpha_2} j (\log n)^{1/k} \bigr)^{+} \bigr) \\
 & \leq \exp\bigl( \alpha_2 \cdot (\log n)^{j/k} \cdot \frac{12.1}{\alpha_1} \cdot (\log n)^{\frac{k - j}{k}} \bigr) \\
 &= \exp\bigl( \frac{12.1 \cdot 0.0002}{0.01} \cdot \log n  \bigr) \leq n^{1/3}.
\end{align*}
For $(iii)$, following the argument in the proof of \cref{clm:psi_potential_poly_implies}, because the potential function is convex, the maximum change is upper bounded by the hypothetical scenario of placing two balls in the heaviest bin, i.e. by $n^{1/3}$.
\end{proof}

The next claim is a simple ``smoothness'' argument showing that the potential cannot decrease quickly within $n/\log^2 n$ steps. The derivation is elementary and relies on the fact that average load does not change by more than $1/\log^2 n$.
\begin{clm} \label{clm:phi_j_does_not_drop_quickly}
For any $s \geq 0$ and any $r \in [s, s + n/\log^2 n]$, we have $\Phi_j^{(r)} \geq 0.99 \cdot \Phi_j^{(s)}$.
\end{clm}
\begin{proof}
The normalized load after $r - s$ steps can decrease by at most $\frac{r - s}{n} \leq \frac{1}{\log^2 n}$. Hence,
\begin{align*}
\Phi_{j, i}^{(r)} 
 & = \exp \left( 0.01 \cdot (\log n)^{j/k} \cdot (x_i^{(s)} - \frac{r-s}{n} - \frac{s}{n}- \frac{2}{\alpha_2} j (\log n)^{1/k} )^{+} \right) \\
 & \geq \exp \left( 0.01 \cdot (\log n)^{j/k} \cdot (x_i^{(s)} - \frac{s}{n} - \frac{2}{\alpha_2} j (\log n)^{1/k} )^+ - \frac{0.01 \cdot (\log n)^{j/k} \frac{n}{\log^2 n}}{n} \right) \\
 & \geq \Phi_{j, i}^{(s)} \cdot e^{-\frac{0.01 \cdot (\log n)^{j/k}}{\log^2 n}} %
 \geq 0.99 \cdot \Phi_{j, i}^{(s)},
\end{align*}
for sufficiently large $n$.
\end{proof}

\subsection{Completing the Proof of Key Lemma \texorpdfstring{(\cref{lem:new_inductive_step})}{}}\label{sec:proof_of_key}

The proof of \cref{lem:new_inductive_step} shares some of the ideas from the proof of \cref{thm:high_prob_beta_potential}. However, there we could more generously take a union bound over the entire time-interval to ensure that the potential is indeed small everywhere with high probability. Here we cannot afford to lose a polynomial factor in the error probability, as the inductive step has to be applied $k=\omega(1)$ times. To overcome this, we will partition the time-interval into consecutive intervals of length $n/\log^2 n$. Then, we will prove that at the end of each such interval the potential is small \Whp, and finally use a simple smoothness argument of the potential to show that the potential is small \Whp~for \emph{all} time steps.

\begin{proof}[Proof of \cref{lem:new_inductive_step}]

The first and second statements in \cref{clm:psi_potential_poly_implies}, imply that if $\Psi_j^{(s)} \leq c n^{12}$ holds, then 
\[
 \left| \tilde{\Phi}_j^{(s+1)} - \tilde{\Phi}_{j}^{(s)} \right| \leq n^{1/3},
\]
and $\tilde{\Phi}_j^{(s)} \leq  n^{4/3}$.

Thus we will next establish that $\Psi_j^{(s)} \leq cn^{12}$ occurs with high probability.

By \cref{lem:exists_s_st_ex_psi_linear}, for all $s \in [\beta_{j-1} +  n \log^3 n, t + n \log^5 n]$, $\ex{\tilde{\Psi}_j^{(s)}} \leq cn$. Using Markov's inequality, we have with probability at least $1-n^{-11}$ that $\tilde{\Psi}_j^{(s)} \leq c n^{12}$. Hence by the union bound it follows that
\begin{align}
 \Pro{\bigcap_{s \in [\beta_{j-1} +  n \log^3 n,t + n \log^5 n]} \lbrace \tilde{\Psi}_j^{(s)} \leq c n^{12} \rbrace } \geq 1 - n^{-9}. \label{eq:badevent}
\end{align}

We now define the intervals
\[
\mathcal{I}_1 := (t_0, t_0 + \Delta], \mathcal{I}_2 = (t_0 + \Delta, t_0 + 2 \Delta],~\ldots~, \mathcal{I}_q := (t_0 + (q-1) \Delta, t + n \log^5 n], 
\]
where $t_0 \in[ \beta_{j-1} + n \log^3 n, \beta_j]$ is arbitrary (but will be chosen later), $\Delta := n/\log^2 n$ and $q := \lfloor \frac{t + n \log^5 n-t_0}{\Delta} \rfloor \leq \log^7 n$. In order to prove that $\Phi_{j}$ is at most $2cn$ over all these intervals, we will use our auxiliary lemmas and the supermartingale concentration inequality~ (\cref{thm:chung_lu_theorem_8_5})
to establish that $\Phi_{j}$ is at most $(c+1) \cdot n$ at the points $t_0 + \Delta, t_0 + 2\Delta, \ldots t_0 \cdot q, t + n \log^5 n$. By using a smoothness argument (\cref{clm:phi_j_does_not_drop_quickly}), this will establish
that $\Phi$ is at most $2cn$ at all points in $[t_0, t + n \log^5 n]$, which is the conclusion of the lemma.

For each interval $i \in [q]$, we define for $s \in (t_0 + (i-1)\Delta, t_0 + i \Delta]$,
\[
X_i^{(s)} := \begin{cases}
\Phi_j^{(s)} & \text{if }\exists u \in (t_0 + (i -1) \cdot \Delta, s)$ such that $\Phi_j^{(u)} > 5 \cdot c \cdot n, \\
5 c n+n^{1/3} & \text{otherwise}.
\end{cases}
\]
Note that whenever the first condition in the definition of $X_i^{(s)}$ is satisfied, it remains satisfied until $t_0+i \cdot \Delta$, again by \cref{clm:phi_j_does_not_drop_quickly}.

Following the notation of \cref{thm:chung_lu_theorem_8_5}, define the event
\[
\overline{B_{i}^{(s)}}:= \left( \bigcap_{u \in [t_0 + (i - 1) \cdot \Delta, s)} \{ \tilde{\Psi}_j^{(u)} \leq c n^{12} \} \right)
\bigcap \left( \bigcap_{u \in [t_0 + (i - 1) \cdot \Delta,s)} \mathcal{E}_{j-1}^{(u)}  \right).
\]
By the inductive hypothesis of  \cref{lem:new_inductive_step} for $j-1$,
\[
 \Pro{\bigcap_{u \in [\beta_{j-1},t+n \log^5 n]} \mathcal{E}_{j-1}^{(u)}} \geq 1 - \frac{(\log n)^{8(j-1)}}{n^4},
\]
and hence by the union bound over this and \cref{eq:badevent}, 
\[
\Pro{ \overline{B}_i^{(s)}} \geq 1 - n^{-9} - \frac{(\log n)^{8(j-1)}}{n^4} \geq 1 - \frac{2 (\log n)^{8(j-1)}}{n^4}.
\]

\begin{clm}\label{clm:x_i_preconditions}
Fix any interval $i \in [q]$. Then the sequence of random variable $X_i^{(s)}$ with  filtration $\mathcal{F}_i^{(s)}$, for $s \in [t_0+(i-1)\Delta, t_0+i\Delta]$ and $B_i^{(s)}$ being the bad set associated, satisfies for all $s$,
\begin{align*}
    \Ex{ X_{i}^{(s)} \, \mid \, \mathcal{F}_i^{(s-1)}} \leq X_{i}^{(s-1)}, 
\end{align*}
and
\[
 \left| \left( X_i^{(s)} - X_i^{(s-1)} \right) \, \mid \, \mathcal{F}_i^{(s-1)} \right| \leq 2 n^{1/3}.
\]
\end{clm}
\begin{proof}[Proof of \cref{clm:x_i_preconditions}]
We begin by noting that, step $s \in [t_0 + (i-1) \Delta, t_0 + i \Delta]$ with $X_{i}^{(s)} \geq 4n$, the first inequality of \cref{lem:rec_inequality_phi_psi} can be relaxed to, \begin{align}
\Ex{\Phi_j^{(s+1)} \,\mid\, \mathcal{E}_{j-1}^{(s)}, \Phi_j^{(s)}} \leq \Big(1 - \frac{1}{2n}\Big) \cdot \Phi_j^{(s)}. \label{eq:drop}
\end{align}
We consider the following three cases:
\begin{itemize} \itemsep0pt
    \item \textbf{Case 1:} Assume that $\Phi_j^{(s)} < 5\cdot c \cdot n$ and this was the case for all previous time steps. Then, the $X_i^{(s+1)} =X_i^{(s)}$, so the two statements hold trivially.
    \item \textbf{Case 2:} Assume that $\Phi_j^{(u)} > 5 \cdot c \cdot n$ for some $u \leq s - 2$, then by~\cref{clm:phi_j_does_not_drop_quickly}, since $s-u \leq n/\log^2 n$,
    \[
     \Phi_{j}^{(s-1)} \geq 0.99 \cdot \Phi_{j}^{(u)} > 2cn,
    \]
    and thus by \cref{eq:drop}, the first statement follows. The second statement follows, since conditional on $\overline{\mathcal{B}}_i^{(s)}$, the precondition of~\cref{clm:psi_potential_poly_implies}~(iii) holds.
    \item \textbf{Case 3:} Assume that $\Phi_j^{(u)} < 5 cn$ for $u \leq s - 2$ and $\Phi_j^{(s-1)} > 5cn$. 
    First, we obtain that  $\Phi_{j}^{(s-1)} \leq \Phi_{j}^{(s-2)} + n^{1/3} < 5 \cdot c \cdot n + n^{1/3}$ (by~\cref{clm:psi_potential_poly_implies}~(ii), using again that $\overline{\mathcal{B}}_i^{(s)}$ holds). Further, by definition,
    $  X_i^{(s-1)} = 5 cn + n^{1/3}$, so $\ex{X_i^{(s)} \, \mid \, \Phi_i^{(s-1)} } = \ex{\Phi_j^{(s)}  \, \mid \, \Phi_i^{(s-1)}}\leq \Phi_j^{(s-1)} < X_i^{(s-1)}$ by \cref{eq:drop}. The second inequality follows by \cref{clm:psi_potential_poly_implies}~(iii), since $|X_i^{(s-1)} - X_i^{(s)}| \leq n^{1/3} + |\Phi_j^{(s)} - \Phi_j^{(s-1)}| \leq 2n^{1/3}$.
\end{itemize}
\end{proof}

Next we claim that $X_{i}^{(s)}$ satisfies the following conditions of \cref{thm:chung_lu_theorem_8_5} (where $\mathcal{F}_i$ are the filtrations associated with the balls allocated at $\mathcal{I}_i$ and $B_i^{(s)}$ is the bad set associated):
\begin{enumerate}
  \item $\ex{X_i^{(s)} \mid \mathcal{F}_{i}^{(s-1)}} \leq X_{i}^{(s-1)}$ by the first statement of~\cref{clm:x_i_preconditions}.
  \item $\var{X_i^{(s)} \mid \mathcal{F}_i^{(s-1)}} \leq n^{2/3}$. This holds, since
  \begin{align*}
  \Var{X_i^{(s)} \mid \mathcal{F}_i^{(s-1)}} &\leq \frac{1}{4} \left| \left( \max X_{i}^{(s)} - \min X_i^{(s)}  \right) \, \mid \, \mathcal{F}_i^{(s-1)} \right|^2 \\ &\leq \frac{1}{4} \left( 2 \cdot \left| X_i^{(s)} - X_i^{(s-1)} \right| \, \mid \, \mathcal{F}_i^{(s-1)} \right)^2 \leq 4 n^{2/3},
  \end{align*}
  where the first inequality follows by Popovicius' inequality, the second by the triangle inequality and the third by~\cref{clm:x_i_preconditions}.
  \item $X_i^{(s)} - \ex{X_i^{(s)} \mid \mathcal{F}_{i}^{(s-1)}} \leq 2 \cdot \bigl( \bigl| X_i^{(s)} - X_i^{(s-1)} \bigr| \, \mid \, \mathcal{F}_i^{(s-1)} \bigr) \leq 4 n^{1/3}$ which follows by the second statement of~\cref{clm:x_i_preconditions}.
\end{enumerate}
Now applying \cref{thm:chung_lu_theorem_8_5} for $i \in [q]$ with $\lambda = \frac{n}{\log^7 n}$, $a_i=4n^{1/3}$, and $M = 0$, we get %
\begin{align*}
\Pro{X_{i}^{(t_0 + i \Delta)} \geq X_{i}^{(t_0 + (i-1) \Delta)} + \lambda} &\leq \exp\left(-\frac{n^2/\log^{14}n}{2(\Delta \cdot (16n^{2/3} + 4 n^{2/3}))} \right) + \frac{2 (\log n)^{8(j - 1)}}{n^4} \\
&\leq \frac{3 (\log n)^{8(j - 1)}}{n^4}.
\end{align*}
Taking the union bound over the $\log^7 n$ intervals $i \in [q]$, it follows that
\begin{align}
\Pro{ \bigcup_{i \in [q]} \{ X_{i}^{(t_0 + i \Delta)} \geq X_{1}^{(t_0)} + i \cdot \lambda \} } \leq \log^7 n \cdot \frac{3 (\log n)^{8(j - 1)}}{n^4}. \label{eq:finalclaim}
\end{align}
It remains to show the existence of a $t_0 \in [\beta_{j-1}, \beta_{j-1} + n \log^3 n ]$ for which $X_{1}^{(t_0)}$ is small.

Since $\Psi_{j}^{(s)} \leq \tilde{\Psi}_{j}^{(s)}$, we can conclude from \cref{eq:badevent} that with probability at least $1-n^{-9}$, for $s=\beta_{j-1} +  n \log^3 n$ we have $\tilde{\Phi}_{j}^{(\beta_{j-1} +  n \log^3 n)} \leq n^{4/3}$. 

Assuming this occurs, then by \cref{lem:exists_s_st_phi_linear_whp}, there exists a time step $t_0 \in [\beta_{j-1}, \beta_j]$ such that $\tilde{\Phi}_j^{(t_0)} \leq cn$ w.p. at least $1 - n^{-4}$. Thus by the union bound over this and \cref{eq:badevent},
\[
 \Pro{ \bigcup_{t_0 \in [\beta_{j-1},\beta_{j}]} \{ \tilde{\Phi}_j^{(t_0)} \leq c n \} } \geq 1 - n^{-4} - n^{-9}.
\]
As \[
\tilde{\Phi}_j^{(t_0)} = \Phi_j^{(t_0)} \cdot \mathbf{1}_{ \cap_{s\in [\beta_{j-1}, t_0]} \Phi_{j-1}^{(s)} \leq 2cn },
\]
and 
$ \Pro{\bigcap_{s=\beta_{j-1}}^{t_0} \{ \Phi_{j-1}^{(s)} \leq 2cn \} } \geq \frac{(\log n)^{8(j-1)}}{n^4}$ by the inductive hypothesis,
a union bound yields
\begin{align*}
 \Pro{ \bigcup_{t_0 \in [\beta_{j-1},\beta_{j}]} \{ \Phi_j^{(t_0)} \leq c n \} } \geq 1 - n^{-4} - n^{-9} - \frac{(\log n)^{8(j-1)}}{n^4} \geq 1 - \frac{2 (\log n)^{8(j-1)}}{n^4}.
\end{align*}
Since $c > 4$, we conclude that
\begin{align*}
 \Pro{ \bigcup_{t_0 \in [\beta_{j-1},\beta_{j}]} \{ X_{1}^{(t_0)} \leq c n \} } 
 = \Pro{ \bigcup_{t_0 \in [\beta_{j-1},\beta_{j}]} \{ X_{1}^{(t_0)} \leq \max\{cn, 4 n + n^{1/3} \} \} }
 \geq 1 - \frac{2 (\log n)^{8(j-1)}}{n^4}.
\end{align*}
Taking the union bound over this and \cref{eq:finalclaim}, we conclude
\begin{align}
\Pro{ \bigcup_{i \in [q]} \{ X_{i}^{(t_0 + i \Delta)} \geq cn + \log^7 n \cdot \frac{n}{\log^ 7 n} \} } \leq \log^7 n \cdot \frac{4 (\log n)^{8(j - 1)}}{n^4}. \label{eq:above}
\end{align}

For the time-step $u=u(i):=t_0 + i \Delta$ at the end of the interval $u$, we cannot deduce anything about $\Phi_j$ from $X_i$ because of the shift-by-one in time-steps. To fix this, recall that by \cref{clm:psi_potential_poly_implies} (third statement), $\Psi_j^{(u)} \leq cn^{12}$ implies $\left| \Phi_j^{(u+1)} -  \Phi_j^{(u)} \right| \leq 2 n^{1/3} $. Hence \cref{eq:badevent} (together with the inductive hypothesis) implies that
\begin{align*}
\Pro{ X_i^{(u)} + 2n^{1/3} \geq \Phi_j^{(u)} } 
 \geq 1-\Pro{ \left| \Phi_j^{(u+1)} -  \Phi_j^{(u)} \right| \leq 2 n^{1/3} }
 \geq 1-n^{-9} - \frac{(\log n)^{8(j - 1)}}{n^4}.
\end{align*}
Using this, \cref{eq:above} and then applying a union bound over $i \in [q]$
\begin{align*}
\Pro{ \bigcup_{i \in [q]} \{ \Phi_j^{(u(i))} \geq cn + \log^7 n \cdot \frac{n}{\log^ 7 n} + 2n^{1/3} \} } \leq \log^7 n \cdot \frac{6 (\log n)^{8(j - 1)}}{n^4}.
\end{align*}

Finally, by \cref{clm:phi_j_does_not_drop_quickly} 
the above statement extends to \emph{all} time-steps at the cost of a slightly larger threshold:
\begin{align*}
\Pro{ \bigcup_{s \in [\beta_j,t + n \log^5 n] } \{ \Phi_{j}^{(s)} \geq 2c \cdot n \} } \leq \log^7 n \cdot \frac{6 (\log n)^{8(j - 1)}}{n^4},
\end{align*}
since $(c+2) \cdot \frac{1}{0.99} \leq 2c$.\end{proof}

\subsection{Proof of Main Theorem~\texorpdfstring{(\cref{thm:new_multiple_quantiles}) using~\cref{lem:new_inductive_step}}{}}\label{sec:proof_of_main_appendix}

\begin{proof}[Proof of \cref{thm:new_multiple_quantiles}]
Consider first the case where $m \geq n \log^5 n$ and let $t = m - n \log^5 n$. We will proceed by induction on the potential functions $\Phi_j$. The base case follows by noting that the probability vector $p$ satisfies the precondition of~\cref{thm:high_prob_beta_potential}, and applying this to all time steps $s \in [t,m]$ and taking the union bound gives,
\[
\Pro{\bigcap_{s \in [t, m]} \{ \Phi_0^{(s)} \leq 2cn \} } \geq 1 - n^{-4}.
\]
For the inductive step, we use \cref{lem:new_inductive_step}. After $k$ applications, we get
\[
\Pro{ \bigcap_{s \in [t+ \beta_{k-1},m]} \{ \Phi_{k-1}^{(s)} \leq 2cn \} } \geq 1-\frac{(\log n)^{8k}}{n^4} \geq 1-\frac{(\log n)^{8 \cdot \frac{1}{\log(10^4)} \log \log n}}{n^4} \geq 1 - n^{-3}.
\]
When this event occurs, the gap at step $m$ cannot be more than $\frac{2}{\alpha_2} \cdot k\cdot (\log n)^{1/k} $, otherwise
\begin{align*}
2cn \geq \Phi_{k-1}^{(m)} & \geq \exp \Big(\alpha_2 (\log n)^{\frac{k-1}{k}}\cdot \Big(\frac{2}{\alpha_2} \cdot k (\log n)^{1/k} - \frac{2}{\alpha_2} \cdot (k-1)(\log n)^{1/k} \Big) \Big) \\
 & = 
 \exp\Big(\alpha_2 (\log n)^{\frac{k-1}{k}}\cdot \Big(\frac{2}{\alpha_2} \cdot (\log n)^{1/k} \Big)\Big) = \exp(2 \cdot \log n) =n^2,
\end{align*}
which leads to a contradiction. 

The other case is $m < n \log^5 n$, when some of the $\beta_j$'s of the analysis above will be negative. To fix this, consider a modified process. The modified process starts at time-step $n \log^5 n -m$ with an empty load configuration. For any time $t \in [n \log^5n,-m]$, it places a ball of fractional weight $\frac{1}{n}$ to each of the $n$ bins. For $t \geq 1$, it works exactly as the original quantile process. Since the load configuration is perfectly balanced at each step $t < 0$, it follows that $\Psi_j^{(t)} = n$ holds deterministically. Since our proof relies only on upper bounds on the potential functions, these are trivially satisfied and hence the above analysis applies for the modified process. Further, as the relative loads of the modified process and the original process behave identically for $t \geq 1$, the statement follows.
\end{proof}

\section{Applications of the Relaxed Quantile Process}\label{sec:applications_abstract}

In this section we present two implications of our analysis in Section~\ref{sec:new_upper_bound_for_k_queries}, exploiting the flexibility of the \emph{relaxed} version of the $k$-quantile process. 
The first implication is based on majorizing the $(1+\beta)$-process by a suitable relaxed $k$-quantile process, where $k$ depends on $\beta$ (see \cref{lem:newthomas_beta}). 

\begin{restatable}{thm}{oneplusbetaapplication}\label{thm:oneplusbeta_application}
Consider a $(1+\beta)$-process with $\beta \geq 1 - 2^{-0.5 (\log n)^{(k-1)/k}}$ for some integer $1 \leq k \leq \kappa \cdot \log \log n$. Then for any $m \geq 1$,
\[
 \Pro{ \Gap(m) \leq 1000 \cdot k \cdot (\log n)^{1/k} } \geq 1-n^{-3}.
\]
In particular, if $\beta \geq 1-n^{-c_1}$, for any (small) constant $c_1 > 0$, then there is a constant $c_2 =c_2(c_1) > 0$ such that the gap is at most $c_2 \cdot \log \log n$ \whp.
\end{restatable}

We can also derive an almost matching lower bound, showing that $1-\beta$ has to be (almost) polynomially small in order to achieve a gap of $\Oh(\log \log n)$ (\cref{rem:lowerbound_beta}).

Our result for $k$ quantiles can be also applied to graphical balanced allocations, where the graph is parameterized by its spectral expansion $\lambda \in [0,1)$. Similar to the derivation of \cref{thm:oneplusbeta_application}, the idea is to show that the graphical balanced allocation process can be majorized by a suitable relaxed $k$-quantile process.
\begin{restatable}[special case of Theorem~\ref{thm:graphical}]{cor}{corthomas}\label{cor:thomas}
Consider graphical balanced allocation on a  $d$-regular graph with spectral expansion $\lambda \leq n^{-c_1}$ for a constant $c_1 > 0$. Then there is a constant $c_2=c_2(c_1) >0$ such that for any $m \geq 1$,
\begin{align*}
    \Pro{ \Gap(m) \leq c_2 \cdot \log \log n } \geq 1-n^{-3}.
\end{align*}
\end{restatable}

As shown in \cite{TY19}, for any $d=\poly(n)$, a random $d$-regular graph satisfies $\lambda=\Oh(1/\sqrt{d})$ \Whp, and thus the gap bound above applies. 
For the lightly loaded case, \cite{KP06} proved that any regular graph with
 degree at least $n^{\Omega(1/\log \log n)}$  achieves a gap $\Oh(\log \log n)$, and they also showed that this density is necessary. For the heavily loaded case, \cite{PTW15} proved a gap bound of $\Oh(\log n)$ for any expander. Hence \cref{cor:thomas} combines these lines of work, and establishes that the $\Oh(\log \log n)$ gap bound extends from complete graphs
 to dense and (strong) expanders.

\subsection{\texorpdfstring{$(1+\beta)$}{(1+beta)}-Process for large \texorpdfstring{$\beta$}{beta}}

We first relate the $(1+\beta)$-process to a relaxed quantile process.
\begin{lem}\label{lem:newthomas_beta}
Consider a $(1+\beta)$-process with $\beta \geq 1 - 2^{-0.5 (\log n)^{(k-1)/k}} = 1 - \tilde{\delta_1}$ for some integer $k \geq 1$. Then this $(1+\beta)$-process is a $\RelaxedQuantile_{\gamma}(\delta_1,\ldots,\delta_k)$ process, where each $\delta_i$ is $\tilde{\delta}_i$ being rounded up to the nearest multiple of $\frac{1}{n}$ and $\gamma=3$.
\end{lem}
\begin{proof}
Let $q$ be the probability vector for the $(1+\beta)$-process, where  $\beta \geq 1 -\tilde{\delta_1}$ for some integer $k \geq 1$.

First, consider any $1 \leq j \leq k$. Note that as $q$ is non-increasing in $i$ and $\beta \geq 1-\delta_1 \geq 1 - \delta_j$, and for any $n \cdot \delta_{j-1} + 1 \leq i \leq n \cdot \delta_{j}$,
\begin{align*}
 q_{i} &\stackrel{(1)}{\leq} q_{n \cdot \delta_j} \\ &\stackrel{(2)}{=} (1-\beta) \cdot \frac{1}{n} + \beta \cdot \frac{2n \cdot \delta_j-1}{n^2} \\ &\stackrel{(3)}{\leq} \delta_{j} \cdot \frac{1}{n} + 1 \cdot 2 \cdot \delta_j \cdot \frac{1}{n} \\ &= 3 \cdot \delta_j \cdot \frac{1}{n} \\ &\leq 3 \cdot (\delta_{j-1} + \delta_{j}) \cdot \frac{1}{n},
\end{align*}
where (1) and (2) hold by definition of $(1+\beta)$-process, and inequality (3) uses $\beta \geq 1 - \tilde{\delta}_1 \geq 1 - \delta_1 \geq 1 -\delta_j $.

Similar to the above, we can upper bound
\begin{align*}
 q_{n/3} &= (1-\beta) \cdot \frac{1}{n} + \beta \cdot \frac{2(n/3) -1}{n^2}  \\
 &\leq \frac{1}{n} + \beta \cdot \left( \frac{2}{3 n} - \frac{1}{n} \right) \\
  &= \frac{1}{n} - \frac{ \frac{1}{3} \beta}{n} \leq \frac{1 - 4 \epsilon}{n},
\end{align*}
for $\epsilon:=\frac{1}{16} \beta$.
Similarly, for the lower bound
\begin{align*}
 q_{2n/3} &= (1-\beta) \cdot \frac{1}{n} + \beta \cdot \frac{2(2n/3) -1}{n^2}  \\
 &\geq \frac{1}{n} + \beta \cdot \left( \frac{4}{3 n} - \frac{1}{n} \right) - \frac{\beta }{n^2} \\
  &= \frac{1}{n} + \frac{ \frac{1}{3} \beta}{n} - \frac{\beta}{n^2} \\ &\geq \frac{1 + \frac{1}{4} \beta}{n}
  = \frac{1+4 \epsilon}{n}.
\end{align*}
\end{proof}

Using the above lemma, majorization (\cref{lem:old_majorisation}) and \thmref{multiple_quantiles_relaxed} yields immediately:
\oneplusbetaapplication*

It is straightforward to derive an almost matching lower bound on the gap, showing that $1-\beta$ has to be (almost) polynomially small in order to achieve a gap of $\Oh(\log \log n)$: %
\begin{rem}\label{rem:lowerbound_beta}
Consider a $(1+\beta)$-process with $\beta \leq 1 - n^{-c_3/\log \log n}$ for some $c_3 >0$ (not necessarily constant). Then,
\[
 \Pro{ \Gap(n) \geq \frac{2}{c_3} \log \log n } \geq 1-o(1).
\]
\end{rem}
\begin{proof}
Consider the allocation of the first $n$ balls into $n$ bins. Then out of the first $n$ balls, at least $(1/2) \cdot n^{1-1/\log \log n}$ balls will be allocated using \OneChoice \whp. If this occurs, then the probability that any fixed bin receives at least $\frac{1}{2c_3} \log \log n$ balls is at least
\begin{align*}
 \lefteqn{\binom{n^{1-c_3/\log \log n}}{\frac{1}{2c_3} \log \log n} \left( \frac{1}{n} \right)^{\frac{1}{2c_3} \log \log n}
 \left(1 - \frac{1}{n} \right)^{n - \frac{1}{2c_3} \log \log n}} \\
 &\geq \left(  \frac{n^{1-c_3/\log \log n}}{
 \frac{1}{2c_3} \log \log n} \right)^{\frac{1}{2c_3} \log \log n} \cdot \left( \frac{1}{n} \right)^{\frac{1}{2c_3} \log \log n} \cdot  \frac{1}{e} \\
 &\geq n^{-\frac{1}{2} - \epsilon},
\end{align*}
for some small constant $\epsilon > 0$. Hence we have a probability of at least $n^{-1/2 - \epsilon} > n^{-1/3}$ for a fixed bin to reach a load of at least $\frac{1}{2c_3} \log \log n$. Using Poissonization for the event that there is a bin with load at least $\frac{2}{c_3} \log \log n$ (similarly to \cref{lem:lower_bound_max_load_whp}), it follows that \whp~at least one bin has a load of at least $\frac{1}{2 c_3} \log \log n$.
\end{proof}

\subsection{Graphical Balanced Allocation}

We now analyze the \emph{graphical balanced allocation process}, with a focus on dense expander graphs. 
To this end, we first recall some basic notation of spectral graph theory and expansion.
For an undirected graph $G$, the normalized Laplacian Matrix of $G$ is an $n \times n$-matrix defined by 
\[
 \mathbf{L} = \mathbf{I} - \mathbf{D}^{-1/2} \cdot \mathbf{A} \cdot \mathbf{D}^{1/2},
\]
where $\mathbf{I}$ is the identity matrix, $\mathbf{A}$ is the adjacency matrix and $\mathbf{D}$ is the diagonal matrix where $D_{u,u} = \deg(u)$ for any vertex $u \in V$. 
Further, let $\lambda_1 \leq \lambda_2 \leq \cdots \leq \lambda_n $ be the $n$ eigenvalues of $\mathbf{L}$, and let $\lambda:= \max_{i \in [2,n]} \left| 1-\lambda_i \right|$ be the spectral expansion of $G$.
Further, for any set $U \subseteq V$ define $\vol(U):=\sum_{v \in U} \deg(v)$.
Note that for a $d$-regular graph, we have $\vol(U)=d \cdot |U|$ and $\vol(V) = d n$.

We now recall the following (stronger) version of the Expander Mixing Lemma (cf.~\cite{Chung}):
\begin{lem}[Expander Mixing Lemma]\label{lem:eml}
For any subsets $X,Y \subseteq V$,
\[
 \left| |E(X,Y)| - \frac{ \vol(X) \cdot \vol(Y)}{\vol(V)} \right| \leq \lambda \cdot \frac{\sqrt{ \vol(X) \cdot \vol(\overline{X}) \cdot \vol(Y) \cdot \vol(\overline{Y})}}{\vol(V)},
\]
where $\vol(\overline{X}) = \vol(V \setminus X)$.
\end{lem}

In the following, we consider $G$ to be a $d$-regular graph. 

\begin{pro}\label{pro:newthomas}
Consider the probability vector $p_i^t$, $1 \leq i \leq n$ of a graphical balanced allocation process on a $d$-regular graph $G$ with spectral expansion $\lambda$. Then this vector satisfies for any load configuration at any time $t$ the following three inequalities.
\begin{enumerate}
    \item For any $1 \leq j \leq \lambda \cdot n$,
    \[
     \sum_{i=1}^j p_i^t \leq 2 \lambda \cdot \frac{j}{n}.
    \]
    \item For any $\lambda \cdot n \leq j $,
    \[
     \sum_{i=1}^j p_i^t \leq 2 \cdot \left( \frac{j}{n} \right)^2.
     \]
     \item For any $1 \leq j \leq n$,
     \[
     \sum_{i=1}^j p_i^t \leq \frac{j}{n} \cdot \left(1 - (1-\lambda) \cdot \frac{n-j}{n} \right).
     \]
\end{enumerate}
\end{pro}
\begin{proof}
Fix $1 \leq j \leq \lambda \cdot n$. and let $S$ be the subset of vertices corresponding to the $j$-th most heavily loaded bins; so in particular, $|S|=j$. Using Lemma~\ref{lem:eml} for $X=Y=S$:
\[
|E(S,S)| \leq d \cdot \frac{|S| \cdot |S|}{n} + \lambda \cdot \frac{d |S| (n-|S|)}{n} \leq d \cdot \frac{|S| \cdot |S|}{n} + \lambda d \cdot |S|.
\]
Since $|S| \leq \lambda \cdot n$, we conclude that
\[
 |E(S,S)| \leq 2 \lambda d \cdot |S|.
\]
Note that the for the graphical balanced allocation process, we  place a ball in one of the $|S|$-th most heavily loaded bins if and only if we pick an edge in $E(S,S)$. Using this and the upper bound on $|E(S,S)|$ from above, it follows that
\[
 \sum_{i=1}^{|S|} p_i^t = \frac{|E(S,S)|}{2 |E|} \leq  \frac{2 \lambda d |S|}{nd} = 2 \lambda \cdot \frac{|S|}{n},
\]
and the first statement follows.

Consider now the case where $\lambda n \leq |S|$. Then,
\begin{align*}
 |E(S,S)| &\leq d \cdot \frac{|S| \cdot |S|}{n} + \lambda d \cdot \frac{ |S| \cdot (n-|S|)}{n} \\
 &\leq d \cdot \frac{|S| \cdot |S|}{n} + d \cdot \frac{|S| \cdot |S|}{n} \cdot \frac{n-|S|}{n} \\
 &\leq 2d \cdot \frac{|S| \cdot |S|}{n}.
\end{align*}
and therefore,
\[
 \sum_{i=1}^{|S|} p_i^t = \frac{|E(S,S)|}{2 |E|} \leq \frac{2d \cdot \frac{|S| \cdot |S|}{n} }{nd} = 2 \cdot \left( \frac{|S|}{n} \right)^2,
\]
which establishes the second statement.

Finally, consider the general case where $1 \leq |S| \leq n$.
Then using Lemma~\ref{lem:eml},
 \begin{align*}
 |E(S,S)| &\leq d \cdot \frac{|S| \cdot |S|}{n} + d \cdot \lambda  \cdot \frac{ |S| (n-|S|)}{n} \\
 &= d \cdot |S| \cdot \left( \frac{|S|}{n}
 + \lambda \cdot \frac{n-|S|}{n}
 \right) \\
  &= d \cdot |S| \cdot \left( \frac{|S|}{n} + \frac{n-|S|}{n}
 - 1 \cdot \frac{n-|S|}{n} + \lambda \cdot \frac{n-|S|}{n}
 \right) \\
  &= d \cdot |S| \cdot \left( 1 - (1 - \lambda) \cdot \frac{n-|S|}{n}
 \right).
 \end{align*}
 
and therefore, 
\[
 \sum_{i=1}^{|S|} p_i = \frac{|E(S,S)|}{2 |E|} \leq \frac{d |S| \cdot \left(1 - (1-\lambda) \cdot \frac{n-|S|}{n} \right)}{n d} = \frac{|S|}{n} \cdot \left(1 - (1-\lambda) \cdot \frac{n-|S|}{n} \right).
 \]
 \end{proof}

\begin{lem}\label{lem:newthomas_graphical}
Consider a graphical balanced allocation process on a connected, $d$-regular graph on $G$ with spectral expansion $\lambda \leq 1/2$. Further, let $2^{-0.5 (\log n)^{(k-1)/k}} \geq \lambda$
for an integer $k \geq 1$.
Then there exists a process in the class $\RelaxedQuantile_{\gamma}(\delta_1,\ldots,\delta_k)$, where each $\delta_i$ is $\tilde{\delta}_i$ being rounded up to the nearest multiple of $\frac{1}{n}$ and $\gamma=2$, which majorises the probability vector of the graphical balanced allocation process in each round $t \geq 1$, for any possible load configuration.
\end{lem}
\begin{proof}

For the graphical balanced allocation process as defined in the statement, for any $1 \leq j \leq n \delta_1$, it follows by \cref{pro:newthomas} (first statement),
\[
  \sum_{i=1}^j p_i^t \leq \frac{j}{n} \cdot 2 \lambda \leq 2 \cdot \frac{j}{n} \cdot \delta_1,
\]
and it is majorized by a $\RelaxedQuantile_{\gamma}(\delta_1,\ldots,\delta_k)$ process provided that $\gamma \geq 2$.

Consider now any prefix sum over $p$, where $n \delta_1 < j $. Then by \cref{pro:newthomas} (second statement) %
\[
 \sum_{i=1}^{j} p_i^t \leq 2 \cdot \left( \frac{j}{n} \right)^2.
\]
Let $q$ be the probability vector of the (original) $k$-quantile. If we take the corresponding sum in the definition of the relaxed quantile process, then, as this is at least the probability that \TwoChoice allocates a ball into one of the $j$-th most heavily loaded bins, we have
\begin{align*}
 \sum_{i=1}^{j} \gamma \cdot q_i &\geq \gamma \cdot \sum_{i=1}^j \frac{2i-1}{n^2} = \gamma \cdot \frac{j^2}{n^2}.
\end{align*}
Hence if $\gamma=2$,  we conclude that the prefix sum of $p$ is smaller than that of a relaxed quantile process.

First, consider any prefix sum until $j \in [0,n/3]$. Then by \cref{pro:newthomas} (third statement),
\begin{align*}
 \sum_{i=1}^{j} p_i^t 
 &\leq 
 \frac{j}{n} \cdot \left(1 - (1- \lambda) \cdot \frac{n-j}{n} \right) \\ &\leq \frac{j}{n} \cdot \left(1 - (1- \lambda) \cdot \frac{2}{3} \right).
\end{align*}
Since by assumption $\lambda \leq 1/2$, we have $1-(1-\lambda) \cdot \frac{2}{3} \leq 1 - 1/3 = 1 - 4 \epsilon$ for $\epsilon = 1/12$. 

Similarly, for any $j \in [(2/3) n,n]$, then by \cref{pro:newthomas} (third statement),
\begin{align*}
 \sum_{i=j}^{n} p_{i}^t &\geq 
 1 - \frac{j}{n} \cdot \left( 1- (1 - \lambda) \cdot \frac{n-j}{n} \right) \\ &\geq \left(1 - \frac{j}{n} \right) + \frac{j}{n} \cdot \frac{n-j}{n} \cdot (1-\lambda) \\
 &\geq \left( 1 - \frac{j}{n} \right) \cdot \left( 1 + \frac{j}{n} \cdot (1-\lambda) \right) \\
 &\geq \left( 1 - \frac{j}{n} \right) \cdot \left( 1 + \frac{2}{3} \cdot (1-\lambda) \right),
\end{align*}
and thus again, since $\lambda \leq 1/2$, we have
$1 + \frac{2}{3} \cdot (1-\lambda) \geq 1 + 4 \epsilon$. Hence there exists a time-invariant process in the class $\RelaxedQuantile_{\gamma}(\delta_1,\ldots,\delta_k)$ for $\gamma=2$, which majorizes the graphical allocation process for any possible load configuration.
\end{proof}

\begin{thm}\label{thm:graphical}
Consider a graphical balanced allocation process on a connected, $d$-regular graph on $G$ with spectral expansion $\lambda \leq 1/2$. 
Further, let $k$ be the largest integer with
$1 \leq k \leq \kappa \cdot \log \log n$ such that
$2^{-0.5 (\log n)^{(k-1)/k}} \geq \tilde{\lambda}:=\max\{\lambda, n^{-0.00005} \}$. Then for any $m \geq 1$, 
\[
 \Pro{ \Gap(m) \leq 1000 \cdot k \cdot \left( \frac{ \log n}{\log (1/\tilde{\lambda}) } \right)^{(k+1)/k} } \geq 1-n^{-3}.
\]
\end{thm}
From the general bound in the above corollary, we can deduce the following two bounds:
\begin{rem}\label{rem:graphical}
Under the assumptions of \cref{thm:graphical}, we have the following more explicit (but slightly weaker) bound for any $2 \leq k \leq \kappa \cdot \log \log n$,
\begin{align*}
    \Pro{ \Gap(m) \leq 1000 \cdot \frac{\log \log n}{\log \log n - \log \log (1/\tilde{\lambda}) + \log(0.5)} \cdot \left( \frac{ \log n}{\log (1/\tilde{\lambda}) } \right)^{3/2} } \geq 1-n^{-3}.
\end{align*}
Also if $\lambda \leq 1/2^{ (\log n)^{c_1}}$ for some constant $0 < c_1 < 1$, then
\begin{align*}
    \Pro{ \Gap(m) \leq 1000 \cdot \frac{1}{\log (10^4)} \log \log n \cdot (\log n)^{ (3/2) \cdot (1 - c_1) } } \geq 1 -n^{-3}.
\end{align*}
\end{rem}
Finally, let us consider the case where $\lambda$ decays polynomially in $n$.
\corthomas*

Note that $\lambda \leq n^{-c_1}$ captures a \emph{relaxed}, \emph{multiplicative} approximation of Ramanujan graphs (it is in fact more relaxed than the existing notion ``weakly Ramanujan''). Recently, \cite{TY19} proved that for any $\poly(n) \leq d \leq n/2$, a random $d$-regular graph satisfies the constraint on $\lambda$ with probability at least $1-n^{-1}$.

Further, we remark that the above result extends one of the main results of \cite{KP06} which states that for any graph with degree $n^{1/\log \log n}$, graphical balanced allocation achieves a gap of at most $\Theta(\log \log n)$ in the lightly loaded case ($m=n$). Our result above also refines a previous result of \cite{PTW15} which states that for any expander graph, a gap bound of $\Oh(\log n)$ holds (even in the heavily loaded case $m \geq n$). In conclusion, we see that the gap bound of $\Oh(\log \log n)$ extends from the complete graph (which is the \TwoChoice process) to other graphs, provided we have a strong expansion and high density.

\begin{proof}[Proof of~\cref{thm:graphical}]
We will use Lemma~\ref{lem:newthomas_graphical}, and then apply the generalized majorization result of \cite[Theorem 3.1]{PTW15} (see Lemma~\ref{lem:new_majorisation}). This way, we can extend the upper bound on the gap \cref{thm:multiple_quantiles_relaxed} to the graphical allocation process. 

First observe that since we require $2^{-0.5(\log n)^{(k-1)/k}} \geq n^{-0.00005}$, the integer $k$ is not larger than $\lceil \frac{1}{\log(10^4)} \log \log n \rceil \leq \frac{1}{\log(10^4)} \log \log n + 1$, so the upper bound on $k$ from \cref{thm:multiple_quantiles_relaxed} is satisfied.

By assumption on $k$, we have
\begin{align}
 -(0.5) \cdot (\log n)^{(k-1)/k} \geq \log ( \tilde{\lambda} ), \label{eq:solvek_two}
\end{align}
but also, as $k$ is chosen as large as possible,
\begin{align}
 -(0.5) \cdot (\log n)^{k/(k+1)} \leq \log ( \tilde{\lambda} ), \notag
\end{align}
or equivalently,
\begin{align}
 (0.5) \cdot (\log n)^{k/(k+1)} \geq \log ( 1/ \tilde{\lambda} ). \label{eq:solvek}
\end{align}
Multiplying both sides by $(\log n)^{1/(k+1)}$ gives
\[
 (0.5) \cdot (\log n) \geq  \log (1/ \tilde{\lambda}) \cdot (\log n)^{1/(k+1)}.
\]
Rearranging, we have
\[
   (\log n)^{1/(k+1)} \leq 0.5 \cdot \frac{ \log n}{\log (1/\tilde{\lambda}) },
\]
and thus
\begin{align}
 (\log n)^{1/k} \leq \left( \frac{ \log n}{\log (1/\tilde{\lambda}) } \right)^{(k+1)/k}. \label{eq:boundonlogn}
\end{align}
Applying the gap bound of $1000 \cdot k \cdot (\log n)^{1/k}$ from \cref{thm:multiple_quantiles_relaxed}, and using 
and \ref{eq:boundonlogn} implies
\[
 \Pro{ \Gap(m) \leq 1000 \cdot k \cdot \left( \frac{ \log n}{\log (1/\tilde{\lambda}) } \right)^{(k+1)/k} } \geq 1-n^{-3}.
\]
\end{proof}

\begin{proof}[Proof of Remark~\ref{rem:graphical}]
We proceed as in the proof of~\cref{thm:graphical}, but we will upper bound the factor of $k$ by a more refined expression (instead of $\Theta(\log \log n)$).
Returning to \cref{eq:solvek_two} and taking logarithms yields
\[
 \log(0.5) + \frac{k}{k-1} \cdot \log \log n \leq \log \log (1/\tilde{\lambda}),
\]
and rearranging,
\[
 1 - \frac{1}{k} \leq \frac{\log \log (1/\tilde{\lambda}) - \log(0.5)}{\log \log n},
\]
and thus 
\begin{align}
 k 
 \leq \frac{\log \log n}{\log \log n - \log \log (1/\tilde{\lambda}) + \log(0.5)}. \label{eq:boundonk}
\end{align}
The rest of the proof is identical to~\cref{thm:graphical}.
\end{proof}

\section{Conclusions}\label{sec:conclusion}

In this work, we introduced a new framework of balls-and-bins with incomplete information. %
The main contributions are as follows (see also Table~\ref{tab:my_label} for a comparison with related work):
\begin{enumerate}\itemsep-1pt
  \item A lower bound of $\Omega(\sqrt{\log n})$ for a fixed $m = \Theta(n \sqrt{\log n})$ for one adaptive query (\cref{thm:new_adaptive_quantile_lower_bound}), disproving Problem 1.3 in~\cite{FG18}. Also, a stronger lower bound of $\Omega(\log n/ \log \log n)$ for ``many'' time-steps in $[1,n \log^2 n]$ (\cref{cor:large_gap_in_nlog2n_interval}), again for one adaptive query. %
  \item Design and analysis of an instance of the $k$-quantile process for any $k \geq 1$. This process achieves \Whp an $\Oh(k \cdot (\log n)^{1/k})$ gap for any $m \geq 1$ and $ k = \Oh(\log \log n)$ (Theorem~\ref{thm:new_multiple_quantiles}). This process has several theoretical applications:
  \begin{itemize}[topsep=0.1pt]\itemsep-1pt
    \item A ``power of two-queries'' phenomenon: reduction of the gap from $\Omega(\log n / \log \log n)$ to $\Oh(\sqrt{\log n})$ by increasing the number of queries from one to two.
    \item For $k = \Theta(\log \log n)$,  a gap bound of $\Oh(\log \log n)$ which matches the gap of the process with full information (\TwoChoice) up to multiplicative constants.
    \item New upper bounds on the gap of the $(1+\beta)$ process with $\beta$ close to $1$ by relating it to a \RelaxedQuantile process. (\cref{thm:oneplusbeta_application})
    \item New upper bounds on the graphical balanced allocation on dense expander graphs, making progress towards Open Question 2 in~\cite{PTW15} (\cref{cor:thomas}).
  \end{itemize}
  \item Several majorizations and reductions between the processes \Quantile, \Threshold, \RelaxedQuantile, \textsc{Thinning}, $(1+\beta)$ and \TwoChoice (see Figure~\ref{fig:process_overview} for a high-level outline, and \cref{sec:relations} for more details).
\end{enumerate}

One natural open question is whether we can prove matching lower bounds, in particular, the case $k \geq 2$ is wide open.
Another interesting direction is to investigate other  allocation processes with limited information, e.g., where a sampled bin reports its actual load perturbed by some random or deterministic noise function.

\section{Experimental Results}\label{sec:experiments}

\begin{figure}[ht]
\begin{subfigure}[t]{0.5\textwidth}
    \centering
    \includegraphics[height=3.7cm]{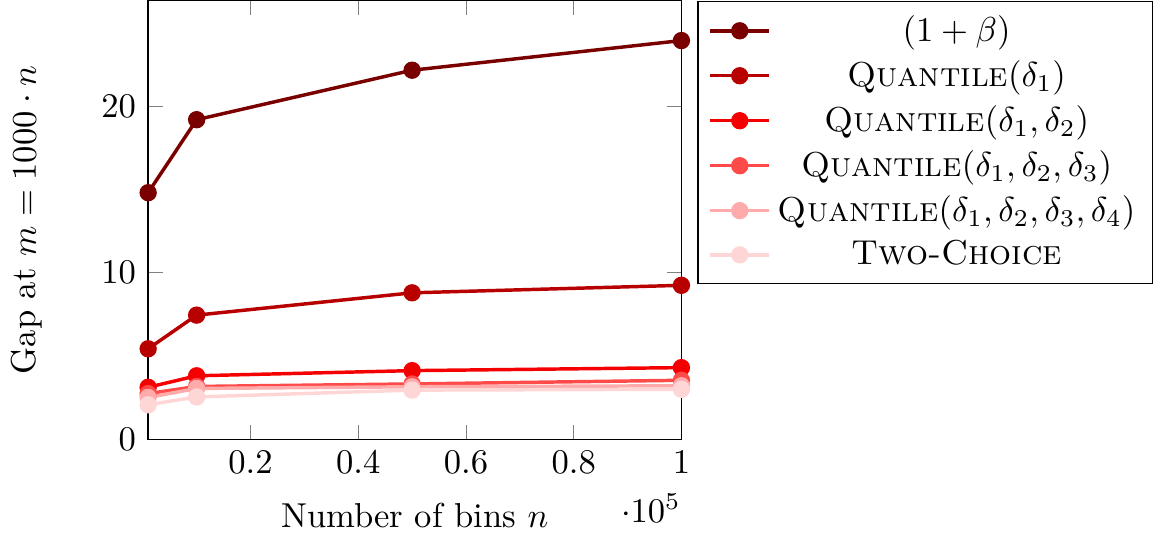}
    \caption{}
    \label{fig:gap_vs_bins}
\end{subfigure}
\quad
\begin{subfigure}[t]{0.5\textwidth}
    \centering
    \includegraphics[height=3.7cm]{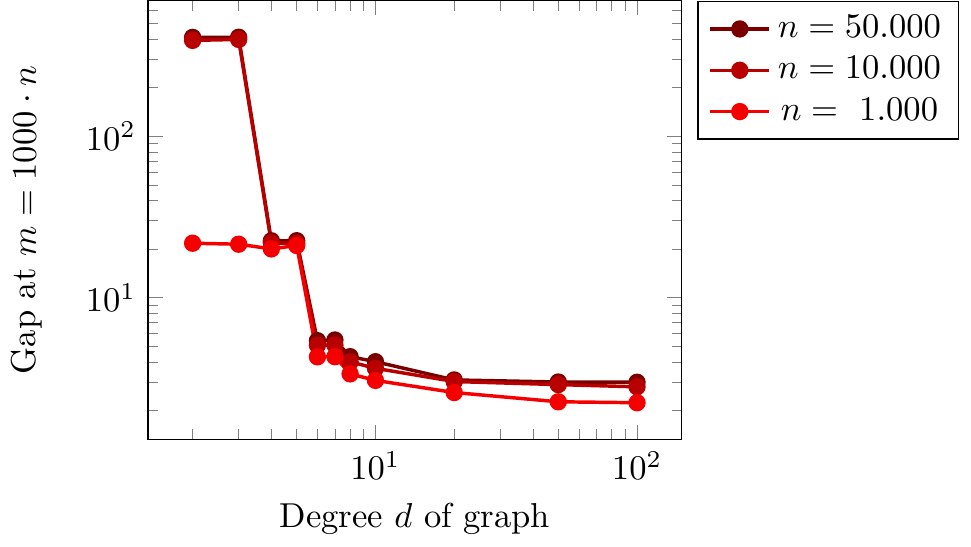}
    \caption{}
    \label{fig:gap_vs_deg}
\end{subfigure}
\caption{(\subref{fig:gap_vs_bins}) Average Gap vs.~$n \in \{ 10^3, 10^4, 5 \cdot 10^4, 10^5\}$ for the experimental setup of \cref{tab:gap_distribution} and (\subref{fig:gap_vs_deg}) Average Gap vs.~$n \in \{ 10^3, 10^4, 5 \cdot 10^4 \}$ for regular graphs generated using~\cite{SW97}.}
\end{figure}

We also recorded the empirical distribution of the gap over $100$ repetitions at $m = 1000 \cdot n$ for the $(1+\beta)$ process with $\beta = 1/2$, the $k$-quantile processes (for $k =1,2,3,4$) of the form defined in \cref{sec:new_upper_bound_for_k_queries}, and the \TwoChoice process. As shown in \cref{tab:gap_distribution} and \cref{fig:gap_vs_bins}, the experiments indicate a superiority of $k$-quantile over $(1+\beta)$ (for $\beta=0.5$), even for $k=1$. The experiments also demonstrate a large improvement of $k=2$ over $k=1$ (``Power of Two Queries''). \cref{fig:gap_vs_deg} shows empirical evidence of how the gap decreases (and reaches values close to the \TwoChoice gap) in regular graphs as the degree increases.

\colorlet{GA}{black!40!white}
\colorlet{GB}{black!70!white}
\colorlet{GC}{black}
\newcommand{\CA}[1]{\textcolor{GA}{#1}} %
\newcommand{\CB}[1]{\textcolor{GB}{#1}} %
\newcommand{\CC}[1]{\textcolor{GC}{#1}} %
\newcommand{\CI}[2]{\FPeval{\result}{min(30 + 3 * #1, 100)} \colorlet{tmpC}{black!\result!white} {\textcolor{tmpC}{#2}}}

\begin{table}[ht]
    \centering
   \footnotesize{
    \begin{tabular}{|c|c|c|c|c|c|c|}
    \hline
$n$ & $(1+\beta)$, for $\beta =0.5$ & $k = 1$& $k = 2$ & $k = 3$ & $k = 4$ & \TwoChoice \\ \hline
        $10^3$ &
\makecell{
\CI{5}{\textbf{12} : \ 5\% } \\
\CI{15}{\textbf{13} : 15\% } \\
\CI{31}{\textbf{14} : 31\% } \\
\CI{21}{\textbf{15} : 21\% } \\
\CI{15}{\textbf{16} : 15\% } \\
\CI{5}{\textbf{17} : \ 5\% } \\
\CI{4}{\textbf{18} : \ 4\% } \\
\CI{2}{\textbf{19} : \ 2\% } \\
\CI{1}{\textbf{20} : \ 1\% } \\
\CI{1}{\textbf{21} : \ 1\% } } &
\makecell{
\CI{1}{\textbf{\ 3} : \ 1\% } \\ 
\CI{11}{\textbf{\ 4} : 11\% } \\ 
\CI{46}{\textbf{\ 5} : 46\% } \\
\CI{33}{\textbf{\ 6} : 33\% } \\
\CI{6}{\textbf{\ 7} : \ 6\% } \\
\CI{2}{\textbf{\ 8} : \ 2\% } \\
\CI{1}{\textbf{10} : \ 1\% } } &
\makecell{
\CI{4}{\textbf{2} : \ 4\% } \\
\CI{80}{\textbf{3} : 80\% } \\ 
\CI{16}{\textbf{4} : 16\% } } &
\makecell{
\CI{24}{\textbf{2} : 24\% } \\
\CI{74}{\textbf{3} : 74\% } \\
\CI{2}{\textbf{4} : 2\% } } &
\makecell{
\CI{50}{\textbf{2} : 50\% } \\
\CI{49}{\textbf{3} : 49\% } \\
\CI{1}{\textbf{4} : \ 1\% } } &
\makecell{
\CI{93}{\textbf{2} : 93\% } \\
\CI{7}{\textbf{3} : \ 7\% }} \\ \hline
        $10^4$ &
\makecell{
\CI{3}{\textbf{16} : \ 3\% } \\
\CI{21}{\textbf{17} : 21\% } \\
\CI{19}{\textbf{18} : 19\% } \\
\CI{10}{\textbf{19} : 10\% } \\
\CI{23}{\textbf{20} : 23\% } \\
\CI{11}{\textbf{21} : 11\% } \\
\CI{10}{\textbf{22} : 10\% } \\
\CI{2}{\textbf{23} : \ 2\% } \\
\CI{1}{\textbf{24} : \ 1\% }} &
\makecell{
\CI{14}{\textbf{\ 6} : 14\% } \\
\CI{42}{\textbf{\ 7} : 42\% } \\
\CI{25}{\textbf{\ 8} : 25\% } \\
\CI{15}{\textbf{\ 9} : 15\% } \\
\CI{2}{\textbf{10} : \ 2\% } \\
\CI{1}{\textbf{11} : \ 1\% } \\
\CI{1}{\textbf{12} : \ 1\% } } &
\makecell{
\CI{27}{\textbf{3} : 27\% } \\
\CI{65}{\textbf{4} : 65\% } \\
\CI{8}{\textbf{5} : \ 8\% } } &
\makecell{
\CI{83}{\textbf{3} : 83\% } \\
\CI{17}{\textbf{4} : 17\% } } &
\makecell{
\CI{95}{\textbf{3} : 95\% } \\
\CI{5}{\textbf{4} : \ 5\% } } &
\makecell{
\CI{46}{\textbf{2} : 46\% } \\
\CI{54}{\textbf{3} : 54\% }} \\ \hline
        $10^5$ &
\makecell{
\CI{2}{\textbf{20} : \ 2\% } \\ 
\CI{7}{\textbf{21} : \ 7\% } \\
\CI{9}{\textbf{22} : \ 9\% } \\
\CI{26}{\textbf{23} : 26\% } \\
\CI{27}{\textbf{24} : 27\% } \\
\CI{14}{\textbf{25} : 14\% } \\
\CI{6}{\textbf{26} : \ 6\% } \\
\CI{3}{\textbf{27} : \ 3\% } \\
\CI{4}{\textbf{28} : \ 4\% } \\
\CI{1}{\textbf{29} : \ 1\% } \\
\CI{1}{\textbf{34} : \ 1\% }} &
\makecell{
\CI{28}{\textbf{\ 8} : 28\% } \\
\CI{42}{\textbf{\ 9} : 42\% } \\
\CI{18}{\textbf{10} : 18\% } \\
\CI{7}{\textbf{11} : \ 7\% } \\
\CI{3}{\textbf{12} : \ 3\% } \\
\CI{1}{\textbf{14} : \ 1\% } \\
\CI{1}{\textbf{15} : \ 1\% }} &
\makecell{
\CI{72}{\textbf{4} : 72\% } \\
\CI{26}{\textbf{5} : 26\% } \\
\CI{2}{\textbf{6} : \ 2\% } } &
\makecell{
\CI{46}{\textbf{3} : 46\% } \\
\CI{54}{\textbf{4} : 54\% } } &
\makecell{
\CI{79}{\textbf{3} : 79\% } \\
\CI{21}{\textbf{4} : 21\% } } &
\makecell{
\CI{100}{\textbf{3} : 100\% }} \\ \hline
    \end{tabular}
    }
    \caption{Summary of our Experimental Results ($m=1000 \cdot n$).
    }
    \label{tab:gap_distribution}
\end{table}

\clearpage
\newpage

\begin{table}[H]
\centering
\tabulinesep=0.8mm
\begin{tabu}{|c|c|c|}
\hline
 & \multicolumn{2}{c|}{\textbf{Lightly loaded case} $m = \Oh(n)$}  \\ \hline
\textbf{Process} & \textbf{LB on Gap} & \textbf{UB on Gap} 
\\ \hline
\OneChoice & \multicolumn{2}{c|}{$\Theta(\frac{\log n}{\log \log n})$~\cite{RS98}}
\\ \hline
$\Threshold(f)$ & \multicolumn{2}{c|}{$(1 + o(1)) \sqrt{\frac{8\log n}{\log \log n}}$~\cite{FG18}} \\ \hline
$\Quantile(\delta)$ & - & $(1 + o(1)) \sqrt{\frac{8\log n}{\log \log n}}$ ~\cite{FG18} \\ \hline
$\Threshold(f_1, \ldots , f_k)$ & - & $\Oh\bigl(\sqrt[k + 1]{(k + 1) \frac{\log n}{\log ((k + 1)\log n)}}\bigr)$ ~\cite{IK04} \\ \hline %
\TwoChoice & \multicolumn{2}{c|}{$\log_2 \log n + \Theta(1)$~\cite{ABK99}} \\ \hline
\end{tabu}

\vspace{1cm}
\begin{tabu}{|c|c|c|}
\hline
 & \multicolumn{2}{c|}{\textbf{Heavily loaded case}  $m \gg n$}  \\ \hline
\textbf{Process} & \textbf{LB on Gap} & \textbf{UB on Gap} \\ \hline
\OneChoice & \multicolumn{2}{c|}{$\Theta(\sqrt{\frac{m}{n} \log n})$~\cite{RS98}} \\ \hline
$(1 + \beta)$ & $\Omega(\log n / \beta)$, $m = \Theta((n \log n)/\beta^2)$ ~\cite{PTW15} & $\Oh(\log n / \beta)$~\cite{PTW15} 
\\ \hline
\multirow{2}{*}{$\Threshold(f)$} & $\Omega(\sqrt{\log n})$, $m = \Theta(n \log^2 n)$~($\star$) & \multirow{2}{*}{-} \\
 & $\Omega(\frac{\log n}{\log \log n})$, $m \in [n \log^2 n]$~($\star$) &  \\ \hline
\multirow{2}{*}{$\Quantile(\delta)$} & $\Omega(\sqrt{\log n})$, $m = \Theta(n \log^2 n)$~($\star$) & \multirow{2}{*}{$\Oh(\log n)$~($\star$)} \\ 
 & $\Omega(\frac{\log n}{\log \log n})$, $m \in [n \log^2 n]$~($\star$) &  \\ \hline
$\Quantile(\delta_1, \ldots , \delta_k)$ & - & $\Oh(k (\log n)^{1/k})$~($\star$)
 \\ \hline 
\TwoChoice & \multicolumn{2}{c|}{$\log_2 \log n + \Theta(1)$~\cite{BCSV06}}  \\ \hline
\end{tabu}
\caption{Lower Bounds (LB) and Upper Bounds (UB) that hold with $1-o(1)$ probability for the gap, for the lightly loaded and heavily loaded cases. Our results are indicated with ($\star$). The $(1+\beta)$ lower bound holds for any $\beta$ bounded away from $1$.}
\label{tab:my_label}
\end{table}

\addcontentsline{toc}{section}{References}

\renewcommand\bibname{}
\bibliographystyle{abbrv}
\bibliography{bibliography}

\appendix

\newpage

\section{Probabilistic Tools} \label{sec:probabilistic_tools}

\subsection{Concentration Inequalities}

\begin{lem}[Multiplicative factor Chernoff Binomial Bound~\cite{MRS01}] %
\label{lem:multiplicative_factor_chernoff_for_binomial}
Let $X_1, \ldots , X_n$ be independent binary random variables with $\Pro{X_i = 1} = p$. Then, 
\[
\Pro{\sum_{i = 1}^n X_i \geq npe } \leq e^{-np},
\]
and
\[
\Pro{\sum_{i = 1}^n X_i \leq \frac{np}{e}} 
\leq e^{\left(\frac{2}{e} - 1 \right)np}.
\]
\end{lem}

\begin{thm}[Berry-Esseen~\cite{E56}] \label{thm:berry_essen}
Let $X_1, \ldots , X_n$ be a sequence of i.i.d random variables with mean $\mu$, variance $\sigma^2$ and central moment $\rho = \ex{|X_i - \mu|^3}$. Then there exists a constant $C > 0$ such that for $\alpha \in \mathbb{R}$
\[
\left\lvert \Pro{\frac{\overline{X}_n - \mu}{\frac{\sigma}{\sqrt{n}}} \leq \alpha} - \tilde{\Phi}(\alpha) \right\rvert \leq C \cdot \frac{\rho}{\sigma^3 \sqrt{n}},
\]
where $\tilde{\Phi}$ is the cumulative distribution of the standard normal distribution.
\end{thm}

\begin{lem}[Berry-Esseen for Poisson r.vs] \label{lem:berry_essen_poisson}
Let $X \sim \mathrm{Po}(m)$, where $m \in \mathbb{N}$, then 
\[
\left| \Pro{X \leq m + \alpha \sqrt{m}} - \tilde{\Phi}(\alpha) \right| \leq C \cdot \frac{\rho}{\sigma^3 \sqrt{m}}.
\]
\end{lem}
\begin{proof}
The sum of $n$ independent Poisson r.vs. with parameters $(k_i)_{i=1}^n$ is a Poisson r.v.~with parameter $\sum_{i = 1}^n k_i$ (e.g.~\cite[Lemma~5.2]{MU2017}). Hence, we can write $X$ as the sum of $m$ r.vs. $X_i \sim \mathrm{Po}(1)$. Then, applying \cref{thm:berry_essen} gives, 
\[
\left| \Pro{\frac{\frac{\sum_{i = 1}^n X_i}{m} - \mu}{\frac{\sigma}{\sqrt{m}}} \leq \alpha} - \tilde{\Phi}(\alpha) \right|
= \left| \Pro{X \leq m + \alpha\sqrt{m}} - \tilde{\Phi}(\alpha) \right| \leq C \cdot \frac{\rho}{\sigma^3 \sqrt{m}}.
\]
\end{proof}

In order to state the concentration inequality for supermartingales conditional on a bad event not occurring, we introduce the following definitions from~\cite{CL06}. Consider any r.v.~$X$ (in our case it will be the potential functions $\Phi_j$ and the $\Gamma_1$) that can be evaluated by a sequence of decisions $Y_1, Y_2, \ldots ,Y_N$ of finitely many outputs (the allocated balls). We can describe the process by a \textit{decision tree} $T$, a complete rooted tree with depth $n$ with vertex set $V(T)$. Each edge $uv$ of $T$ is associated with a probability $p_{uv}$ depending on the decision made from $u$ to $v$. 

We say $f : V (T) \to \mathbb{R}$ satisfies an \textit{admissible condition} $P$ if $P = \{P_v\}$ holds for every vertex $v$. For an admissible condition $P$, the associated bad set $B_i$ over the $X_i$ is defined to be
\[
B_i = \{ v \mid \text{the depth of $v$ is $i$, and $P_u$ does not hold for some ancestor $u$ of $v$} \}.
\]

\begin{thm}[Theorem 8.5 from~\cite{CL06}] \label{thm:chung_lu_theorem_8_5}
For a filter $\mathcal{F}$, $\{\emptyset, \Omega \} = \mathcal{F}^{(0)} \subset \mathcal{F}^{(1)} \subset \ldots \subset \mathcal{F}^{(N)} = \mathcal{F}$,
suppose that a random variable $X^{(s)}$ is $\mathcal{F}^{(s)}$-measurable, for $0 \leq s \leq N$. Let $B$ be the
bad set associated with the following admissible conditions:
\begin{align*}
\Ex{X^{(s)} \mid \mathcal{F}^{(s-1)}} & \leq X^{(s - 1)}, \\
\Var{X^{(s)} \mid \mathcal{F}^{(s-1)}} & \leq \sigma_s^2, \\
X^{(s)} - \Ex{X^{(s)} \mid \mathcal{F}^{(s-1)}} & \leq a_s + M,
\end{align*}
for fixed $\sigma_s > 0$ and $a_s > 0$. Then, we have for any $\lambda >0$,
\[
\Pro{X^{(N)} \geq X^{(0)} + \lambda} \leq \exp\left( - \frac{\lambda^2}{2(\sum_{s = 1}^N (\sigma_s^2 + a_s^2)+M \lambda /3) } \right) + \Pro{B}.
\]
\end{thm}

\subsection{Majorization Tools}

In \cref{sec:relations}, we prove some majorizations between processes. The following lemma in \cite{PTW15}, shows that probability vector majorization implies majorization of the load distribution (under an appropriate coupling).

\begin{lem}[{see~\cite[Theorem 3.1]{PTW15}}]\label{lem:old_majorisation}
Let $p$ and $q$ be probability vectors associated with processes $P$ and $Q$ respectively. If $p$ is majorized by $q$, then there is a coupling such that for all rounds $t \geq 1$, $y^{(t)}(P)$ is majorized by $y^{(t)}(Q)$.
\end{lem}

In \cref{sec:applications_abstract}, we make use of the following extension.

\begin{lem}[{see~\cite[Section 3]{PTW15}}]\label{lem:new_majorisation}
Consider two allocation processes $P$ and $Q$.  The allocation process $P$ uses a time-dependent probability vector $p^{(t)}$, which may depend on $\mathcal{F}^{(t-1)}$. The allocation process $Q$ uses at each time a fixed probability vector $q$, such that at each time $t \geq 1$, $p^{(t)}$ is majorized by $q$. Then there is a coupling such that for all rounds $t \geq 1$, $y^{(t)}(P)$ is majorized by $y^{(t)}(Q)$.
\end{lem}

\subsection{Facts about the One-Choice Process}

The following facts about the (very) lightly loaded region of \OneChoice, follow from the concentration inequalities stated before. The results by Raab and Steger~\cite{RS98} do not cover the region $m \ll n/\polylog(n)$, do not provide an estimate for the number of balls with height at least $k$ and also the bounds are not derived for at least $1 - n^{-c}$ probability.

\begin{lem} \label{lem:lower_bound_max_load_whp} %
Consider the \OneChoice process with $m = \frac{n}{\log^c{n}}$ balls into $n$ bins, where $c>0$ is an arbitrary constant. Then, for any constant $\alpha > 0$ and for sufficiently large $n$,
\[
\Pro{\Gap(m) > \frac{1}{c + 1} \cdot \frac{\log{n}}{\log{ \log{ n}}}} \geq 1 - \frac{2}{n^{\alpha}}.
\]
\end{lem}
\begin{proof}
We will bound the probability of event $\mathcal{E}$, that the maximum load is less than $M = \frac{1}{c + 1} \cdot \log{n}/ \log{ \log{ n}}$. The maximum load is a function that is increasing with the number of balls. 

The technique of Poissonization~\cite[Theorem 12]{ACM98} states that for \OneChoice, the probability of a monotonically increasing event (in this case $\mathcal{E}$) is bounded by twice the probability that the event holds for independent Poisson r.vs. in place of the load r.vs.

We define $\mathcal{E}'$ to be the event that the maximum load is less than $M$, for $n$ Poisson r.vs. Thus, $\Pro{\mathcal{E}} \leq 2 \cdot \Pro{\mathcal{E}'}$. We bound $\Pro{\mathcal{E}'}$ by bounding the probability that no bin has load exactly $M$. We want
\[
\Pro{\mathcal{E}'} \leq \left( 1 - \frac{e^{-\frac{1}{\log^c{n}}} \left( \frac{1}{\log^c{n}}\right)^M}{M!} \right)^n \leq \exp{ \left(-n \frac{e^{-\frac{1}{\log^c{n}}} \left( \frac{1}{\log^c{n}}\right)^M}{M!} \right)} \leq \frac{1}{n^\alpha}.
\]
This is equivalent to showing that

\begin{align*}
-n \frac{e^{-\frac{1}{\log^c{n}}} \left( \frac{1}{\log^c{n}}\right)^M}{M!} < -\alpha\log{n} & \Longleftrightarrow 
\log{n} -\frac{1}{\log^c{n}} - Mc \log{\log{n}} -\log{M!}> \log{(\alpha\log{n})} \\
& \Longleftrightarrow 
\log{n} - \log{(\alpha\log{n})} -\frac{1}{\log^c{n}} > Mc \log{\log{n}} + \log{M!}.
\end{align*} 

Using Stirling's upper bound~\cite[Equation 9.1]{F08},
\begin{align*}
Mc \log{\log{n}} + \log{M!} & < Mc \log{\log{n}} + M(\log{M} - 1) + \log{M} \\
& = M(c \log{\log{n}} - \log{(c+1)} + \log{\log{n}} - \log{\log{\log{n}}} - 1) + \log{M} \\
& = M(c+1) \log{\log{n}} - CM - M \log{\log{\log{n}}} + \log{M} \\
& = \log{n} - CM - M \log{\log{\log{n}}} + \log{M} \\
& < \log{n} - \log{(\alpha\log{n})} - \frac{1}{\log^c{n}},
\end{align*}
for sufficiently large $n$, since $\log{(\alpha\log{n})} + \frac{1}{\log^c{n}} = o(M \log{\log{\log{n}}} - M)$ for any constant $\alpha > 0$. Hence, we get the desired lower bound.
\end{proof}

We now extend \cref{lem:lower_bound_max_load_whp} to a case with fewer balls.
\begin{lem} \label{lem:new_lower_bound_max_load_whp}
(cf.~\cref{lem:lower_bound_max_load_whp}) Consider the \OneChoice process with $m = \frac{n}{e^{u\log^c{n}}}$ (for constants $0 < c < 1$ and $u>0$) balls into $n$ bins. Then, for any constant $k > 0$ with $u\cdot k < 1$, for any constant $\alpha > 0$ and for sufficiently large $n$,
\[
\Pro{\Gap(m) \geq  k \cdot (\log n)^{1- c}} \geq 1 - \frac{2}{n^{\alpha}}.
\]
\end{lem}
\begin{proof}
We define $\mathcal{E}$ and $\mathcal{E}'$ as in \cref{lem:lower_bound_max_load_whp}. We bound $\Pro{\mathcal{E}'}$ by bounding the probability that no bin has load exactly $M = k \cdot (\log n)^{1- c}$. We claim
\[
\Pro{\mathcal{E}'} \leq \left( 1 - \frac{e^{-e^{-u\log^c{n}}} \left( e^{-u\log^c{n}}\right)^M}{M!} \right)^n 
\leq \exp{ \left(-n \frac{e^{-e^{-u\log^c{n}}} \left( e^{-u\log^c{n}}\right)^M}{M!} \right)} < \frac{1}{n^\alpha},
\]
which is equivalent to showing that
\begin{align*}
-n \frac{e^{-e^{-u\log^c{n}}} \left( e^{-u\log^c{n}}\right)^M}{M!} < -\alpha\log{n} 
& \Leftrightarrow 
\log{n} -e^{-u\log^c{n}} - M u\log^c{n} -\log{M!}> \log{(\alpha\log{n})} \\
& \Leftrightarrow 
\log{n} - \log{(\alpha\log{n})} -e^{-u\log^c{n}} > M u\log^c{n} + \log{M!}.
\end{align*} 
Using Stirling's upper bound~\cite[Equation 9.1]{F08},
\begin{align*}
Mu\log^c{n} + \log{M!} & < M u\log^c{n} + M(\log{M} - 1) + \log{M} \\
& = M(u\log^c{n} + \log{k} + (1-c) \log{\log{n}} - 1) + \log{M} \\
& = ku \cdot \log{n} + M(\log{k} + (1-c) \log{\log{n}} - 1) + \log{M} \\
& < \log{n} - \log{(\alpha\log{n})} - e^{-u\log^c{n}},
\end{align*}
for sufficiently large $n$, since $\log{(\alpha\log{n})} + e^{-u\log^c{n}} + M(\log{k} + (1-c) \log{\log{n}} - 1) + \log{M} = o((1 -u\cdot k)\log{n})$ for any constant $\alpha > 0$ and $u \cdot k < 1$. Hence, we get the desired lower bound.
\end{proof}

\begin{lem} \label{lem:one_choice_close_to_max_load}
Consider the \OneChoice process for $m =n \log^2 n$. With probability at least $1-o(n^{-2})$, there are at least $c n\log n$ balls with at least $\frac{m}{n} + \frac{a}{2} \log n$ height for $a = 0.4$ and $c = 0.25$.
\end{lem}
\begin{proof}
Consider the event $\mathcal{E}$ that the number of balls with load above $\frac{a}{2} \log n$ is at most $\frac{1}{5} \log n$. Since $\mathcal{E}$ is monotonically increasing in the number of balls, its probability is bounded by twice the probability of the event occurring for independent Poisson random variables~\cite[Theorem 12]{ACM98}.

By Berry-Esseen inequality for Poisson random variables (\cref{lem:berry_essen_poisson}), for sufficiently large $n$ and since $\epsilon = (\log n)^{-4}$,
\[
\lvert \Pro{Y\geq a} - \tilde{\Phi}(a) \rvert \leq \epsilon \Rightarrow \tilde{\Phi}(a) - \epsilon \leq \Pro{X \geq \log^2 n + a \log n} \leq \tilde{\Phi}(a) + \epsilon.
\]
For $a = 0.4$, we get $\tilde{\Phi}(a) \leq 0.35$. Let $X_i := \mathbf{1}(Y_i \geq \log^2 n + a \log n)$ and let $X := \sum_{i=1}^n X_i$, then $X$ is a Binomial distribution with $p \leq 0.35$. Using the lower tail Chernoff bound for the Binomial distribution (\cref{lem:multiplicative_factor_chernoff_for_binomial}),
\[
\Pro{\sum_{i = 1}^n X_i \leq \frac{np}{e}} \leq e^{-\Omega(n)}.
\]
For sufficiently large $n$, the RHS can be made $o(1/n^2)$, hence there are at least $np/e$ bins with load at least $\frac{m}{n} + a \log n$ \wp~$1-o(1/n^2)$. This means that \Whp~at least $np/e \cdot a \log n = \frac{npa}{e} \log n \leq 0.26 \cdot n \log n$ balls have height $\frac{m}{n} + \frac{a}{2} \log n = \frac{m}{n} + 0.4 \log n$.
\end{proof}

\begin{lem} \label{lem:new_one_choice_close_to_max_load}
(cf. \cref{lem:one_choice_close_to_max_load}) \sloppy In the \OneChoice process, with $m = K n \sqrt{\log n} - \Oh(K n \sqrt{\log n} \cdot e^{-\sqrt{\log n}})$ with probability at least $1-o(n^{-2})$, for sufficiently large $n$, there are at least $e^{-0.21 \sqrt{\log n}} \cdot C n\sqrt{\log{n}}$ balls with height at least $(K + C)\cdot \sqrt{\log n}$, for $K = 1/10$ and for $C = 1/20$.
\end{lem}
\begin{proof}
Note that $m = K (1 - o(1)) n \sqrt{\log n}$. Using Poissonization~\cite[Theorem 12]{ACM98},  the probability that the statement of the lemma does not hold is upper bounded by twice the probability for the corresponding event with $n$ independent Poisson random variables $X_1,X_2,\ldots,X_n$ with parameter $\lambda = \frac{m}{n} = K (1 - o(1))\sqrt{\log n}$. For a single Poisson random variable $X$, we lower bound the probability that $X \geq u$ for $u = (K + 2 \cdot C) \sqrt{\log n}$,
\begin{align*}
\Pro{X \geq u} & \geq \Pro{X = u} = \frac{e^{-\lambda} \lambda^u}{u!} \geq \frac{e^{-\lambda} \lambda^u}{eu(u /e)^u} = e^{-\lambda + u - 1 - \log{u} } \left(\frac{\lambda}{u} \right)^u \\
 & \geq \exp{\left( (K+2\cdot C)\sqrt{\log n} \cdot \log{\left( \frac{K (1 - o(1))}{K+2\cdot C} \right)} \right)} \\
 & \geq \exp{(-0.8 (K+2\cdot C)\sqrt{\log n} )} > \exp{(-0.2\sqrt{\log n} )},
\end{align*}
where the penultimate inequality used $\log{\left( \frac{K (1 - o(1))}{(K+2\cdot C)} \right)} > -0.8$. %
Using \cref{lem:multiplicative_factor_chernoff_for_binomial}, this implies that w.p.\ $1 - o(n^{-2})$ at least $n e^{-0.20 \sqrt{\log n} - 1} \geq n e^{-0.21 \sqrt{\log n}}$ bins have load at least $(K + 2\cdot C)\sqrt{\log{n}}$, so at least $e^{-0.21 \sqrt{\log n}} \cdot C n\sqrt{\log{n}}$ balls have height at least $(K + C)\sqrt{\log{n}}$.
\end{proof}

\end{document}